\documentclass[aps, tightenlines, floatfix, twocolumn, amsmath, superscriptaddress, showpacs, showkeys]{revtex4-2}
\usepackage{graphicx}
\usepackage{dcolumn}
\usepackage{bm}
\usepackage{float}
\usepackage{hyperref}
\usepackage{bigints}
\usepackage{xcolor}
\usepackage{threeparttable}

\begin{document}


\title{Cosmic Colliders in Binary Blazars Probing Ultra-High-Energy Particle Physics}

\author{Giorgio Galanti}
\email{gam.galanti@gmail.com}
\affiliation{INAF, Istituto di Astrofisica Spaziale e Fisica Cosmica di Milano, Via Alfonso Corti 12, I -- 20133 Milano, Italy}
\affiliation{INFN, Sezione di Pavia, Via Agostino Bassi 6, I -- 27100 Pavia, Italy}

\date{\today}
\begin{abstract}
While terrestrial colliders and fixed-target cosmic-ray interactions are bounded at center-of-mass (CM) energies $\sqrt{s} \lesssim \mathcal{O}(10^5)\,\text{GeV}$, we demonstrate that sub-parsec binary blazars form broad-band hadronic ``{\it Cosmic Colliders}'', overcoming this limitation. In the compressed overlap region of two nearly aligned jets ($\theta_{\rm mis} \sim 1^\circ - 5^\circ$), relativistic proton populations collide at CM energies reaching $\sqrt{s} \sim 10^{10}\,\text{GeV}$ -- six orders of magnitude above the Large Hadron Collider. Operating in an optically thin regime ($\tau_{pp} \ll 1$), the system performs a continuous CM energy scan, probing asymptotic QCD scaling, gluon saturation, and Froissart unitarity, while opening a possible high-sensitivity window into physics beyond the Standard Model. Secondary TeV-EeV neutrinos and GeV-TeV $\gamma$-rays fall within the reach of current and planned observatories such as LHAASO, CTAO, IceCube, and GRAND, establishing binary blazars as laboratories for physics at the ultimate energy limit.
\end{abstract}
 
\maketitle


{\it Introduction} -- Progress in fundamental particle physics is severely bottlenecked by the energy limits of terrestrial technology~\cite{LHCPhysics,FCCCDR}. With the Large Hadron Collider (LHC) operating at the center-of-mass (CM) energy $\sqrt{s} \simeq 13.6 \, \rm TeV$ and proposed future circular colliders constrained by immense technical and financial hurdles~\cite{FCCCDR}, directly probing subatomic physics beyond the ${\cal O}(10^4) \, \rm GeV$ scale appears impossible on Earth. While cosmic rays provide particles, and in particular protons, with energies up to $E_p \sim 10^{20} \, \rm eV$~\cite{AugerSpectrum,TASpectrum}, their interactions with stationary target protons of mass $m_p$ produce a severe CM energy suppression ($\sqrt{s} \simeq \sqrt{2m_p c^2 E_p}$), restricting observables to $\sqrt{s} \lesssim {\cal O}(10^5) \, \rm GeV$~\cite{BlockHadronic,Auger:2012}.

In this Letter, we demonstrate that nature circumvents technological limits and fixed-target kinematic penalties inside sub-parsec binary blazars, where co-propagating relativistic jets -- composed of protons, leptons, hadronic nuclei, and radiation~\cite{jet1,Blandford2019} -- form broad-band ``{\it Cosmic Colliders}''. In the compressed overlap region of two nearly aligned jets ($\theta_{\rm mis} \sim 1^\circ-5^\circ$), quasi-isotropic ultra-high-energy (UHE) proton populations collide at large relative angles -- a process inefficient in isolated, low-density jets -- accessing CM energies up to $\sqrt{s} \sim 10^{10} \, \rm GeV$, {\it six orders of magnitude} beyond the LHC and five orders beyond fixed-target cosmic-ray interactions.

Supermassive black hole binaries (SMBHBs) with masses $M \sim (10^6-10^9) \, M_\odot$ are a natural end-state of galaxy mergers~\cite{Begelman1980,Colpi2014,DeRosa2019}. When both supermassive black holes (SMBHs) launch relativistic jets pointing near our line of sight, a binary blazar is established. Dual nuclear jet structures recently resolved in blazar Mrk~501~\cite{Mrk501dualjet} show that such configurations exist in the local Universe. When these highly collimated jets overlap at sub-parsec scales, we consider a benchmark interaction zone extending from $r_0 = {\cal O}(0.1-1) \, \rm pc$ to $r_f = {\cal O}(10) \, \rm pc$ and a proton density profile $n_p(r) = n_{p,0} (r/r_0)^{-0.5}$, with a baseline density $n_{p,0} = {\cal O}(10^4) \, \rm cm^{-3}$ (justified below), wherein both jets' proton populations collide at modest $\theta_{\rm mis}$.

Assuming a representative proton-proton ($pp$) inelastic cross section $\sigma_{pp}^{\rm inel} \simeq 5 \times 10^{-26} \, \rm cm^2$~\cite{PDG}, the hadronic optical depth integrates to $\tau_{pp} = {\cal O}(10^{-3})$. This establishes a unique collision regime: interactions are efficient yet deeply optically thin ($\tau_{pp} \ll 1$), ensuring that the system acts as a non-calorimetric collider where pion-decay products ($p + p \to p + p + k\pi$, $k \ge 1$)~\cite{PDG,Kelner2006} escape the source without internal particle cascading or quenching.

Because the colliding populations are multi-energetic rather than monochromatic, the system operates as a broad-spectrum CM energy scanner across nine orders of magnitude in $\sqrt{s}$. This broad-band collider transforms high-energy astrophysics into a direct laboratory for fundamental physics, enabling tests of asymptotic QCD scaling~\cite{AltarelliScaling}, ultra-low-$x$ parton distribution functions~\cite{PDF}, gluon saturation regimes~\cite{gluonSaturation}, hadronization models~\cite{hadronization}, and Froissart-bound unitarity~\cite{froissart,Martin:1966,BlockHadronic} (see Appendix). By accessing CM energies up to $\sqrt{s} \sim 10^{10}\,\mathrm{GeV}$, this framework also offers a unique window to probe physics beyond the Standard Model~\cite{PDG}, including possible signatures of String Theory~\cite{string1,string2,string3}, Loop Quantum Gravity~\cite{LQG1,LQG2,LQG3}, and Extra Dimensions~\cite{KK1,KK2}. The resulting secondary multi-messenger fluxes -- broad-band TeV-EeV neutrinos and GeV-TeV gamma rays -- fall directly within the reach of current and planned observatories ({\it Fermi}-LAT~\cite{Fermi}, CTAO~\cite{CTAO}, LHAASO~\cite{LHAASO}, HAWC~\cite{HAWC}, IceCube~\cite{IceCube}, KM3NeT~\cite{KM3NeT}, IceCube-Gen2~\cite{IceCubeGen2}, GRAND~\cite{GRAND}), establishing an immediate observational bridge between blazar astrophysics and fundamental physics at the ultimate energy frontier.

\begin{figure}[t]
\begin{center}
\includegraphics[width=.49\textwidth]{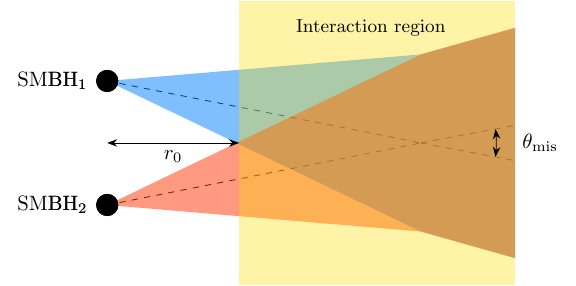}
\end{center}
\caption{\label{CCfig} Schematic of the Cosmic Collider: interaction of two relativistic jets in a sub-parsec SMBHB. The interaction onset distance $r_0$, the interaction region, and the misalignment angle $\theta_{\rm mis}$ are indicated.}
\end{figure}

{\it The binary blazar system} -- UHE $pp$ collisions occur within the overlap region of two relativistic jets launched by a bound SMBHB with comparable masses $M_1 \sim M_2 = {\cal O}(10^8) \, M_\odot$, as shown in Fig.~\ref{CCfig}.
We focus on the hard-binary stage at separation $d_{\rm bin} = \mathcal{O}(0.1) \, \rm pc$. With an orbital period $P_{\rm bin} \sim \mathcal{O}(10^2) \, \rm yr$ and gravitational-wave inspiral timescales $t_{\rm GW} \gtrsim \mathcal{O}(10^{8}) \, \rm yr$, the configuration remains effectively stationary over multi-year observational campaigns (see Supplemental Material~\cite{SupMat}).

The jets propagate with bulk Lorentz factors $\Gamma_{b,1} \simeq \Gamma_{b,2} = \Gamma_b \sim 10-20$~\cite{footnotegammab}. Given an effective semi-aperture angle $\theta_j^{\rm eff} \sim 0.1 \, \rm rad$~\cite{Pushkarev2009}, the lateral expansion causes the jets to intersect at an onset distance $r_0 \simeq d_{\rm bin} /(2 \tan \theta_j^{\rm eff}) = {\cal O}(0.1 - 1) \, \rm pc$, where protons have already achieved maximum acceleration via standard shock~\cite{Drury1983,BlandfordEichler1987,Marcowith2016} or magnetic reconnection~\cite{SironiSpitkovsky2014} mechanisms in the stronger upstream magnetic fields (see Supplemental Material~\cite{SupMat}). Circumbinary accretion drives the SMBH spins toward alignment via Bardeen-Petterson torques, naturally sustaining a modest residual jet misalignment $\theta_{\rm mis} \sim 1^\circ - 5^\circ$~\cite{BardeenPetterson1975,Bogdanovic2007,Dotti2010} (see Supplemental Material~\cite{SupMat}). Beyond the interaction region ($r > r_f$), kinematic consistency can yield a single merged flow with an enhanced-density core, an effective bulk Lorentz factor $\Gamma_{b,f} \simeq \Gamma_b$, and a semi-aperture angle $\theta_{j,f} \simeq 1/\Gamma_b + \theta_{\rm mis}$.

At $r \ge r_0$, the overlapping jets undergo mutual hydrodynamic compression. Building upon the physics of oblique relativistic shocks~\cite{Daly1988,KomissarovFalle1997,Komissarov2007,AsadaNakamura2012}, the jet-jet interaction forms a compressed core of radius $R_{pp}(r_0) = R_{\rm jet}(r_0)/\xi_{\rm jet}$ with jet radius $R_{\rm jet}$ and compression factor $\xi_{\rm jet} = {\cal O}(10)$ (see Supplemental Material~\cite{SupMat}), mitigating rapid geometric dilution ($n_p \propto r^{-2}$) over $\mathcal{O}(10) \, \rm pc$ scales.

Within the interaction core of radius $R_{pp}$ over $r_0 \le r \le r_f$, for each jet $i \in \{1,2\}$ the proton spectral number density $n_{p,i}^{(i)}$ (see~\cite{footNotation} for notation) -- representing the differential number of protons ${\rm d}^3N_{p,i}^{(i)}$ per unit comoving energy $E_{p,i}^{(i)}$, volume $V_i^{(i)}$, and solid angle $\Omega_i^{(i)}$ -- is isotropic in its rest frame and modeled as
\begin{equation}
\begin{aligned}
n_{p,i}^{(i)}(E_{p,i}^{(i)},r) &\equiv \frac{{\rm d}^3N_{p,i}^{(i)}}{{\rm d}E_{p,i}^{(i)}{\rm d}V_i^{(i)}{\rm d}\Omega_i^{(i)}} = \frac{K_{p,i}}{4\pi}\left(\frac{E_{p,i}^{(i)}}{m_p c^2}\right)^{-\alpha_{p,i}} \\
& \quad \times \exp\left(- \frac{E_{p,i}^{(i)}}{\gamma_{p,i}^{\rm max}m_pc^2} \right)
\left(\frac{r}{r_0}\right)^{-a}~.
\end{aligned}
\label{np}
\end{equation}
The normalization $K_{p,i}$ relates to the total integrated comoving proton number density $n_{p,i,0}^{(i)}$ at $r_0$ via $n_{p,i,0}^{(i)} = K_{p,i} m_p c^2 \int_{\gamma_{p,i}^{\rm min}}^{\infty} \gamma_{p,i}^{-\alpha_{p,i}} \exp(-\gamma_{p,i}/\gamma_{p,i}^{\rm max})\, {\rm d}\gamma_{p,i}$, while $\alpha_{p,i}$ is the spectral index, and $\gamma_{p,i}^{\rm min}=1$ and $\gamma_{p,i}^{\rm max}$ are the minimum and maximum proton Lorentz factors, respectively. The index $a$ in Eq.~\eqref{np} models jet-jet mutual hydrodynamic compression within $R_{pp}$ (see Supplemental Material~\cite{SupMat}). We adopt $a = 0.5$, $\alpha_{p,i} = 2.2$, and $\gamma_{p,i}^{\rm max} = 10^{10}$ as benchmark reference and scan $a \in [0, 1.2]$~\cite{footnoteAequal2}, $\alpha_{p,i} \in [2.1, 2.4]$, and $\gamma_{p,i}^{\rm max} \in [10^8, 10^{11}]$ to evaluate uncertainties. For baseline calculations, we assume symmetric jets and proton distributions, exploring asymmetric configurations in the Supplemental Material~\cite{SupMat}.


{\it Proton-proton collisions} -- We evaluate the efficiency of $pp$ collisions within the interaction core, where each projectile proton targets the protons of both jets~\cite{footnoteSingleJet}. In the jet 1 comoving frame, the four-momenta of a projectile from jet $i$ and a target from jet $j$ are $k_i^{(1)}=(E_{p,i}^{(1)},{\bf p}_i^{(1)}c)$ and $\tilde{k}_j^{(1)}=(\tilde{E}_{p,j}^{(1)},\tilde{\bf p}_j^{(1)}c)$, respectively, with $i,j \in \{1,2\}$. The Mandelstam invariant $s \equiv (k_i^{(1)}+\tilde{k}_j^{(1)})^2$ then reads
\begin{equation}
s = 2\left[ (m_p c^2)^2 + E_{p,i}^{(1)}\tilde{E}_{p,j}^{(1)} - (p_i^{(1)}c) (\tilde{p}_j^{(1)}c) \cos\theta^{(1)} \right]~,
\label{ss}
\end{equation}
where $\theta^{(1)}$ is the collision angle in the comoving frame of jet 1, and we used the dispersion relations $\big(E_{p,i}^{(1)}\big)^2 = m_p^2 c^4 + (p_i^{(1)}c)^2$. The CM energy $\sqrt{s}$ determines the inelastic cross section $\sigma_{pp}^{\rm inel}(s)$~\cite{footnoteSigmaHeaviside}. We anchor $\sigma_{pp}^{\rm inel}(s)$ to experimental accelerator and cosmic-ray data up to $\sqrt{s} \sim 10^5\,\mathrm{GeV}$~\cite{PDG} and extrapolate it to the extreme CM energies attained by the Cosmic Collider -- reaching $\sqrt{s} \sim 10^{10}\,\mathrm{GeV}$ (see below) -- via a Froissart-Martin-inspired scaling $\sigma_{pp}^{\rm inel}(s)\propto \ln^2(s/s_{\rm th})$ that respects unitarity bounds~\cite{froissart,Martin:1966} (see Supplemental Material~\cite{SupMat}).

Using $\sigma_{pp}^{\rm inel}(s)$ and Eq.~\eqref{np}, we compute the optical depth $\tau_{pp}$ for a projectile proton of jet 1 with energy $E_{p,1}^{(1)}$. In the comoving frame of jet 1, the target distribution of jet 2 follows from phase-space invariance~\cite{RybickiLightman1979}: $n_{p,2}^{(1)}(\tilde{E}_{p,2}^{(1)}, \mu_t, r) = [\tilde{p}_2^{(1)}\!\tilde{E}_{p,2}^{(1)}/(\tilde{p}_2^{(2)}\!\tilde{E}_{p,2}^{(2)})] \, n_{p,2}^{(2)}(\tilde{E}_{p,2}^{(2)}, r)$, with $\tilde{E}_{p,2}^{(2)} = \Gamma_{\rm rel}(\tilde{E}_{p,2}^{(1)} - \beta_{\rm rel}\, \tilde{p}_2^{(1)}c\, \mu_t)$, $\tilde{p}_2^{(2)}$ the corresponding momentum, $\mu_t$ the cosine of the angle between the target momentum and the jet 2 velocity, and $\Gamma_{\rm rel}(\theta_{\rm mis})$ the relative Lorentz factor of the two jets ($\beta_{\rm rel} = \sqrt{1 - 1/\Gamma_{\rm rel}^2}$)~\cite{footnoteGammaRel}. Since the projectile distribution is isotropic, only the angle-averaged densities $\tilde{n}_{p,j}^{(1)} = \frac{1}{2}\int_{-1}^{1}{\rm d}\mu_t\, n_{p,j}^{(1)}$ enter, with $\tilde{n}_{p,1}^{(1)} = n_{p,1}^{(1)}$. The interaction region is stationary in the host-galaxy frame, with the jet plasma flowing through it: a proton crossing ${\rm d}r$ covers a comoving path length ${\rm d}r/\Gamma_b$ within it~\cite{SteadyJet2021}, so that
\begin{equation}
\begin{aligned}
\tau_{pp}(E_{p,1}^{(1)}) &= \frac{2\pi}{\Gamma_b} \sum_{j=1}^{2} \int_{r_0}^{r_f} \! {\rm d}r \! \int_{-1}^1 \! {\rm d}(\cos\theta^{(1)}) \! \int_{m_pc^2}^{\infty} \! {\rm d}\tilde{E}_{p,j}^{(1)} \, \tilde{n}_{p,j}^{(1)} \\ 
& \quad \times \sigma_{pp}^{\rm inel}(s) \, (1-\beta_1^{(1)}\tilde{\beta}_j^{(1)} \cos \theta^{(1)})~,
\end{aligned}
\label{tau}
\end{equation}
where the comoving proton velocities are $\beta_1^{(1)} = \beta(E_{p,1}^{(1)})$ and $\tilde{\beta}_j^{(1)} = \beta(\tilde{E}_{p,j}^{(1)})$ (in units of $c$), and the factor $(1-\beta_1^{(1)}\tilde{\beta}_j^{(1)}\cos\theta^{(1)})$ accounts for their relative motion.

\begin{figure}
\begin{center}
\includegraphics[width=.47\textwidth]{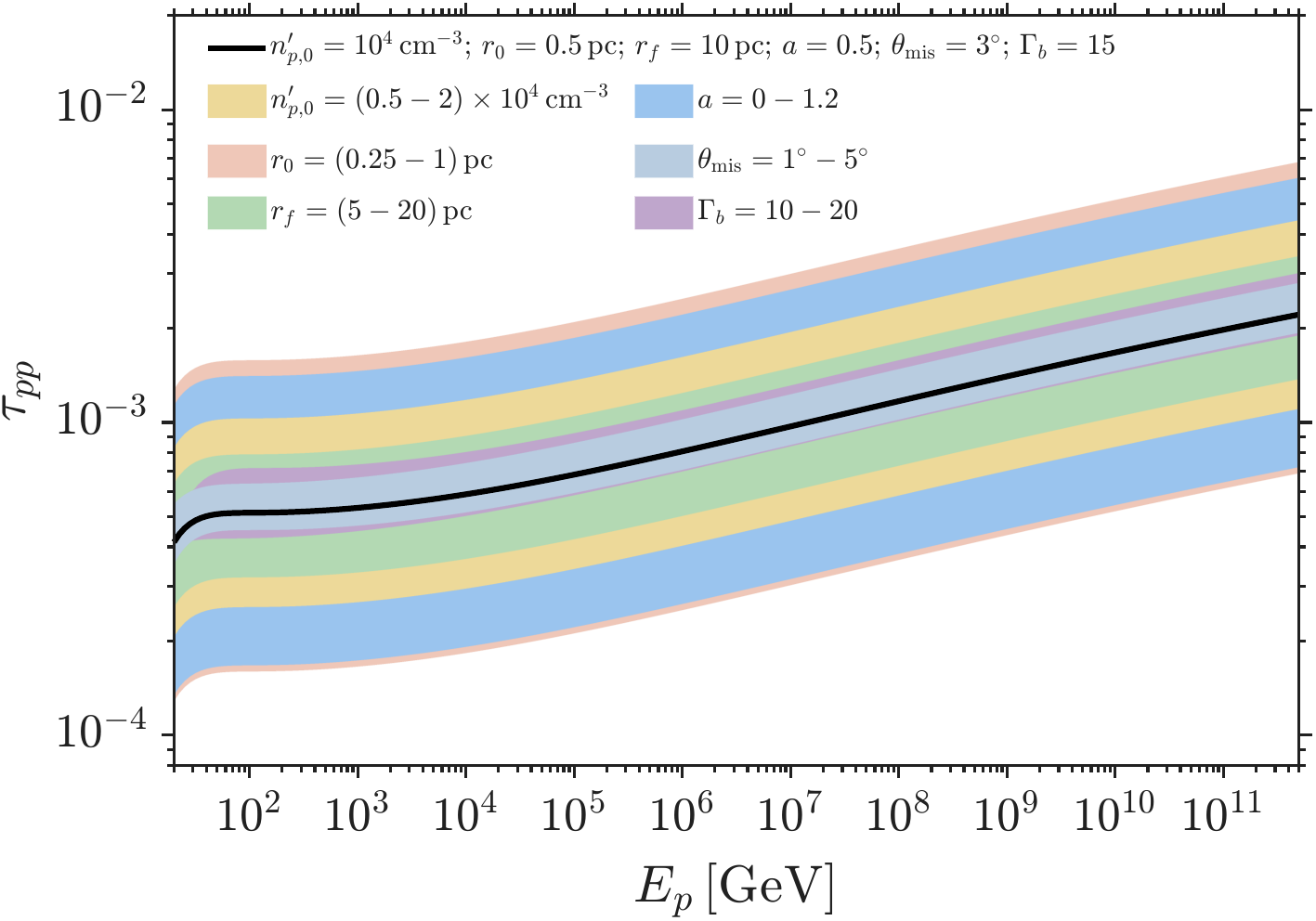}
\end{center}
\caption{\label{taufig} Proton-proton optical depth $\tau_{pp}$ as a function of the host-galaxy frame proton energy $E_p$. The solid black curve depicts our benchmark model, while shaded bands show sensitivity to individual variations in $n_{p,0}'$, $r_0$, $r_f$, $a$, $\theta_{\rm mis}$, and $\Gamma_b$ (see legend). For visual clarity, sensitivity bands are displayed with full opacity and stacked by decreasing width.}
\end{figure}

Figure~\ref{taufig} shows $\tau_{pp}$ as a function of the host-galaxy frame proton energy $E_p = \mathcal{D}_D E_{p,1}^{(1)} \simeq \Gamma_b E_{p,1}^{(1)}$~\cite{footnoteDoppler,footnote2}.
For our benchmark setup ($n_{p,0}' \equiv n_{p,1,0}^{(1)} = n_{p,2,0}^{(2)} = 10^4 \,\rm{cm^{-3}}$, $r_0=0.5 \,\rm{pc}$, $r_f= 10 \,\rm{pc}$, $a=0.5$, $\theta_{\rm mis}=3^\circ$, and $\Gamma_b=15$), we find $5 \times 10^{-4} \lesssim \tau_{pp} \lesssim 2 \times 10^{-3}$ for $E_p \gtrsim 30\,\mathrm{GeV}$, with a mild logarithmic increase at high energies. Within the astrophysical uncertainty ranges -- $n_{p,0}' = (0.5 - 2) \times 10^4 \, \rm cm^{-3}$, $r_0 = (0.25 - 1) \, \rm pc$, $r_f = (5 - 20) \, \rm pc$, $a = 0 - 1.2$, $\theta_{\rm mis}=1^{\circ} - 5^{\circ}$, and $\Gamma_b = 10 - 20$ -- the optical depth remains robust, spanning $10^{-4} \lesssim \tau_{pp} \lesssim 7 \times 10^{-3}$, while variations in the injection parameters ($\alpha_{p,i} \in [2.1, 2.4]$, $\gamma_{p,i}^{\rm max} \in [10^8, 10^{11}]$)~\cite{protonAcc1,protonAcc2,protonAcc3} leave it unaffected (see Supplemental Material~\cite{SupMat}). Across the whole parameter space, $pp$ interactions sustain an active hadronic collider while $\tau_{pp} \ll 1$, so that secondary multi-messenger signatures escape without intra-source cascades or reprocessing.

Competing energy-loss channels -- $p\gamma$ interactions and proton synchrotron -- do not hamper $pp$ interactions. For binary systems composed of BL Lac objects (BL Lacs), $pp$ interactions govern the proton energy-loss budget up to host-galaxy frame proton energies $E_p \sim 10^{11}\,\mathrm{GeV}$, with photohadronic losses negligible (integrated energy-loss fraction $f_{p\gamma} < 0.1$) at all energies. Even for luminous flat spectrum radio quasars (FSRQs), where dense external fields (broad-line region, dusty torus) boost $p\gamma$ above $pp$ for $E_p \gtrsim 3 \times 10^7\,\mathrm{GeV}$, the condition $f_{p\gamma} \lesssim 0.1$ holds up to $E_p \sim 10^9 \,\mathrm{GeV}$. No severe beam depletion thus occurs, keeping the $pp$-driven Cosmic Collider fully functional up to CM energies far exceeding terrestrial limits (see Supplemental Material~\cite{SupMat}).


{\it Particle production rate} -- Inelastic $pp$ interactions produce pions (of mass $m_\pi$) via $p + p \to p + p + k\pi$ [$k \ge 1$ and threshold $s_{\rm th} = (2m_p c^2 + m_\pi c^2)^2$], which decay into secondary photons ($\gamma$) and neutrinos ($\nu$). To quantify the Cosmic Collider microscopic mechanics, we evaluate particle production rates differential in both secondary comoving energy ($E_{\gamma,\nu}' \equiv E_{\gamma,\nu}^{(1)}$ in the jet 1 frame, with $E_\gamma$ and $E_\nu$ denoting photon and neutrino energies in the host-galaxy or Earth frame, as specified) and CM energy $\sqrt{s}$. Assuming the benchmark source setup established above, the double-differential production rate~\cite{footnoteRate} reads
\begin{equation}
\begin{aligned}
\frac{{\rm d}^2N_{\gamma,\nu}}{{\rm d}E_{\gamma,\nu}^{\prime}{\rm d}\sqrt{s}} &= 4\pi^2 c \sum_{i,j=1}^{2}\int_{r_0}^{r_f} A(r) {\rm d}r \int_{-1}^1 {\rm d}(\cos\theta^{(1)}) \\
& \quad \times \int_{m_pc^2}^{\infty} {\rm d}E_{p,i}^{(1)}\int_{m_pc^2}^{\infty} {\rm d}\tilde{E}_{p,j}^{(1)} \, n_{p,i}^{(1)} \, \tilde{n}_{p,j}^{(1)} \\
& \quad \times \frac{{\rm d}\sigma_{pp \to \gamma,\nu}}{{\rm d}E_{\gamma,\nu}^{\prime}}(s, E_{\gamma,\nu}^{\prime}) \, (1-\beta_i^{(1)}\tilde{\beta}_j^{(1)} \cos \theta^{(1)}) \\
& \quad \times \delta\left(\sqrt{s} - \sqrt{s(E_{p,i}^{(1)}, \tilde{E}_{p,j}^{(1)}, \cos\theta^{(1)})}\right)~,
\end{aligned}
\label{rate}
\end{equation}
where the sum runs over projectile and target jets~\cite{footnoteSingleJet}, and $A(r) = \pi R_{pp}^2(r_0)(r/r_0)^a$ is the cross-sectional area of the compressed interaction core [$R_{pp}(r_0)=R_{\rm jet}(r_0)/\xi_{\rm jet}$ with benchmark value $\xi_{\rm jet} = 10$ (see Supplemental Material~\cite{SupMat})], ensuring particle flux conservation~\cite{footnote3}. The differential cross sections ${\rm d}\sigma_{pp \to \gamma,\nu}/{\rm d}E_{\gamma,\nu}^{\prime}$ enforce the threshold via $\Theta(s-s_{\rm th})$ \cite{footnoteSigmaHeaviside} and adapt standard parameterizations \cite{kafexhiu} to dual-relativistic proton distributions (see Appendix). The Dirac delta function isolates specific CM energy slices, with $s$ given by Eq.~\eqref{ss}.

\begin{figure}[h]
\begin{center}
\includegraphics[width=.47\textwidth]{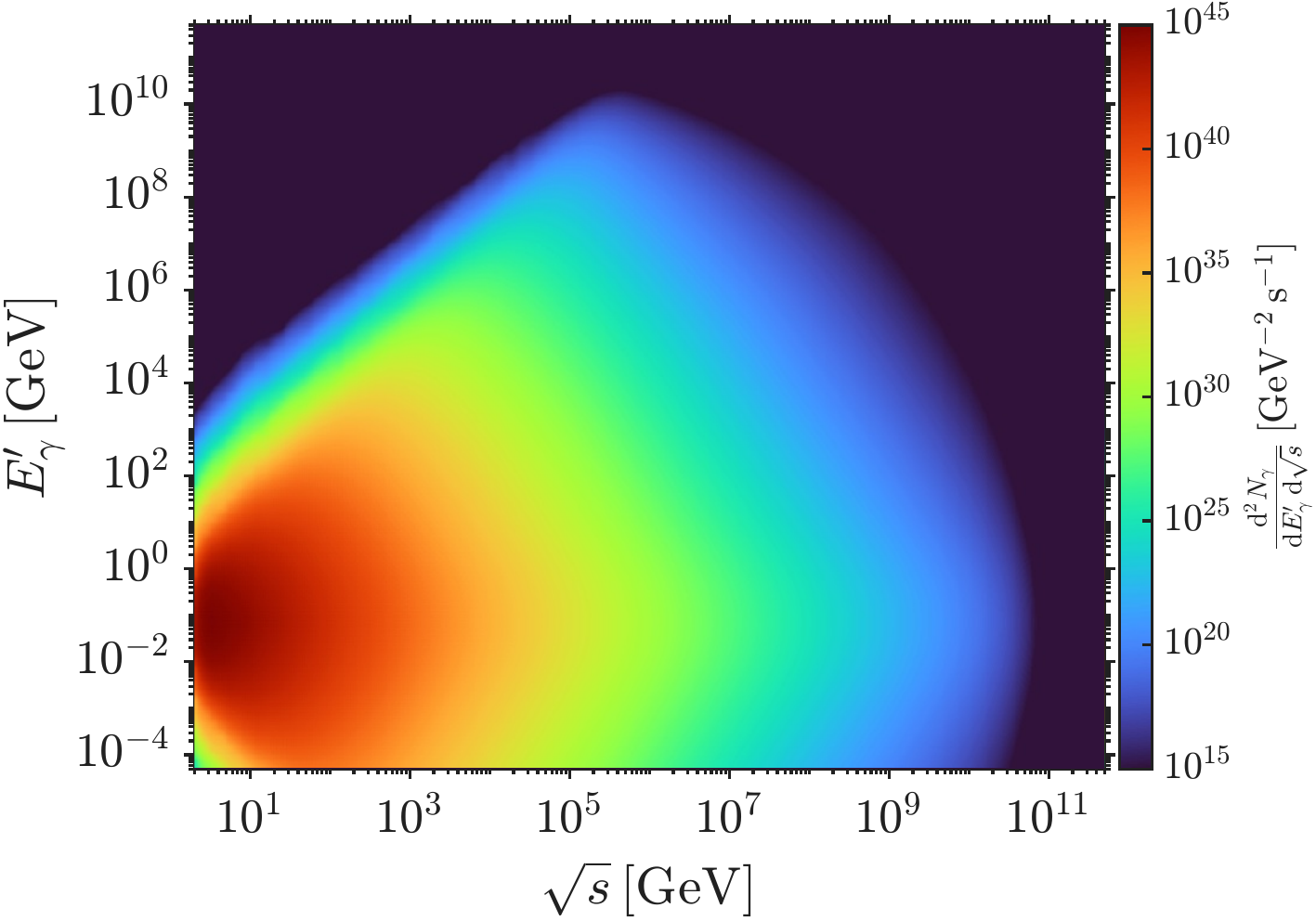}
\end{center}
\caption{\label{HeatMapFig} Double-differential photon production rate ${\rm d}^2N_{\gamma}/{\rm d}E_{\gamma}^{\prime}{\rm d}\sqrt{s}$ as a function of the photon energy $E_{\gamma}^{\prime}$ in the jet 1 comoving frame and of the CM energy $\sqrt{s}$.}
\end{figure}

Figure~\ref{HeatMapFig} maps the double-differential photon rate ${\rm d}^2N_{\gamma}/{\rm d}E_{\gamma}^{\prime}{\rm d}\sqrt{s}$ (neutrinos exhibit an analogous distribution), revealing the operational phase space of the Cosmic Collider. The production rate peaks in the region $3 \, {\rm GeV} \lesssim \sqrt{s} \lesssim 5 \times 10^3 \, {\rm GeV}$ and $E_{\gamma}^{\prime} \lesssim 10^2 \, \rm GeV$. This dominant yield is driven by the power-law proton distributions $n_{p,i}^{(i)} \propto (E_{p,i}^{(i)})^{-2.2}$ [Eq.~\eqref{np}], which favor collisions from $s_{\rm th}$ through intermediate $\sqrt{s}$. At low $\sqrt{s} \sim 10 \, {\rm GeV}$, photons can still reach energies $E_{\gamma}^{\prime} \sim 10^4 \, {\rm GeV}$. This stems from quasi-collinear scatterings where highly energetic protons ($E_{p,i}^{(1)}, \tilde{E}_{p,j}^{(1)} \gtrsim 10^5\,\mathrm{GeV}$) collide at grazing angles ($\theta^{(1)} \lesssim 0.01^\circ$), suppressing $\sqrt{s}$ while boosting the secondary Lorentz factor. The sharp diagonal boundary rising toward higher $E_\gamma'$ reflects the expanding kinematic phase space: as $\sqrt{s}$ increases, larger collision angles become available, significantly enhancing the phase-space volume for producing UHE secondaries. 

Remarkably, the maximum energy of secondaries is $E_{\gamma}' \sim 10^{10}\, \mathrm{GeV}$ around $\sqrt{s} \sim (10^5 - 10^6) \,\mathrm{GeV}$. This occurs because a proton carrying the cutoff energy $\sim 10^{10}\,\mathrm{GeV}$ collides with the abundant low-energy proton population of both jets. At $\sqrt{s} \gtrsim 10^7\,\mathrm{GeV}$, reaching such large CM energies requires both colliding protons to originate from the depleted high-energy spectral tail [see Eq.~(\ref{np})], naturally reducing the interaction rate. Despite this spectral suppression, the Cosmic Collider actively populates the phase space up to CM energies of $\sqrt{s} \sim 10^{10}\,\mathrm{GeV}$, generating UHE secondary photons and neutrinos that directly probe physics {\it six orders of magnitude} beyond terrestrial colliders.


{\it Multi-messenger particle spectra} -- Integrating the differential rate ${\rm d}^2N_{\gamma, \nu}/{\rm d}E_{\gamma, \nu}^{\prime}{\rm d}\sqrt{s}$ [Eq.~\eqref{rate}] over CM energy or secondary energy yields the intrinsic luminosity distributions of the Cosmic Collider: the secondary particle luminosity $E_{\gamma, \nu}^{\prime} L'_{E_{\gamma, \nu}^{\prime}} \equiv E_{\gamma, \nu}^{\prime 2} {\rm d}N_{\gamma, \nu}/{\rm d}E_{\gamma, \nu}^{\prime}$ and collider luminosity $\sqrt{s} L_{\sqrt{s}} \equiv s \, {\rm d}N_{\gamma, \nu}/{\rm d}\sqrt{s}$. Notably, $\sqrt{s} L_{\sqrt{s}}$ at extreme CM energies ($\sqrt{s} \sim 10^9\,\mathrm{GeV}$) drops by only about four orders of magnitude compared to its peak value at $\sqrt{s} \sim 10\,\mathrm{GeV}$, demonstrating remarkable production efficiency across the entire phase space (see Appendix).

The observed spectral energy distribution (SED) $E_{\gamma, \nu}^2{\cal F}_{\gamma, \nu}$ accounts for cosmological expansion at redshift $z$, luminosity distance $d_L$, extragalactic propagation, and relativistic jet beaming ($\Gamma_b \simeq 15$) as
\begin{equation}
E_{\gamma, \nu}^2 {\cal F}_{\gamma, \nu} = \frac{\mathcal{T}_{\gamma, \nu}}{4\pi d_L^2} {\cal D}_D^3 E_{\gamma, \nu}^{\prime 2} \frac{{\rm d}N_{\gamma, \nu}}{{\rm d}E_{\gamma, \nu}^{\prime}} \Bigg|_{E_{\gamma, \nu}^{\prime} = \frac{1+z}{{\cal D}_D} E_{\gamma, \nu}}~,
\label{obsSED}
\end{equation}
where ${\cal D}_D^3$ reflects the stationary emission region~\cite{Sikora1997,SteadyJet2021}, and we take ${\cal D}_D \simeq \Gamma_b$~\cite{footnoteDoppler}. The transmission factor $\mathcal{T}_{\gamma, \nu}$ encodes intergalactic attenuation. While the cosmos is fully transparent to neutrinos ($\mathcal{T}_{\nu} = 1$), gamma rays suffer pair-production attenuation [$\mathcal{T}_{\gamma} = \exp(-\tau_{\gamma\gamma})$] against the cosmic microwave background (CMB), radio, and extragalactic background light (EBL)~\cite{dwekkrennrich,footnote4}.
\begin{figure}[t]
\begin{center}
\includegraphics[width=.47\textwidth]{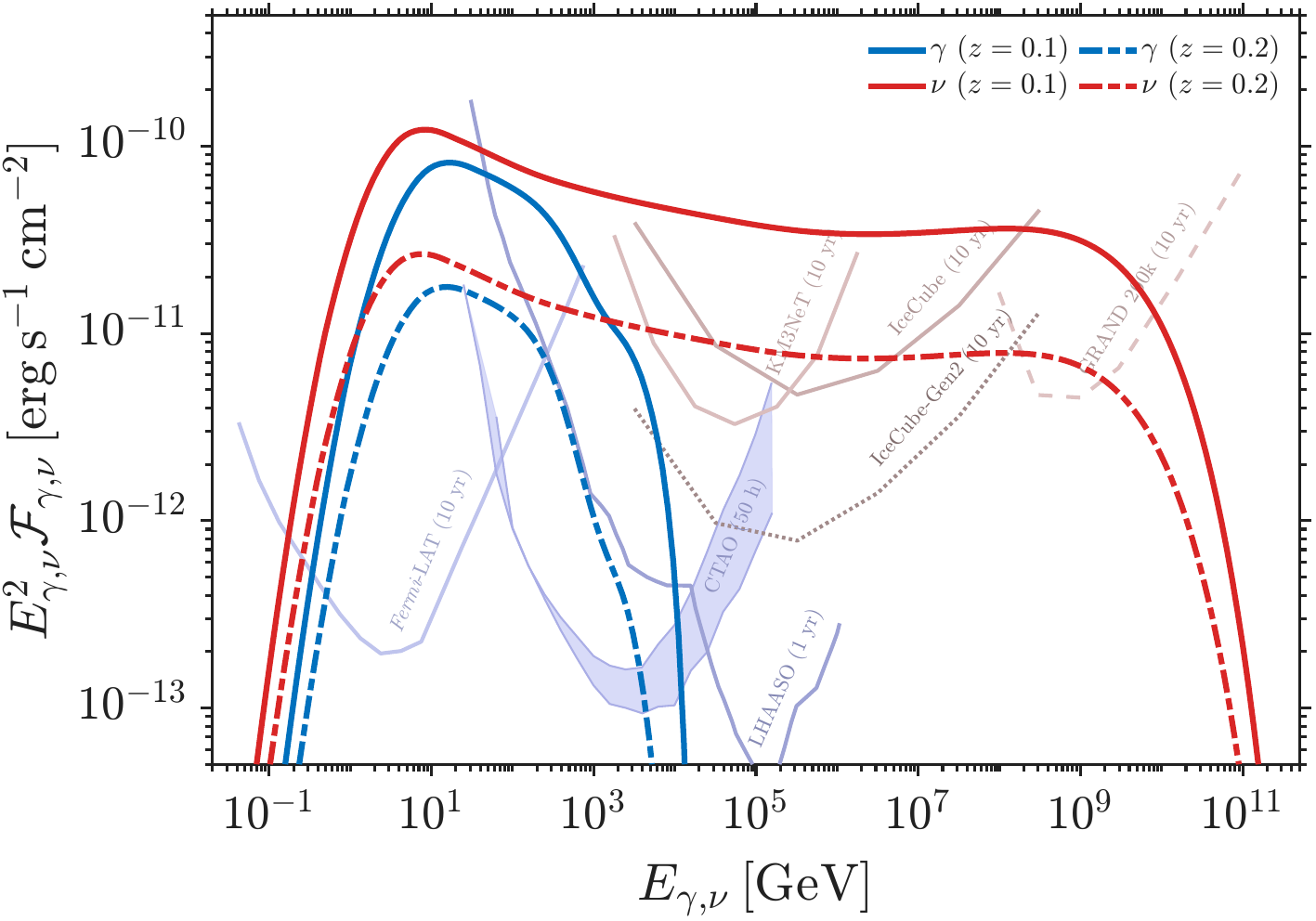}
\end{center}
\caption{\label{GammaNuSEDFig} Projected multi-messenger SEDs for photons (blue) and neutrinos (red) produced by the Cosmic Collider at redshifts $z = 0.1$ (solid) and $z = 0.2$ (dashed), compared against differential sensitivities of leading and planned observatories.}
\end{figure}

Keeping all parameters fixed at their benchmark values, Fig.~\ref{GammaNuSEDFig} shows the projected multi-messenger SEDs for a Cosmic Collider placed at redshifts $z = 0.1$ and $z = 0.2$, overlaid with differential sensitivities of current and future $\gamma$-ray ({\it Fermi}-LAT~\cite{Fermi}, CTAO~\cite{CTAO}, LHAASO~\cite{LHAASO}) and neutrino detectors (IceCube~\cite{IceCube,footnoteIceCube,footnoteFlavor}, KM3NeT~\cite{KM3NeT,footnoteFlavor}, IceCube-Gen2~\cite{IceCubeGen2,footnoteFlavor}, GRAND~\cite{GRAND,footnoteGRAND}).

For $z = 0.1$, very-high-energy photons (solid blue) are detectable by LHAASO and CTAO in the $(10^2 - 10^4)\,\mathrm{GeV}$ window, prior to a sharp EBL-induced exponential cutoff, and by {\it Fermi}-LAT at lower energies. At $z = 0.2$ (dashed blue), EBL absorption shrinks the gamma-ray horizon, leaving sub-TeV observations with CTAO and {\it Fermi}-LAT as the main electromagnetic channel. Conversely, neutrinos (red curves) propagate unattenuated: at $z = 0.2$ the flux lies above the IceCube and KM3NeT sensitivities in the range $(10^4 - 10^{7})\,\mathrm{GeV}$, and above the GRAND sensitivity in the band $(10^8 - 10^{10})\,\mathrm{GeV}$. IceCube-Gen2 will extend the neutrino horizon to $z \sim 0.5$, and to $z \sim 1$ for sources observed close to the jet axis (see Supplemental Material~\cite{SupMat}), making neutrino astronomy the primary probe of UHE particle physics.


{\it Discussion and Conclusions} -- In this Letter, we introduce the ``Cosmic Collider'' framework, establishing sub-parsec SMBHBs hosting dual relativistic jets as UHE $pp$ particle reactors. Far from a drawback, the multi-energetic nature of this system is its greatest asset: rather than scanning energies in incremental steps like monochromatic terrestrial facilities, the Cosmic Collider continuously operates across the entire CM energy spectrum, populating the interaction phase space seamlessly up to $\sqrt{s} \sim 10^{10}\,\mathrm{GeV}$ -- exceeding the LHC by {\it six orders of magnitude}. The required jet power, $L_{\rm jet} = {\cal O}(10^{47})\,\mathrm{erg\,s^{-1}}$ (see Supplemental Material~\cite{SupMat}), is consistent with physically realistic blazar energy budgets.

Experimentally, our framework predicts a striking multi-messenger complementarity. While $\gamma$-rays suffer exponential intergalactic attenuation via pair production ($\gamma\gamma \to e^+e^-$) off the EBL, CMB, and radio backgrounds at high redshifts and energies, secondary neutrinos escape completely unimpeded across cosmological distances. Current detectors like IceCube~\cite{IceCube} and KM3NeT~\cite{KM3NeT} can probe Cosmic Colliders in the energy range $(10^4 - 10^7)\,\mathrm{GeV}$ up to $z \sim 0.25-0.3$, whereas IceCube-Gen2~\cite{IceCubeGen2} will extend this horizon to $z \sim 0.5$, and GRAND~\cite{GRAND} will probe the $(10^8 - 10^{10})\,\mathrm{GeV}$ window up to $z \sim 0.25$.

While systems meeting these collider conditions are rare, they are a natural outcome of galaxy mergers~\cite{Begelman1980,Colpi2014,DeRosa2019}: our demographic analysis indicates $\mathcal{O}(1 - 10)$ active Cosmic Colliders within the reach of IceCube-Gen2 under conservative benchmark assumptions, with higher rates toward the upper limits of jet power and efficiency (see Supplemental Material~\cite{SupMat}). Crucially, even a single active system suffices to establish an unprecedented natural laboratory for fundamental particle physics. Mild asymmetries in jet parameters or proton distributions further imprint distinct high-energy neutrino spectral signatures that set these systems apart from standard single-jet blazars (see Supplemental Material~\cite{SupMat}), while spin precession in SMBHBs (e.g., Lense-Thirring~\cite{BardeenPetterson1975}) can induce long-term modulations of $\theta_{\rm mis}$, providing an additional observational signature. Dual-jet structures, as resolved in Mrk~501 ($z \simeq 0.034$)~\cite{Mrk501dualjet}, and SMBHB candidates such as OJ~287 ($z \simeq 0.306$)~\cite{OJ287bin1,OJ287bin2} and PG~1553+113 ($z \simeq 0.43$)~\cite{PG1553bin1,PG1553bin2} show that the building blocks of the Cosmic Collider exist in nature.

Finally, by opening an unprecedented natural laboratory, the Cosmic Collider bridges blazar astrophysics with UHE particle physics, probing asymptotic QCD regimes~\cite{AltarelliScaling,PDF,hadronization}, gluon saturation~\cite{gluonSaturation}, and Froissart unitarity~\cite{froissart,Martin:1966,BlockHadronic} (as discussed in the Appendix), alongside potential signatures of String Theory~\cite{string1,string2,string3}, Loop Quantum Gravity~\cite{LQG1,LQG2,LQG3}, and Extra Dimensions~\cite{KK1,KK2}. Unveiling these unique multi-messenger signatures with current and next-generation observatories will mark the transition of blazar astrophysics into direct UHE particle physics, testing fundamental interactions at the ultimate energy frontiers of the Universe.

\

{\it Acknowledgments} -- The author would like to thank Fabrizio Tavecchio for crucial discussions and parameter suggestions, Giacomo Bonnoli for his support, and Massimo Dotti and Paola Severgnini for helpful comments. The work of G.G. is supported by a contribution from Grant No. ASI-INAF 2023-17-HH.0.

\

{\it Data Availability} -- The data that support the findings of this study are available from the corresponding author upon request.

\



{\it Appendix} -- We outline the formalism used to evaluate the secondary photon and neutrino production rates originating from hadronic $pp$ interactions ($p + p \to p + p + k\pi$, with $k \ge 1$) occurring within and between the two relativistic proton distributions in the Cosmic Collider binary blazar jets.

Assuming approximate isospin symmetry at high energies, pions are produced with equal ratios, $(\pi^0 : \pi^+ : \pi^-) \approx (1 : 1 : 1)$. The subsequent decay chains ($\pi^0 \to \gamma\gamma$, $\pi^+ \to \mu^+ \nu_\mu \to e^+ \nu_e \nu_\mu \bar{\nu}_\mu$, and $\pi^- \to \mu^- \bar{\nu}_\mu \to e^- \bar{\nu}_e \nu_\mu \bar{\nu}_\mu$) couple the photon and total neutrino production via kinematic scaling relations. In the high-energy limit, secondary particles inherit average energies following $E_\nu^{\prime} \simeq E_\gamma^{\prime} / 2$. Since charged pions (2/3 of the total) transfer $\sim 3/4$ of their energy to neutrinos, while neutral pions transfer all their energy to photons, the secondary production rates are related by
\begin{equation}
E_\nu^{\prime\,2}\frac{{\rm d}^2N_{\nu}}{{\rm d}E_{\nu}^{\prime}{\rm d}\sqrt{s}}\bigg|_{E_\nu^\prime = E_\gamma^\prime/2} \simeq \eta_{\nu\gamma}\, E_\gamma^{\prime\,2}\frac{{\rm d}^2N_{\gamma}}{{\rm d}E_{\gamma}^{\prime}{\rm d}\sqrt{s}}~,
\label{eq:rate_nu_gamma}
\end{equation}
where $\eta_{\nu\gamma} \simeq 3/2$ is the all-flavor neutrino-to-photon energy ratio~\cite{Kelner2006}, accounting for all neutrino species ($\nu_e, \nu_\mu, \bar{\nu}_e, \bar{\nu}_\mu$) from charged pion decays. Standard vacuum oscillations over cosmological distances average the neutrino flavor ratio to an isotropic partition $(\Phi_{\nu_e} : \Phi_{\nu_\mu} : \Phi_{\nu_\tau}) = (1 : 1 : 1)$ at Earth, meaning that each individual neutrino flavor carries one-third of the total flux generated by Eq.~(\ref{eq:rate_nu_gamma}).

This link allows us to extend the parameterization of photon production from $pp$ interactions existing in the literature~\cite{kafexhiu} -- appropriately modified to account for non-stationary target kinematics (see below) -- directly to the total neutrino sector. This mapping remains highly accurate up to UHEs, where alternative or competing decay channels introduce at most a $\sim 10\%$ uncertainty in the overall neutrino yield~\cite{PDG}.

Because the optical depth satisfies $\tau_{pp} \ll 1$ throughout the parameter space of the Cosmic Collider, the interaction zone operates in a strictly optically thin regime. Consequently, internal particle cascades remain negligible, guaranteeing that high-energy secondary photons escape the system without spectral reprocessing, while neutrinos stream out completely unimpeded.

{\it Generalized $pp$ rates for relativistic beams} -- We now extend the photon production parameterization of Ref.~\cite{kafexhiu} -- originally derived for a relativistic proton beam striking a stationary target -- to the general configuration where both colliding proton distributions possess ultra-relativistic velocities.

In the original formulation, the differential cross section ${\rm d}\sigma_{pp \to \gamma}/{\rm d}E_{\gamma}^{\prime}$ is parametrized as a function of the projectile kinetic energy $T_p$ in the rest frame of a stationary target. To apply this parametrization to two moving proton beams, we map the dual-beam collision onto a fictitious reference frame in which one of the colliding protons is at rest. In this effective rest frame, the other proton arrives with an equivalent kinetic energy $T_{p,\text{eq}}$ defined via the Lorentz-invariant CM energy squared $s$ of Eq.~(\ref{ss}) as
\begin{equation}
T_{p, {\rm eq}} \equiv \frac{s - 4 m_p^2 c^4}{2 m_p c^2}~.
\label{eq:Tequiv}
\end{equation}

By exploiting the Lorentz invariance of the cross section, evaluating the interaction in this effective rest frame allows us to directly employ the Kafexhiu et al. parametrization~\cite{kafexhiu}, recasting the differential cross section as a function of the invariant $s$ of Eq.~(\ref{ss}) as
\begin{equation}
\frac{{\rm d}\sigma_{pp \to \gamma}}{{\rm d}E_{\gamma}^{\prime}}(s, E_{\gamma}^{\prime}) \equiv \frac{{\rm d}\sigma_{pp \to \gamma}}{{\rm d}E_{\gamma}^{\prime}}\left( T_{p,{\rm eq}}(s), E_{\gamma}^{\prime} \right)~,
\label{eq:sigma_equiv}
\end{equation}
as used in our double-differential production rates of Eq.~(\ref{rate}) (evaluated in the rest frame of jet 1), where the spectral shape is evaluated with the CM kinematics of the jet 1 frame and its normalization is rescaled to preserve the (Lorentz-invariant) photon yield per collision of the fixed-target parametrization; we verified that this rescaling has a negligible impact on our results.

Strictly speaking, while the double-differential cross section ${\rm d}^2\sigma_{pp \to \gamma} /{\rm d}E_\gamma^{\prime} {\rm d}\Omega^{\prime}$ specifies the angular distribution of the emitted photons, integrating over the solid angle ${\rm d}\Omega^{\prime}$ to yield ${\rm d}\sigma_{pp \to \gamma}/{\rm d}E_\gamma^{\prime}$ as in Ref.~\cite{kafexhiu} introduces a mild reference-frame dependence. In our scenario, this choice is fully justified: at high collision energies ($\sqrt{s} \gg m_p c^2$), secondary hadron production is strongly forward-boosted, and the multi-angular, turbulent nature of the colliding flows effectively isotropizes the emission over the interaction volume. Consequently, the angle-integrated differential cross section ${\rm d}\sigma_{pp \to \gamma}/{\rm d}E_\gamma^{\prime}$ evaluated at $T_{p, {\rm eq}}(s)$ accurately describes the total secondary emissivity in the reference frame of jet 1.

Finally, because the fit in Ref.~\cite{kafexhiu} is calibrated against several Monte Carlo codes~\cite{geant42003,pythia82008,sibyll212009,qgsjetII2006} up to $T_{p,{\rm eq}} \sim 10^5~\text{GeV}$, a naive extrapolation to UHEs introduces an unphysical hyper-logarithmic growth driven by the $A_{\rm max} \propto \exp(b_3 \ln^2\theta_p)$ fitting factor~\cite{kafexhiu}. In the parametrization of Ref.~\cite{kafexhiu}, the amplitude $A_{\rm max}$ acts as an effective normalization factor that tightly couples the growth of the inelastic cross section $\sigma_{pp}^{\rm inel}$ with the average secondary pion multiplicity $\langle n_\pi \rangle$. At UHEs ($T_{p,{\rm eq}} > 10^5\,\mathrm{GeV}$), asymptotic QCD regimes~\cite{AltarelliScaling,PDF,hadronization} and gluon saturation~\cite{gluonSaturation} may suppress both cross-section growth and pion multiplicity. To enforce asymptotic Feynman scaling and naturally incorporate QCD saturation effects, while satisfying the Froissart-Martin unitarity bound~\cite{froissart,Martin:1966} on $\sigma_{pp}^{\rm inel}$ for $T_{p,{\rm eq}} > 10^5\,\mathrm{GeV}$, we apply a smooth power-law damping factor $A_{\rm max} \to A_{\rm max} (T_{p,{\rm eq}} / 10^5 \, {\rm GeV})^{-\eta_{\rm damp}}$, where $\eta_{\rm damp}$ serves as an effective asymptotic index. We adopt $\eta_{\rm damp} = 0.1$ as our baseline benchmark and systematically vary $\eta_{\rm damp} \in [0.05, 0.15]$ to encompass QCD asymptotic uncertainties.

\begin{figure}
\begin{center}
\includegraphics[width=.47\textwidth]{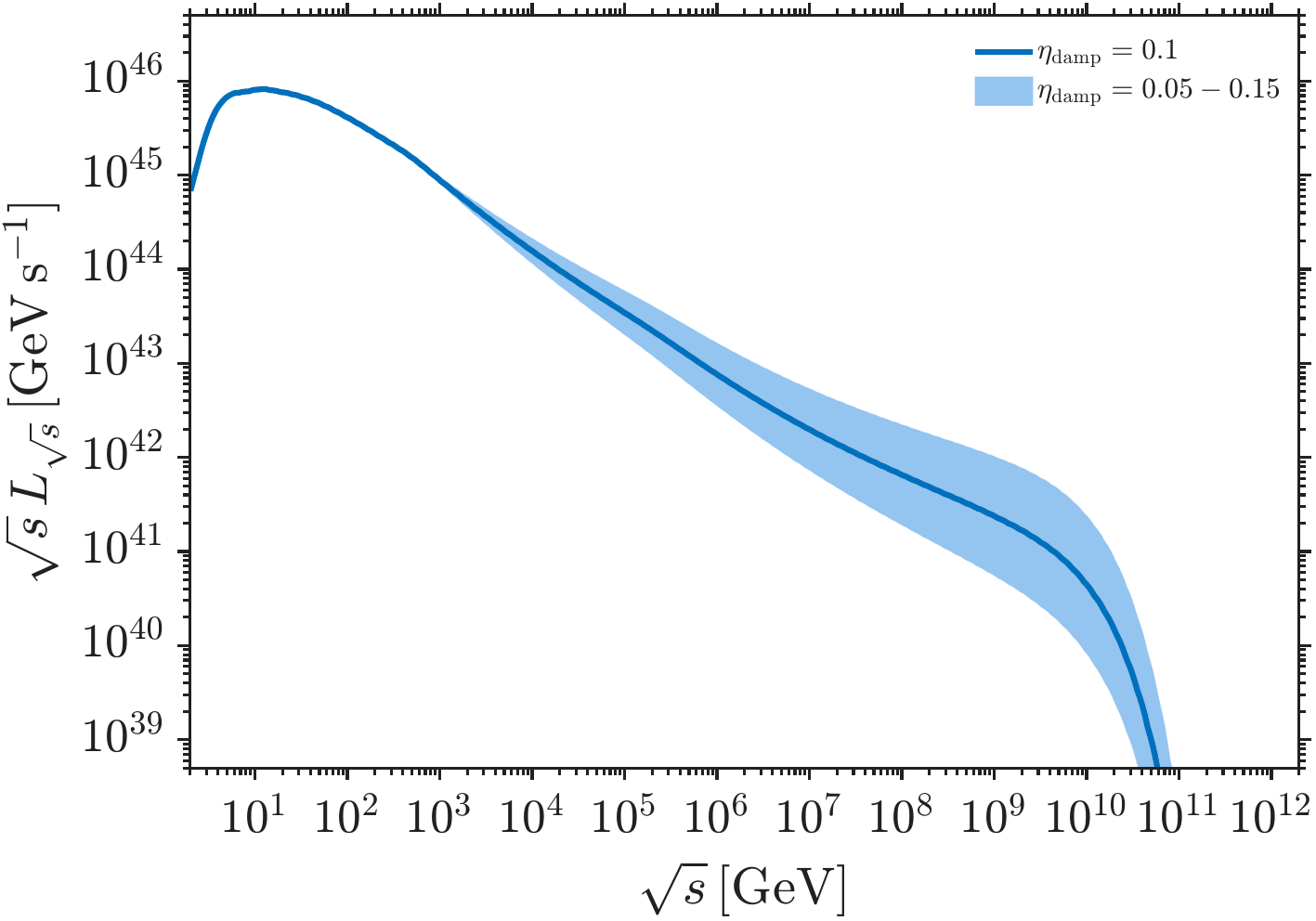}
\end{center}
\caption{\label{fig:machine_lum} Differential collider luminosity $\sqrt{s} L_{\sqrt{s}}$ as a function of the CM energy $\sqrt{s}$. The solid line denotes our baseline ($\eta_{\rm damp} = 0.1$), and the shaded band covers the QCD uncertainty corridor for $\eta_{\rm damp} \in [0.05, 0.15]$.}
\end{figure}

{\it Cosmic Collider luminosity} -- To evaluate the differential luminosity budget of the colliding jets as a function of the CM energy $\sqrt{s}$, we integrate the double-differential production rate of Eq.~(\ref{rate}) over the secondary particle energy $E_{\gamma, \nu}^{\prime}$ as
\begin{equation}
\sqrt{s}L_{\sqrt{s}} \equiv s \, \frac{{\rm d}N_{\gamma, \nu}}{{\rm d}\sqrt{s}} = s \int \frac{{\rm d}^2N_{\gamma, \nu}}{{\rm d}E_{\gamma, \nu}^{\prime}{\rm d}\sqrt{s}} \, {\rm d}E_{\gamma, \nu}^{\prime}~.
\label{eq:luminosity_s_app}
\end{equation}
This quantity, plotted in Fig.~\ref{fig:machine_lum}, represents the total luminosity processed by the Cosmic Collider as a function of the CM energy $\sqrt{s}$. 

As shown in Fig.~\ref{fig:machine_lum}, the Cosmic Collider luminosity peaks at $\sqrt{s} \sim 10 \, \rm GeV$, close to the inelastic threshold $\sqrt{s}_{\text{th}} = 2 m_p c^2 + m_\pi c^2 \simeq 2.01 \, \rm GeV$. The distribution then exhibits a remarkably shallow power-law decline toward ultra-high CM energies: between $\sqrt{s} \sim 10 \, \rm GeV$ and $\sqrt{s} \sim 10^9 \, \rm GeV$, the output luminosity drops by only about four orders of magnitude. This shallow profile confirms that the Cosmic Collider operates as a highly resilient hadronic accelerator, efficiently populating the entire available phase space up to the highest kinematically allowed scale ($\sqrt{s} \sim 10^{10}~\text{GeV}$). The shaded band in Fig.~\ref{fig:machine_lum} illustrates the structural sensitivity corridor obtained by scanning $\eta_{\rm damp} \in [0.05, 0.15]$. The onset of this band around $\sqrt{s} \sim 10^3 \, {\rm GeV}$ highlights how relativistic kinematic transformations map the $T_{p,{\rm eq}} > 10^5 \, {\rm GeV}$ threshold onto the CM frame, confirming that high-energy collider luminosity profiles carry a direct, observable signature of the asymptotic QCD interaction scaling.

\begin{figure}
\begin{center}
\includegraphics[width=.47\textwidth]{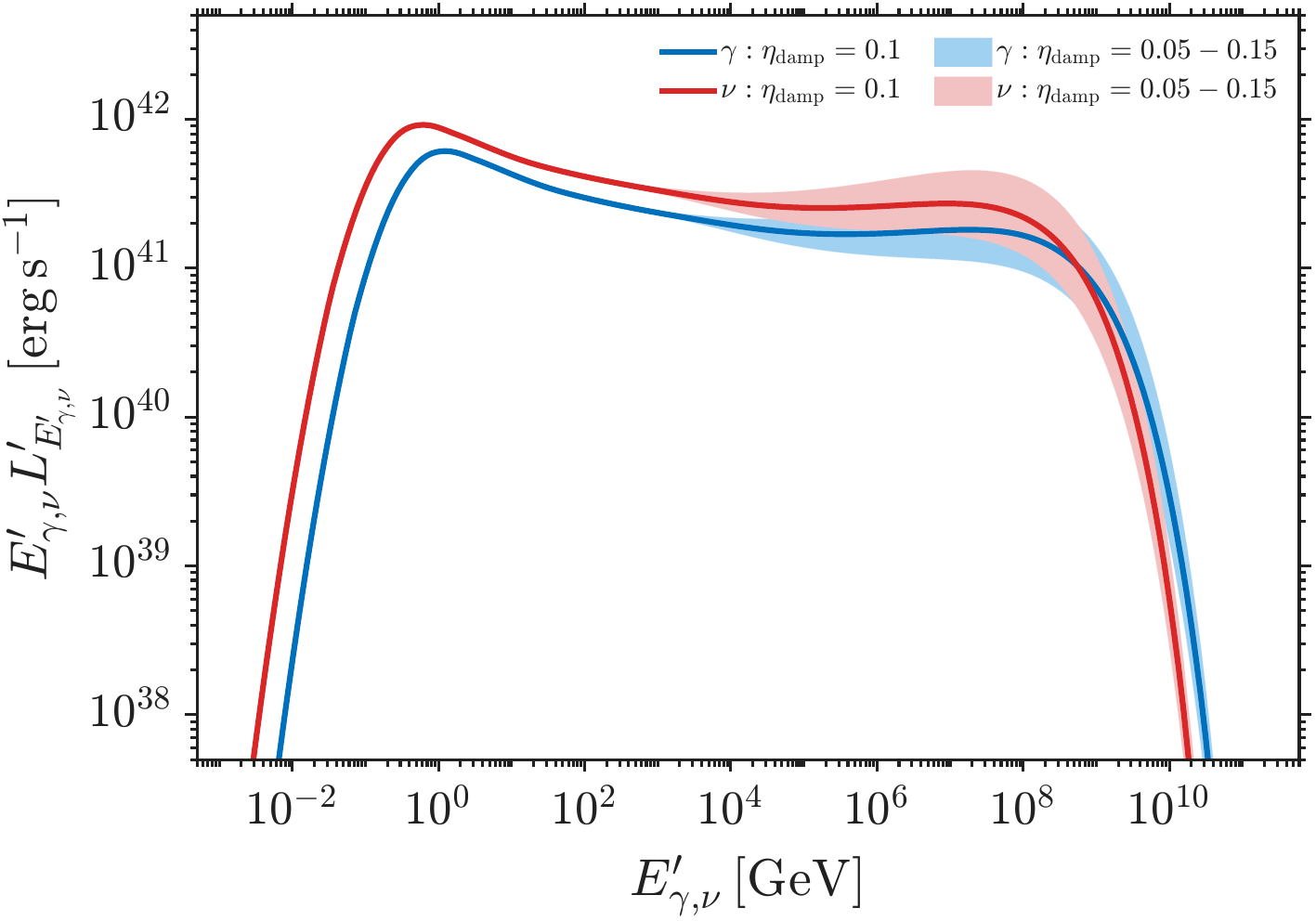}
\end{center}
\caption{\label{fig:intrinsic_sed} Secondary photon ($E'_\gamma L'_{E'_{\gamma}}$, solid blue) and neutrino ($E'_\nu L'_{E'_{\nu}}$, solid red) spectral luminosities in the jet frame as a function of the energy $E'_{\gamma,\nu}$. Solid curves denote the baseline model ($\eta_{\rm damp} = 0.1$), while the shaded bands span the UHE QCD sensitivity range $\eta_{\rm damp} \in [0.05, 0.15]$.}
\end{figure}

{\it Multi-messenger spectral luminosities} -- Complementarily, integrating Eq.~(\ref{rate}) over the CM energy $\sqrt{s}$ yields the spectral luminosities for secondary photons and neutrinos before extragalactic propagation and bulk beaming, given by
\begin{equation}
E_{\gamma, \nu}^{\prime} L'_{E_{\gamma, \nu}^{\prime}} \equiv E_{\gamma, \nu}^{\prime \, 2} \frac{{\rm d}N_{\gamma, \nu}}{{\rm d}E_{\gamma, \nu}^{\prime}} = E_{\gamma, \nu}^{\prime \, 2} \int \frac{{\rm d}^2N_{\gamma, \nu}}{{\rm d}E_{\gamma, \nu}^{\prime} {\rm d}\sqrt{s}} \, {\rm d}\sqrt{s}~.
\label{eq:sed_intrinsic_app}
\end{equation}
The resulting spectral luminosities $E_{\gamma,\nu}^{\prime} L'_{E_{\gamma,\nu}^{\prime}}$ are depicted in Fig.~\ref{fig:intrinsic_sed}.

The luminosities span over twelve orders of magnitude in energy, $10^{-2} \, {\rm GeV} \lesssim E_{\gamma,\nu}^{\prime} \lesssim 10^{10} \, {\rm GeV}$, displaying a broad, quasi-flat plateau between $1 \, \rm GeV$ and $10^8 \, \rm GeV$. This shape reflects a slight spectral hardening relative to the nominal proton power-law index ($\alpha_{p,1}=\alpha_{p,2}=2.2$, which would yield a nominal slope of $-0.2$ in a SED versus energy plot) driven by the logarithmic growth of $\sigma_{pp}^{\rm inel}$ with energy. Comparing the two messengers, the total neutrino luminosity (red curve) remains systematically higher than the secondary photon luminosity (blue curve) across the entire central plateau due to the energy ratio $\eta_{\nu\gamma} \simeq 3/2$ derived in Eq.~(\ref{eq:rate_nu_gamma}). Finally, at the highest energies ($E_{\gamma,\nu}^{\prime} \gtrsim 10^{8}~\text{GeV}$), both spectra steepen smoothly due to parent proton distribution cutoffs, with neutrinos systematically less energetic and experiencing a slightly lower cutoff energy due to the relation $E_\nu^{\prime} \simeq E_\gamma^{\prime} / 2$, as discussed above. As shown by the shaded bands in Fig.~\ref{fig:intrinsic_sed}, varying $\eta_{\rm damp} \in [0.05, 0.15]$ demonstrates that secondary multi-messenger spectral slopes provide an effective observational handle on asymptotic QCD dynamics -- gluon saturation and Froissart-Martin unitarity -- at energies currently unreachable by terrestrial facilities.

Since variations of $\alpha_{p,i}$ modify the secondary spectra already from $E'_{\gamma,\nu} \sim 1\,\mathrm{GeV}$, while $\eta_{\rm damp}$ only acts above $T_{p,{\rm eq}} \sim 10^5\,\mathrm{GeV}$, and $\gamma_{p,i}^{\rm max}$ induces an exponential rather than a power-law suppression, broadband spectra allow astrophysical and QCD effects to be disentangled. A refined QCD treatment beyond the effective parametrization adopted here is expected to further enhance this distinction, and future high-statistics spectra could be unfolded with accurate hadronic models to reconstruct $\sqrt{s}L_{\sqrt{s}}$ directly from observations.

\begin{widetext}

\vskip 1 cm

\section*{SUPPLEMENTAL MATERIAL}

\medskip
\medskip

\section{Cross Sections}
\label{sec:cross_sections}

Evaluating the efficiency of binary supermassive black hole jet collisions as relativistic {\it Cosmic Colliders} requires accurate analytical descriptions of the fundamental interaction cross sections over an unprecedented energy range. While proton-proton ($pp$) interactions constitute the primary hadronic beam-collision process governing the secondary gamma-ray and neutrino production, photo-hadronic ($p\gamma$) interactions with ambient radiation fields represent a competing energy-loss mechanism that determines the relative dominance of each channel. In this section, we present global, smooth analytical parametrizations for both the inelastic proton-proton cross section $\sigma_{pp}^{\rm inel}(s)$ and the total photo-hadronic cross section $\sigma_{p\gamma}^{\rm tot}(s)$. Both models cover the complete kinematic range from their physical thresholds up to ultra-high center-of-mass (CM) energies ($\sqrt{s} \sim 10^{10} \, \rm GeV$). This approach integrates low-energy laboratory measurements with high-energy collider and cosmic-ray extractions from the Particle Data Group (PDG)~\cite{PDG}, while strictly respecting kinematic threshold conditions and Froissart-Martin asymptotic bounds~\cite{froissart, Martin:1966}.

\subsection{Inelastic Proton-Proton Cross Section Parametrization}
\label{sec:cross_section_pp}

To compute the optical depth $\tau_{pp}$, the related cooling time $t_{pp}$, the energy-loss fraction $f_{pp}$, and the production rate of secondary particles from proton-proton interactions, we employ a theoretically motivated analytical fit to the inelastic cross section $\sigma_{pp}^{\rm inel}(s)$ spanning from the kinematic threshold up to ultra-high-energy (UHE) cosmic rays ($\sqrt{s} \sim 10^5 \, \rm GeV$). The dataset used for the fit includes comprehensive accelerator measurements (e.g., TOTEM~\cite{TOTEM:2013} and ATLAS Collaborations~\cite{ATLAS:2016}) alongside UHE cosmic-ray air-shower extractions (e.g., Pierre Auger Collaboration~\cite{Auger:2012}) compiled by the PDG~\cite{PDG}.

Motivated by Regge theory~\cite{Donnachie:1992} and the asymptotic behavior permitted by the Froissart-Martin bound~\cite{froissart, Martin:1966}, we adopt a parametrization of $\sigma_{pp}^{\rm inel}(s)$ that combines near threshold power-law terms with a quadratic logarithmic high-energy behavior, given by
\begin{equation}
\sigma_{pp}^{\rm inel}(s) = \left[ \sigma_0 + R_1 \left(\frac{s}{s_{\rm th}}\right)^{-\eta_1} + R_2 \left(\frac{s}{s_{\rm th}}\right)^{-\eta_2} + P_1 \ln\left(\frac{s}{s_{\rm th}}\right) + P_2 \ln^2\left(\frac{s}{s_{\rm th}}\right) \right]\Theta(s-s_{\rm th})~,
\label{eq:sigma_inel}
\end{equation}
where $s$ is the CM energy squared and $s_{\rm th} = (2m_p c^2 + m_{\pi} c^2)^2 \simeq 4.047 \, \rm GeV^2$ represents the kinematic threshold for single pion production, with $m_p \simeq 0.938 \, \rm GeV$ and $m_\pi \simeq 0.135 \, \rm GeV$ denoting the proton and neutral pion rest masses, respectively, while $c$ is the speed of light. The Heaviside step function $\Theta(s-s_{\rm th})$ is present in Eq.~(\ref{eq:sigma_inel}) to ensure exact physical truncation below threshold. 

The physical interpretation of each component in Eq.~(\ref{eq:sigma_inel}) is structured as follows:
\begin{itemize}

\item {\bf Constant Offset ($\sigma_0$):} This term represents the constant baseline cross section amplitude, ensuring the correct normalization of $\sigma_{pp}^{\rm inel}(s)$ at intermediate energies where Reggeon contributions (see below) subside before the onset of asymptotic growth.

\item {\bf Reggeon Exchange Terms ($R_1, R_2$):} The power-law components $(s/s_{\rm th})^{-\eta_{1,2}}$ account for effective secondary meson exchanges (e.g., $f_2, a_2, \rho, \omega$). The large effective exponents ($\eta_1 \simeq 2.521, \eta_2 \simeq 3.463$) describe the steep threshold onset, and their combined contribution becomes negligible above $\sqrt{s} \sim 3 \, \rm GeV$.

\item {\bf Pomeron Logarithmic Asymptotics ($P_1, P_2$):} The $\ln(s/s_{\rm th})$ and $\ln^2(s/s_{\rm th})$ terms model Pomeron exchange dynamics. In particular, the quadratic term $P_2 \ln^2(s/s_{\rm th})$ is consistent with the asymptotic behavior permitted by the Froissart-Martin bound~\cite{froissart, Martin:1966}, accurately reproducing the high-energy growth observed at the Large Hadron Collider and in cosmic-ray air shower observations.

\item {\bf Kinematic Threshold Constraint:} We impose the condition $R_2 = -(\sigma_0 + R_1)$ such that the parametrized cross section $\sigma_{pp}^{\rm inel}(s)$ vanishes continuously at $s = s_{\rm th}$, while the Heaviside factor $\Theta(s-s_{\rm th})$ enforces the physical truncation below threshold. We have verified that the combined threshold dynamics of the Reggeon exponents yield a strictly positive [$\sigma_{pp}^{\rm inel}(s) > 0$] and monotonically increasing cross section in the immediate neighborhood above $s_{\rm th}$, absorbing the negative linear Pomeron coefficient $P_1$ without generating unphysical sub-threshold dips or negative artifacts.

\end{itemize}

While the parameter values are anchored to laboratory and cosmic-ray data up to $\sqrt{s} \sim 10^5 \, \rm GeV$, the $\ln^2(s/s_{\rm th})$ functional form, together with the fitted value of $P_2$, ensures that the extrapolation to the maximum Cosmic Collider scales investigated in this work ($\sqrt{s} \sim 10^{10} \, \rm GeV$) remains consistent with unitarity. The Froissart-Martin bound~\cite{froissart, Martin:1966} limits the asymptotic growth of the total cross section as $\sigma \lesssim \pi (\hbar/m_\pi c)^2 \ln^2(s/s_0)$, where $\pi (\hbar/m_\pi c)^2 \sim 60\,\rm mb$ and $s_0$ is an unspecified scale. Since a change of scale ($s_0$ or $s_{\rm th}$) only modifies subleading terms [$\ln^2(s/s_0) = \ln^2 s + {\cal O}(\ln s)$], for $s \to \infty$ both the bound and Eq.~(\ref{eq:sigma_inel}) are dominated by their $\ln^2 s$ terms, and the condition reduces to $P_2 \le \pi (\hbar/m_\pi c)^2$. Our fitted coefficient $P_2 \simeq 0.2\,\rm mb$ lies more than two orders of magnitude below this limit, so the parametrization is free from the divergent behavior that typically affects unconstrained empirical or polynomial fits.

The best-fit parameters appearing in Eq.~(\ref{eq:sigma_inel}) obtained from the global optimization are summarized in Table~\ref{tab:cross_section_params}.

\begin{table}[h!]
\centering
\begin{threeparttable}
\setlength{\tabcolsep}{12pt} 
\caption{Best-fit parameters for the inelastic proton-proton cross section $\sigma_{pp}^{\rm inel}(s)$ in Eq.~(\ref{eq:sigma_inel}). Amplitudes are given in millibarns ($\rm mb$), while exponents are dimensionless.}
\label{tab:cross_section_params}
\begin{tabular}{@{}ccc@{}} 
\hline\hline
\noalign{\vspace{2pt}}
Parameter & Value & Description / Physical Meaning \\[3pt]
\hline
\noalign{\vspace{2pt}}
$\sigma_0$   & $31.730 \, \rm mb$   & Constant baseline amplitude \\
$R_1$        & $27.391 \, \rm mb$   & Reggeon threshold amplitude \\
$R_2$        & $-59.121 \, \rm mb$  & Constrained parameter ($R_2 = -\sigma_0 - R_1$) \\
$P_1$        & $-0.850 \, \rm mb$   & Linear Pomeron coefficient \\
$P_2$        & $0.203 \, \rm mb$    & Froissart-Martin $\ln^2 s$ Pomeron coefficient \\
$\eta_1$     & $2.521$              & First threshold Regge exponent \\
$\eta_2$     & $3.463$              & Second threshold Regge exponent \\
$s_{\rm th}$ & $4.047 \, \rm GeV^2$ & Kinematic threshold $(2m_p c^2 + m_{\pi} c^2 )^2$ \\[3pt]
\hline\hline
\end{tabular}
\end{threeparttable}
\end{table}

\subsection{Photo-Hadronic Cross Section Parametrization}
\label{sec:cross_section_pg}

We construct a global analytical fit for the total photo-hadronic cross section $\sigma_{p\gamma}^{\rm tot}(s)$ in order to compute the associated cooling time $t_{p\gamma}$ and the energy-loss fraction $f_{p\gamma}$, allowing a direct comparison between $p\gamma$ interactions and the $pp$ process. The parametrization covers the relevant kinematical range, extending smoothly from the single-pion photoproduction threshold $s_{\rm th}^\gamma = (m_p c^2 + m_{\pi} c^2)^2 \simeq 1.152 \, \rm GeV^2$ up to ultra-high CM energies. The dataset incorporates all total photo-hadronic cross section measurements compiled by the PDG~\cite{PDG}, including low-energy resonance measurements and high-energy collider extractions (e.g., ZEUS~\cite{ZEUS:2002} and H1 Collaborations~\cite{H1:1995}).

To provide a flexible description of the resonance region near threshold, dominated by the $\Delta(1232)$ isobar and higher nucleon excitations, while ensuring the asymptotic behavior allowed by the Froissart-Martin bound~\cite{froissart, Martin:1966}, we parametrize $\sigma_{p\gamma}^{\rm tot}(s)$ using a combined Rational-Pad\'e and Froissart-type formulation reading
\begin{equation}
\sigma_{p\gamma}^{\rm tot}(s) = \left[\sigma_0^\gamma \exp \left( \frac{\sum\limits_{n=0}^{3} p_n \ln^n(s/s_{\rm th}^\gamma)}{\sum\limits_{m=0}^{4} q_m \ln^m(s/s_{\rm th}^\gamma)} \right) + P_2^\gamma \ln^2\left(\frac{s}{s_{\rm th}^\gamma}\right) \right] \left( 1 - \frac{s_{\rm th}^\gamma}{s} \right)^{\eta^\gamma}\Theta(s-s_{\rm th}^\gamma)~,
\label{eq:sigma_gamp}
\end{equation}
where $\sigma_0^\gamma \simeq 0.090 \, \rm mb$ is the fitted physical baseline normalization scale, $q_4 = 1$ is fixed by construction, and $s_{\rm th}^\gamma = (m_p c^2 + m_{\pi} c^2)^2 \simeq 1.152 \, \rm GeV^2$ represents the exact physical threshold for neutral pion photoproduction. The kinematic factor $(1 - s_{\rm th}^\gamma/s)^{\eta^\gamma}$ combined with the Heaviside step function $\Theta(s - s_{\rm th}^\gamma)$ rigorously guarantees smooth kinematic truncation at threshold [$\sigma_{p\gamma}^{\rm tot}(s_{\rm th}^\gamma) = 0$] without introducing non-physical sub-threshold artifacts.

The structural components of Eq.~(\ref{eq:sigma_gamp}) are detailed as follows:
\begin{itemize}

\item {\bf Kinematic Threshold Suppression Factor $(1 - s_{\rm th}^\gamma/s)^{\eta^\gamma}$:} To enforce the correct vanishing of the cross section at the kinematic threshold, we multiply the parametrization by the factor $(1-s_{\rm th}^\gamma/s)^{\eta^\gamma}$. At $s=s_{\rm th}^\gamma$, this factor vanishes identically, ensuring $\sigma_{p\gamma}^{\rm tot}(s_{\rm th}^\gamma)=0$. The exponent ${\eta^\gamma}$ controls the effective steepness of the threshold turn-on. At high energies, $(1-s_{\rm th}^\gamma/s)^{\eta^\gamma}\to1$, leaving the resonance and asymptotic components unaffected.

\item {\bf Threshold and Resonance Region ($P_3/Q_4$ Pad\'e Approximant Ratio):} Low-energy photo-hadronic interactions exhibit prominent resonant structure, most notably the $\Delta(1232)$ resonance peak at $E_\gamma^{\rm lab} \sim 0.34 \, \rm GeV$. We use an exponentiated $P_3/Q_4$ Pad\'e rational approximant $\sum_n p_n \ln^n(s/s_{\rm th}^\gamma) / \sum_m q_m \ln^m(s/s_{\rm th}^\gamma)$ to provide a flexible description of the measured resonance-region structure above threshold. The fitted denominator $Q_4$ possesses no real roots over the physical domain considered, ensuring a smooth profile free of non-physical pole singularities.

\item {\bf Asymptotic Decoupling of Low-Energy Resonances:} The degree of the denominator polynomial ($M=4$) exceeds that of the numerator polynomial ($N=3$) by one. As a result, $\lim_{s \to \infty} \sum_n p_n \ln^n(s/s_{\rm th}^\gamma) / \sum_m q_m \ln^m(s/s_{\rm th}^\gamma) = 0$, and hence the exponential factor approaches unity, yielding the asymptotic baseline $\sigma_0^\gamma \exp(0) = \sigma_0^\gamma \simeq 0.090 \, \rm mb$. This ensures that the Pad\'e component approaches a constant baseline at high energies and does not introduce additional power-law growth.

\item {\bf Froissart-Martin Asymptotic Behavior ($P_2^\gamma \ln^2(s/s_{\rm th}^\gamma)$):} At high energies ($s \gg s_{\rm th}^\gamma$), the total photo-hadronic cross section is expected to approach a slowly rising asymptotic behavior associated with Pomeron exchange and vector-meson dominance (VMD)~\cite{Bauer:1978}. We therefore include an explicit $P_2^\gamma \ln^2(s/s_{\rm th}^\gamma)$ term, consistent with the asymptotic energy dependence allowed by the Froissart-Martin bound~\cite{froissart, Martin:1966}. The best-fit coefficient is $P_2^\gamma \simeq 6.091 \times 10^{-4} \,\rm mb$, substantially smaller than the corresponding proton-proton coefficient, $P_2 \simeq 0.203 \,\rm mb$.

\end{itemize}

Although the original axiomatic derivation of the Froissart-Martin bound~\cite{froissart, Martin:1966} strictly applies to purely hadronic scattering involving massive fields, the extension of the quadratic logarithmic scaling ($\ln^2 s$) to photo-hadronic channels is motivated by the framework of VMD~\cite{Bauer:1978}. Under VMD, high-energy hadronic fluctuations of the real photon (predominantly into $\rho$, $\omega$, and $\phi$ mesons) govern the total interaction cross section, effectively inheriting the unitarity constraints of strong interactions. Our coefficient $P_2^\gamma \simeq 6 \times 10^{-4}\,\rm mb$ lies far below the corresponding bound $\pi(\hbar/m_\pi c)^2 \sim 60\,\rm mb$. As a consistency check, the two independently fitted parametrizations preserve the ratio $\sigma_{p\gamma}^{\rm tot}/\sigma_{pp}^{\rm inel} \sim 1/300$ expected from VMD and observed at collider energies, both at $\sqrt{s} \sim 10^{10}\,\rm GeV$ ($\sigma_{p\gamma}^{\rm tot} \simeq 1.4\,\rm mb$ versus $\sigma_{pp}^{\rm inel} \simeq 0.4\,\rm b$) and asymptotically ($P_2^\gamma/P_2 \simeq 1/330$). Our model thus provides a theoretically consistent and well-behaved extrapolation across the vast energy gap separating accelerator data from the extreme Cosmic Collider limits ($\sqrt{s} \sim 10^{10} \, \rm GeV$) studied in this work.

The best-fit parameters appearing in Eq.~(\ref{eq:sigma_gamp}) obtained from the global optimization are provided in Table~\ref{tab:cross_section_params_gamp}.

\begin{table}[h!]
\centering
\begin{threeparttable}
\setlength{\tabcolsep}{12pt}
\caption{Best-fit parameters for the total photo-hadronic cross section $\sigma_{p\gamma }^{\rm tot}(s)$ defined in Eq.~(\ref{eq:sigma_gamp}). Coefficients $p_n$, $q_m$, and $\eta^\gamma$ are dimensionless parameters, while $P_2^\gamma$ and $\sigma_0^\gamma$ are expressed in millibarns ($\rm mb$).}
\label{tab:cross_section_params_gamp}
\begin{tabular}{@{}ccc@{}}
\hline\hline
\noalign{\vspace{2pt}}
Parameter & Value & Description / Physical Meaning \\[3pt]
\hline
\noalign{\vspace{2pt}}
$\sigma_0^\gamma$ & $0.090 \, \rm mb$ & Continuum baseline normalization amplitude \\
$p_0$ & $0.023$ & Numerator Pad\'e coefficient (order 0) \\
$p_1$ & $0.021$ & Numerator Pad\'e coefficient (order 1) \\
$p_2$ & $-0.408$ & Numerator Pad\'e coefficient (order 2) \\
$p_3$ & $0.610$ & Numerator Pad\'e coefficient (order 3) \\
$q_0$ & $0.024$ & Denominator Pad\'e coefficient (order 0) \\
$q_1$ & $-0.186$ & Denominator Pad\'e coefficient (order 1) \\
$q_2$ & $0.692$ & Denominator Pad\'e coefficient (order 2) \\
$q_3$ & $-1.289$ & Denominator Pad\'e coefficient (order 3) \\
$q_4$ & $1$ & Fixed leading denominator coefficient (order 4) \\
$\eta^\gamma$ & $0.696$ & Effective threshold turn-on exponent \\
$P_2^\gamma$ & $6.091 \times 10^{-4} \, \rm mb$ & Froissart-Martin $\ln^2 s$ Pomeron coefficient \\
$s_{\rm th}^\gamma$ & $1.152 \, \rm GeV^2$ & Photoproduction threshold $(m_p c^2 + m_{\pi} c^2)^2$ \\[3pt]
\hline\hline
\end{tabular}
\end{threeparttable}
\end{table}

\section{Active galactic nuclei}
\label{sec:supp_agn_framework}

Active galactic nuclei (AGN) are powered by supermassive black holes (SMBHs) with masses $M \sim (10^6-10^9) \, M_\odot$ and are among the most efficient particle accelerators in the Universe. In roughly 10\% of the AGN population~\cite{Kellermann1989}, a fraction of the gravitational energy released by accretion fuels two collimated, highly relativistic jets propagating in opposite directions along the rotation axis of the accretion disk via magnetic extraction processes~\cite{BlandfordZnajek1977,Romero2017,Blandford2019}. When one of these jets happens to be oriented at a small viewing angle with respect to our line of sight, relativistic beaming drastically boosts the observed flux and compresses the variability timescales, defining the blazar class. According to the AGN unified model~\cite{UrryPadovani1995}, blazars are the beamed counterparts of radio-loud AGN, while the same parent population observed at larger viewing angles appears as radio galaxies.

Across the entire electromagnetic spectrum -- from radio waves up to very-high-energy (VHE, $E > 100\,\text{GeV}$) $\gamma$-rays -- the spectral energy distribution (SED) of blazars is characteristically non-thermal and displays a distinct \textit{two-hump} structure~\cite{Fossati1998,Ghisellini1998}: a lower-energy peak spanning the infrared to X-ray bands and a higher-energy peak located in the $\gamma$-ray domain. Two competing frameworks account for this phenomenology:
\begin{itemize}
\item \textbf{Leptonic models:} The low-energy hump is produced by synchrotron radiation from relativistic electrons, while the high-energy hump stems from Inverse Compton (IC) scattering off internal synchrotron photons (Synchrotron Self-Compton, SSC) or external photon fields (External Compton, EC)~\cite{Maraschi1992,Sikora1994,Bloom1996}.
\item \textbf{Hadronic models:} Relativistic protons co-accelerated with leptons undergo synchrotron emission or interact with target fields~\cite{Mannheim1993a,Mannheim1993b,Mucke2003}. In single-jet scenarios, photohadronic ($p\gamma$) reactions typically dominate over $pp$ processes, yielding VHE $\gamma$-rays and neutrinos through pion decay cascades.
\end{itemize}

Blazars are conventionally divided into two main classes based on their optical spectral properties and intrinsic jet power:
\begin{itemize}
\item \textbf{Flat Spectrum Radio Quasars (FSRQs):} These are highly luminous and powerful blazars fueled by standard, radiatively efficient accretion disks. Their optical spectra show intense, Doppler-broadened emission lines originating from high-velocity photoionized gas clouds orbiting the SMBH within the broad-line region (BLR, with radius $r_{\rm BLR} \sim 0.1 - 0.3 \,{\rm pc}$)~\cite{Kaspi2000,Bentz2013}. Additional external photon targets include the infrared radiation field of a surrounding dusty torus (DT) at parsec scales (with radius $r_{\rm DT} \sim 1-10\,{\rm pc}$)~\cite{Nenkova2008,Galanti2020}.
\item \textbf{BL Lacertae objects (BL Lacs):} These are intrinsically less powerful sources, often associated with radiatively inefficient accretion flows (RIAF) onto the central SMBH~\cite{GhiselliniMar2009,Righi2019}. They exhibit weak or entirely absent optical emission lines, indicating a negligible external photon environment (no prominent BLR or dusty torus). Consequently, the target photon field inside the jet is overwhelmingly dominated by the internal non-thermal synchrotron radiation field (SSC regime).
\end{itemize}

{\it Notation note} -- Throughout this section, primed quantities denote variables evaluated in a generic jet rest frame, whereas unprimed quantities refer to the host galaxy or terrestrial laboratory frame, as specified. When evaluating the interaction between the two proton populations ($pp$), the notation explicitly identifies both the jet of origin (jet 1 or 2) and the reference frame in which the variable is measured. In particular, the superscript $(1)$ denotes quantities evaluated in the rest frame of jet 1, and a tilde denotes target-proton quantities (see Main Text for details).

\subsection{Jet Structure}
\label{sec:jet_structure}

To quantitatively model the competing energy-loss mechanisms for relativistic protons with the studied $pp$ interaction, we adopt representative jet profiles for both source classes, BL Lacs and FSRQs. In the rest frame of the jet (moving with a bulk Lorentz factor $\Gamma_b$), the spatial profiles of the magnetic field $B'$ and the isotropic spectral electron number density $n_e'$ along the jet axis $r$ in the jet-jet interaction region are modeled as
\begin{equation}
B'(r) = B'_0 \left( \frac{r}{r_0} \right)^{-a}~,
\label{eq:jet_profiles_B}
\end{equation}
\begin{equation}
n_e'(E_e',r) = \frac{\mathrm{d}^3 N_e'}{\mathrm{d}E_e' \mathrm{d}V' \mathrm{d}\Omega'} = \frac{K_e}{4 \pi} \left(\frac{E_e'}{m_e c^2}\right)^{-\alpha_e} \exp\left(- \frac{E_e'}{\gamma_e^{\mathrm{max}}m_e c^2} \right)\left( \frac{r}{r_0} \right)^{-a}~,
\label{eq:jet_profiles_ne}
\end{equation}
where $B'_0$ is the magnetic field strength at $r_0$, which denotes the distance where the two jets start to interact, $a$ is a parameter measuring the jet-jet compression (see Main Text and Sec.~\ref{sec:jet_jet_compression}), $E_e'$ and $m_e$ are the electron comoving energy and rest mass, respectively, $\alpha_e$ is the electron spectral index, and $c$ is the speed of light. Here, $N_e'$ represents the total number of electrons in the comoving frame, $V'$ is the comoving volume, and $\Omega'$ is the comoving solid angle. The normalization factor $K_e$ is directly related to the total integrated comoving electron number density $n_{e,0}'$ at $r_0$, via $K_e m_e c^2 \int_{\gamma_e^{\rm min}}^{\infty} \gamma_e^{-\alpha_e} \exp(-\gamma_e/\gamma_e^{\rm max}) {\rm d}\gamma_e = n_{e,0}'$, where $\gamma_e = E_e'/(m_e c^2)$ is the electron Lorentz factor between a minimum Lorentz factor $\gamma_e^{\rm min}$, and a maximum cutoff $\gamma_e^{\rm max}$. For the particle densities in the jet, we adopt a purely proton-electron plasma ($n_{e,0}' \simeq n_{p,0}'$, with $n_{p,0}'$ the total integrated comoving proton number density defined in the Main Text), ensuring local charge neutrality in the absence of significant positron loading ($n_{e^+} \ll n_{e^-}$)~\cite{CelottiFabian1993}. This assumption provides a clean, conservative baseline for the total hadronic energy budget without introducing unconstrained free parameters for pair production efficiency.

\subsection{Target Photon Fields}
\label{sec:target_photon_fields}

To model photohadronic interactions ($p\gamma$), the target photon spectral number density $n_\gamma'(\epsilon', r)$ must be evaluated in the jet comoving frame, where $\epsilon'$ is the corresponding photon energy. Depending on the blazar class, this field is composed of different physical contributions: internal synchrotron radiation $n_{e\text{-sync}}'(\epsilon', r)$, external radiation fields from the Broad Line Region (BLR) $n_{\rm BLR}'(\epsilon', r)$ and Dusty Torus (DT) $n_{\rm DT}'(\epsilon', r)$, and the Cosmic Microwave Background (CMB) $n_{\rm CMB}'(\epsilon', r)$.

\begin{itemize}

\item {\it Internal Synchrotron Radiation Field ($n_{e\text{-sync}}'$):} The local synchrotron photon spectral number density is calculated by integrating the single-electron synchrotron power $P_s(\epsilon', \gamma_e, B')$, evaluated at a representative pitch angle of $\pi/4$, over the primary electron distribution $n_e'(E_e', r)$ defined in Eq.~(\ref{eq:jet_profiles_ne}). Here, $B'$ is given by Eq.~(\ref{eq:jet_profiles_B}). To account for synchrotron self-absorption (SSA) at lower frequencies, the local specific intensity $I_\nu'(r)$ is evaluated through the radiative transfer equation for a spherical region of radius $R_{e\text{-sync}}(r) = R_{e\text{-sync}}(r_0)\,(r/r_0)^{a/2}$, following the expansion of the compressed core, as~\cite{RybickiLightman1979}
\begin{equation}
I_\nu'(r) = \frac{j_\nu'(r)}{\alpha_\nu'(r)} \left( 1 - e^{-\alpha_\nu'(r) R_{e\text{-sync}}(r)} \right)~,
\end{equation}
where $j_\nu'(r)$ is the total synchrotron emissivity and $\alpha_\nu'(r)$ is the SSA absorption coefficient. The synchrotron photon spectral number density is then given by $n_{e\text{-sync}}'(\epsilon', r) = 4\pi I_\nu'(r) / (h\,c \, \epsilon')$, with $h$ the Planck constant.

\item {\it External Radiation Fields ($n_{\rm BLR}'$ and $n_{\rm DT}'$):}
For FSRQ-like environments, external photon fields originating from the BLR and DT are driven directly by the total accretion disk luminosity $L_{\rm disc}$. The underlying primary continuum emitted by the accretion disk is modeled adopting a standard multi-temperature Shakura-Sunyaev disk spectrum~\cite{ShakuraSunyaev1973}, where the differential spectral luminosity $L_{\epsilon, \rm disc}$ is characterized by a power-law index of $1/3$ combined with an exponential energy cutoff reading
\begin{equation}
L_{\epsilon, \rm disc} = L_{0, \text{disc}} \left( \frac{\epsilon}{\epsilon_{\text{ref}}} \right)^{1/3} \exp \left( -\frac{\epsilon}{\epsilon_{\rm disc, max}} \right)~,
\label{eq:L_eps_disc_shape}
\end{equation}
where $\epsilon_{\text{ref}}$ is a reference energy scale and $\epsilon_{\rm disc, max} = 20 \, {\rm eV}$ represents the peak energy of the thermal disk emission. The normalization constant $L_{0, \text{disc}}$ is uniquely fixed by enforcing that the energy-integrated spectrum matches the total accretion disk luminosity as
\begin{equation}
\int_0^\infty L_{\epsilon, \rm disc} \, {\rm d}\epsilon = L_{\rm disc}~.
\end{equation}

The external fields are then derived separately for the BLR and DT components as follows:
\begin{enumerate}

\item {\bf Broad Line Region:} The BLR reprocesses UV/optical disk radiation. Its differential spectral luminosity in the laboratory frame, $L_{\epsilon, \rm BLR}$, inherits the characteristic Shakura-Sunyaev disk continuum shape, scaled by the covering factor $\eta_{\rm BLR}$ as
\begin{equation}
L_{\epsilon, \rm BLR} = \eta_{\rm BLR} \, L_{\epsilon, \rm disc}~,
\end{equation}
where typically $\eta_{\rm BLR} \sim 0.1$ for radiatively efficient FSRQs~\cite{GhiselliniTavecchio2008}.
    
\item {\bf Dusty Torus:} The dust grains absorb the UV disk continuum and re-emit it thermally in the infrared. The differential spectral luminosity $L_{\epsilon, \rm DT}$ is modeled as a Planckian blackbody spectrum at a temperature $T_{\rm DT} \sim 1000 \, {\rm K}$, representative of the bulk of the hot dust emission, normalized to the total reprocessed power via $\int_0^\infty L_{\epsilon, \rm DT} \, {\rm d}\epsilon = \eta_{\rm DT} \, L_{\rm disc}$, where $\eta_{\rm DT}$ is the DT covering factor. Due to the expected higher covering fraction of dust in AGN environments~\cite{Nenkova2008b, Ricci2017} and to minimize free parameters, we set $\eta_{\rm DT} = 5 \, \eta_{\rm BLR}$ (yielding $\eta_{\rm DT} \sim 0.5$ for radiatively efficient FSRQs).

\end{enumerate}

Inside their characteristic radii ($r \le R_{\rm ext}$, with ${\rm ext} \in \{{\rm BLR, DT}\}$) -- which scale with disk luminosity as $R_{\rm BLR} \simeq 0.032 \times (L_{\rm disc} / 10^{45} \, {\rm erg \, s^{-1}})^{1/2}\, {\rm pc}$~\cite{GhiselliniTavecchio2008} and $R_{\rm DT} \simeq 0.4 \times (L_{\rm disc} / 10^{45} \, {\rm erg \, s^{-1}})^{1/2}(T_{\rm DT}/1500 \, {\rm K})^{-2.8} \, {\rm pc}$~\cite{Barvainis1987} -- the spectral energy density of the external photon field in the laboratory frame reads
\begin{equation}
u_{\rm ext}(\epsilon) = \frac{L_{\epsilon, \rm ext}}{2 \pi R_{\rm ext}^2 c}~.
\end{equation}
For positions outside these emission regions ($r > R_{\rm ext}$), the spectral energy density decays geometrically as $(r / R_{\rm ext})^{-2}$, yielding $u_{\rm BLR}(\epsilon,r)$ and $u_{\rm DT}(\epsilon,r)$. These stationary spectral fields are transformed to the comoving jet frame via the Lorentz boost, $u_{\rm ext}'(\epsilon', r) \simeq \Gamma_b \, u_{\rm ext}(\epsilon, r)$ (with $\epsilon' = \Gamma_b \epsilon$), providing the comoving spectral photon number densities $n_{\rm ext}'(\epsilon', r) = u_{\rm ext}'(\epsilon', r) / \epsilon'$. Note that since $\mathrm{d}\epsilon' = \Gamma_b \mathrm{d}\epsilon$, integrating $u_{\rm ext}'(\epsilon', r)$ over comoving energies naturally recovers the expected relativistic transformation for the total integrated energy density, $U_{\rm ext}'(r) = \int u_{\rm ext}'(\epsilon',r)\mathrm{d}\epsilon' \approx \Gamma_b^2 \, U_{\rm ext}(r)$~\cite{Sikora1994, Dermer1994}.

In the limit $\eta_{\rm BLR} = \eta_{\rm DT} = 0$, we naturally recover the BL Lac regime where all external photon fields vanish.

\item {\it Cosmic Microwave Background ($n_{\rm CMB}'$):}
The CMB is modeled as an isotropic blackbody spectrum with temperature $T_{\rm CMB} = 2.73 \, {\rm K}$ in the laboratory frame, transformed into the comoving jet frame via the bulk Lorentz factor $\Gamma_b$.

\end{itemize}

Combining these components, the total target photon number density in the comoving jet frame, $n_\gamma'(\epsilon', r)$, reads
\begin{equation}
n_\gamma'(\epsilon', r) = 
\begin{cases} 
n_{e\text{-sync}}'(\epsilon', r) + n_{\rm CMB}'(\epsilon', r) , & \text{for BL Lacs} , \\[6pt]
n_{e\text{-sync}}'(\epsilon', r) + f_{\rm BLR} \, n_{\rm BLR}'(\epsilon', r) + f_{\rm DT} \, n_{\rm DT}'(\epsilon', r) + n_{\rm CMB}'(\epsilon', r) , & \text{for FSRQs} ,
\end{cases}
\label{eq:n_gamma_tot}
\end{equation}
where $f_{\rm BLR}$ and $f_{\rm DT}$ are geometric anisotropy factors accounting for external photons entering the moving jet frame at non-isotropic incidence angles~\cite{Dermer2009, GhiselliniTavecchio2009}. Since the interaction region lies beyond the BLR and, for most of its extent, beyond the DT, external photons enter the jet mostly from behind and are de-boosted, so that realistic values are $f_{\rm BLR}, f_{\rm DT} < 1$. As the main objective of this work is to evaluate the relative hierarchy between $pp$ and $p\gamma$ cooling rates rather than to perform a detailed angle-dependent photohadronic transport calculation, we conservatively adopt $f_{\rm BLR} = f_{\rm DT} = 1$, i.e., we treat the external fields as isotropic in the host-galaxy frame. This choice maximizes $p\gamma$ losses and thus provides an upper limit on their impact relative to $pp$ interactions: for realistic values $f_{\rm BLR}, f_{\rm DT} \sim 0.2$, $p\gamma$ losses on external fields are reduced by a factor $\sim 5$.

\subsection{Proton Cooling and Interaction Timescales}
\label{sec:proton_cooling_time}

Relativistic protons with energy $E_p'$ and Lorentz factor $\gamma_p = E'_p / (m_p c^2)$ in the jet comoving frame undergo three primary competing energy-loss mechanisms: inelastic proton-proton collisions ($pp$), photohadronic interactions ($p\gamma$), and proton-synchrotron radiation ($p\text{-sync}$). The respective comoving cooling rates $t_{\chi}^{\prime -1} \equiv 1/t_{\chi}^\prime$ (where $t_{\chi}^\prime$ denotes the characteristic comoving cooling time for channel $\chi \in \{pp, p\gamma, p\text{-sync}\}$) are formulated as follows:

\begin{itemize}

\item{\it Hadronic collisions ($pp$):}
For two colliding relativistic proton populations, the total hadronic optical depth $\tau_{pp}$ experienced by a test proton traversing the interaction zone is defined in Eq.~(3) of the Main Text. The comoving $pp$ cooling rate $t_{pp}^{\prime-1}(E'_p, r)$ is obtained by considering the local interaction differential rate weighted by the average inelasticity factor $K_{pp} \sim 0.5$~\cite{Kelner2006,Gaisser2016} as
\begin{equation}
{t_{pp}^{(1)}}^{-1}(E_{p,1}^{(1)}, r) = 2\pi c \, K_{pp} \sum_{j=1}^{2} \int_{-1}^1 {\rm d}(\cos\theta^{(1)}) \int_{m_pc^2}^{\infty} {\rm d}\tilde{E}_{p,j}^{(1)} \, \tilde{n}_{p,j}^{(1)} \, \sigma_{pp}^{\rm inel}(s) \, (1-\beta_1^{(1)}\tilde{\beta}_j^{(1)} \cos \theta^{(1)})~,
\label{eq:t_pp_explicit}
\end{equation}
where $E_{p,1}^{(1)}$ is the energy of a test proton belonging to jet 1 and evaluated in the jet 1 frame, $\tilde{n}_{p,j}^{(1)}$ are the angle-averaged spectral number densities of target protons belonging to jet $j$ evaluated in the jet 1 frame (see Main Text) with energy $\tilde{E}_{p,j}^{(1)}$, $\sigma_{pp}^{\rm inel}(s)$ is given by Eq.~(\ref{eq:sigma_inel}), $\theta^{(1)}$ is the collision angle in the jet 1 frame, and $\beta_1^{(1)}$, $\tilde{\beta}_j^{(1)}$ are the proton speeds in units of $c$. In what follows, since no ambiguity arises when comparing the different cooling channels, we adopt the simplified notation $E_{p,1}^{(1)} \equiv E_p'$ to denote the comoving proton energy in the jet 1 frame, so that $t_{pp}^{\prime -1}(E_p', r) \equiv t_{pp}^{(1)-1}(E_{p,1}^{(1)}, r)$.

\item{\it Photohadronic interactions ($p\gamma$):}
The comoving interaction rate off target photons with energy $\epsilon'$ (in the jet frame) and spectral number density $n_\gamma'(\epsilon',r)$ is evaluated as~\cite{Stecker1968,Berezinskii1990}
\begin{equation}
t_{p\gamma}^{\prime -1}(E'_p,r) = \frac{c}{2} \int_{\frac{\epsilon_{\rm th}}{2\gamma_p}}^{\infty} \mathrm{d}\epsilon' \frac{n_\gamma'(\epsilon',r)}{\gamma_p^2 \epsilon^{\prime 2}} \int_{\epsilon_{\rm th}}^{2\gamma_p \epsilon'} \epsilon_r \, \sigma_{p\gamma}^{\rm tot}(\epsilon_r) \, K_{p\gamma}(\epsilon_r) \, \mathrm{d}\epsilon_r~,
\label{eq:t_pgamma}
\end{equation}
where $\epsilon_r$ is the target photon energy in the proton rest frame, $\epsilon_{\rm th} \simeq 145\,{\rm MeV}$ is the pion production threshold, $\sigma_{p\gamma}^{\rm tot}(\epsilon_r)$ is given by Eq.~(\ref{eq:sigma_gamp}) with $s = m_p^2c^4 + 2m_pc^2\epsilon_r$, and $K_{p\gamma}(\epsilon_r)$ is the photohadronic inelasticity parameter~\cite{AtoyanDermer2003}. For BL Lacs, $n_\gamma'(\epsilon',r)$ includes the internal electron-synchrotron photon spectrum $n_{e\text{-sync}}'$ and that of the CMB $n_{\rm CMB}'$, whereas for FSRQs, the Doppler-boosted external fields ($n_{\rm BLR}', n_{\rm DT}'$) are also accounted for, as detailed in Sec.~\ref{sec:target_photon_fields}.

\item {\it Proton-Synchrotron losses ($p\text{-sync}$):}
The comoving synchrotron cooling rate for ultra-relativistic protons in the magnetic field $B'(r)$ of Eq.~(\ref{eq:jet_profiles_B}) reads~\cite{RybickiLightman1979}
\begin{equation}
t_{p\text{-sync}}^{\prime -1}(E'_p,r) = \frac{4}{3} \sigma_T \left( \frac{m_e}{m_p} \right)^2 \frac{U_B'(r)}{m_p^2 c^3} \, E_p' ~,
\label{eq:t_psyn}
\end{equation}
where $\sigma_T$ is the Thomson cross section, $m_e/m_p$ is the electron-to-proton mass ratio, and $U_B'(r) = B^{\prime 2}(r) / (8\pi)$ is the comoving magnetic energy density.

\end{itemize}

\subsection{Energy-Loss Fractions}
\label{sec:fChi}

To evaluate the cumulative energy loss experienced by a relativistic proton traversing the finite interaction region $[r_0,r_f]$ along the axis $r$, we calculate the energy-loss fractions $f_{\chi}(E_p')$ for each channel $\chi \in \{pp, p\gamma, p\text{-sync}\}$ by integrating the comoving cooling rate over the time spent by the proton within it, as
\begin{equation}
f_{\chi}(E_p') = \int_{r_0}^{r_f} \frac{\mathrm{d}r}{\Gamma_b \, c \, t_{\chi}'(E_p', r)}~.
\label{eq:f_integrated}
\end{equation}
Since the interaction region is stationary in the host-galaxy frame, with the jet plasma flowing through it, a proton crossing ${\rm d}r$ spends a comoving time ${\rm d}r/(\Gamma_b c)$ within it~\cite{SteadyJet2021} (see also Main Text). All quantities entering Eqs.~(\ref{eq:t_pp_explicit}--\ref{eq:t_psyn}) are parameterized as functions of the host-frame position $r \in [r_0, r_f]$. These dimensionless parameters measure the total fractional energy extracted from a proton of comoving energy $E_p'$ by mechanism $\chi$ during its passage through the interaction site.

By comparing the standard hadronic optical depth $\tau_{pp}$ of Eq.~(3) in the Main Text and Eq.~(\ref{eq:t_pp_explicit}), the hadronic energy-loss efficiency $f_{pp}$ is directly related to $\tau_{pp}$ by:
\begin{equation}
f_{pp}(E_p') = K_{pp} \, \tau_{pp}(E_p') \simeq 0.5 \, \tau_{pp}(E_p')~.
\label{eq:f_vs_tau_pp}
\end{equation}
Analogously, for photohadronic interactions, $f_{p\gamma}(E_p') = \langle K_{p\gamma} \rangle \, \tau_{p\gamma}(E_p')$, where $\langle K_{p\gamma} \rangle \approx 0.2$ is the effective photohadronic inelasticity near the single-pion threshold, which increases up to $\sim 0.6$ at higher energies where multi-pion production dominates~\cite{AtoyanDermer2003}.

We adopt $f_\chi \sim 0.1$ as a practical boundary between two regimes. For $f_\chi \lesssim 0.1$, the channel $\chi$ acts as a minor perturbation, extracting less than $10\%$ of the primary proton energy without depleting the beam, so that the different cooling processes effectively operate independently. Conversely, for $f_\chi \gtrsim 0.1$ (and approaching unity), the primary proton population suffers significant energy depletion, channeling a substantial fraction of the proton energy into secondary particles and directly suppressing the energy available for competing interaction channels.

\section{Supermassive black hole binaries}
\label{sec:BSMBH}

Supermassive black hole binaries (SMBHBs) are a natural and inevitable outcome of hierarchical galaxy formation, in which host galaxies undergo repeated mergers over cosmic time~\cite{Begelman1980}. Following the collision of two massive galaxies, their central SMBHs dynamically pair up and sink toward the common center of potential through dynamical friction, eventually forming a bound sub-parsec binary~\cite{Milosavljevic2001}. In gas-rich environments, these systems can sustain dual AGN activity, where both black holes simultaneously accrete material and launch relativistic outflows~\cite{VanWassenhove2012}. The recent detection of a second jet within the nuclear core of the blazar Mrk 501 provides direct observational evidence that dual-jet configurations exist~\cite{Mrk501dualjet}.

While the microphysics of hadronic interactions and radiative loss rates within relativistic jets dictates the local particle production (as detailed in the Main Text and in previous sections), the geometric viability and longevity of the Cosmic Collider framework fundamentally depend on the dynamics of the SMBHB system. In this section, we provide an astrophysical justification for the sub-parsec binary configuration, supporting its dynamical stability, examining the angular self-correction mechanisms for jet alignment, and evaluating the expected demographic abundance of Cosmic Colliders in the observable Universe.

\subsection{Orbital Kinematics}
\label{sec:SMBHB_orbit}

The geometry of the Cosmic Collider is anchored by the orbital parameters of a central SMBHB system. We consider two SMBHs of comparable masses $M_1 \sim M_2 ={\cal O}(10^8) \, M_\odot$ -- yielding a total binary mass $M = M_1 + M_2$ -- separated by a distance $d_{\rm bin} = {\cal O}(0.1) \, \rm pc$ (see the Main Text). Assuming a quasi-circular Keplerian orbit, the orbital period of the binary system is
\begin{equation}
P_{\rm bin} = 2\pi \sqrt{\frac{d_{\rm bin}^3}{G M}} \simeq 3 \times 10^2  \left( \frac{d_{\rm bin}}{0.1\text{ pc}} \right)^{3/2} \left( \frac{M}{ 10^8\,M_\odot} \right)^{-1/2} \, \rm yr~,
\end{equation}
where $G$ is the gravitational constant. Since $P_{\rm bin} \sim \mathcal{O}(10^2 - 10^3) \, \rm yr$, the binary completes only a small fraction of an orbit over typical multi-messenger observing campaigns of $\sim (10-20) \, \rm yr$. The orbital configuration and, consequently, the orientation of the interaction region starting at $r_0$ may therefore be regarded as effectively stationary on observational timescales.

At separations of $d_{\rm bin} = {\cal O}(0.1) \, \rm pc$, the binary is expected to reside in, or close to, the hard-binary regime, where its orbital evolution is governed primarily by dynamical interactions and, where present, torques exerted by a circumbinary gaseous disk~\cite{Begelman1980,Milosavljevic2001, Cuadra2009}. These mechanisms can keep the binary at sub-parsec separations for astrophysically long timescales before gravitational-wave (GW) radiation becomes the dominant channel of orbital energy and angular momentum loss. The possible stalling of the binary at these separations is commonly referred to as the ``Final Parsec Problem''~\cite{Begelman1980,Milosavljevic2003}.

If GW emission alone were responsible for the orbital decay, the inspiral timescale would be given by the Peters formula \cite{Peters1964}
\begin{equation}
t_{\rm GW} = \frac{5}{256} \frac{c^5 d_{\rm bin}^4}{G^3 \eta_{\rm bin} M^3} \simeq 5.8 \times 10^{10} \, \eta_{\rm bin}^{-1} \left( \frac{d_{\rm bin}}{0.1\text{ pc}} \right)^4 \left( \frac{M}{10^8\,M_\odot} \right)^{-3} \rm yr~,
\end{equation}
where $\eta_{\rm bin} \equiv M_1 M_2 / M^2$ is the symmetric mass ratio. Across the considered mass range, this inspiral timescale spans from $\sim 10^8 \, \rm yr$ to beyond the Hubble time, exceeding by many orders of magnitude both multi-year observational campaigns and the binary orbital period. Binaries at $d_{\rm bin} = {\cal O}(0.1) \, \rm pc$ therefore constitute a long-lived evolutionary stage, ensuring the long-term stability of the Cosmic Collider configuration.

\subsection{Dynamical Regulation of Small Jet Misalignment}
\label{sec:jet_spin_and_misalignment}

The efficiency of the Cosmic Collider mechanism requires two relativistic jets with a small relative inclination angle, typically in the degree-scale regime $\theta_{\rm mis}\sim1^\circ-5^\circ$ (see Main Text). Such a configuration does not require fine tuning, since gas-rich SMBHBs are naturally expected to evolve toward nearly aligned spin-orbit configurations.

The presence of powerful relativistic jets is itself associated with an active accretion phase. Therefore, the gas reservoir required to launch and sustain the jets also provides the physical channel through which angular momentum exchange and alignment torques operate. The jet-producing phase of a SMBHB is thus naturally coupled to the same gas-driven processes that regulate the relative orientation of the SMBH spins, the binary orbit, and the jet axes.

In the presence of a circumbinary gas reservoir, gravitational and viscous torques tend to align the angular momentum of the accretion flow with the binary orbital angular momentum~\cite{Milosavljevic2003,Bogdanovic2007,Dotti2010}. In addition, Lense-Thirring torques and viscous dissipation within the inner accretion flows lead to Bardeen-Petterson alignment between the SMBH spin axes and their surrounding disks~\cite{BardeenPetterson1975}. Over longer timescales, coherent accretion and gas torques can further drive the individual SMBH spins toward alignment with the angular momentum of the circumbinary disk and the binary orbit~\cite{Bogdanovic2007,Dotti2010}.

Assuming a spin-powered jet launching mechanism, such as the Blandford-Znajek process~\cite{BlandfordZnajek1977}, the relativistic jet axes are expected to be coupled to the SMBH spin directions. Therefore, gas-driven spin alignment naturally translates into a nearly aligned large-scale jet configuration. The resulting jet-jet inclination is thus expected to remain small throughout the gas-rich evolutionary phase of the binary.

The final degree of alignment is not uniquely determined, as it depends on the accretion history, gas morphology, and angular momentum transport in the circumbinary environment. Stochastic accretion episodes, turbulent gas inflows, and disk inhomogeneities naturally introduce small residual angular offsets between the SMBH spins and the large-scale angular momentum axis. Consequently, residual misalignment angles of order $\theta_{\rm mis}\sim1^\circ-5^\circ$ represent a physically plausible regime for the Cosmic Collider mechanism, rather than a finely tuned configuration.

Finally, the Cosmic Collider mechanism does not require a strictly non-zero misalignment angle. Because the proton distributions $n_p'(E_p',r)$ in the two jets are approximately isotropic [see Eq.~(1) of the Main Text, where the full two-jet notation is employed instead of the simplified one introduced in Sec.~\ref{sec:supp_agn_framework}], the proton-proton ($pp$) interaction remains efficient also in the limiting case of nearly co-axial jets. The relevant physical requirement is therefore the existence of a long-lived, quasi-aligned dual-jet configuration naturally produced by gas-rich SMBHB evolution.

\subsection{Mutual Hydrodynamic Compression of the Interaction Core}
\label{sec:jet_jet_compression}

At the onset interaction distance $r_0 = {\cal O}(0.1-1) \, \rm pc$, the two relativistic jets begin to overlap while remaining nearly co-aligned, with a relative misalignment angle of only a few degrees (see Main Text and Sec.~\ref{sec:jet_spin_and_misalignment}). As a reference fiducial setup, we adopt $r_0 = 0.5 \, \rm pc$. Owing to the small relative transverse velocity between the two jets, the interaction fundamentally differs from both head-on collisions and classical single-jet recollimation against an external ambient medium~\cite{Daly1988,KomissarovFalle1997,Komissarov2007,AsadaNakamura2012}. Instead, building upon the physics of oblique relativistic shock interfaces, the overlapping flows undergo mutual hydrodynamic compression: each jet acts as a high-pressure dynamic boundary for the other, forming a compressed interaction core bounded by oblique shock structures. Transversely, this dynamic compression restricts the primary hadronic interaction volume to a central core radius $R_{pp}(r_0) \equiv R_{\rm jet}(r_0)/\xi_{\rm jet}$, where $R_{\rm jet}(r_0)$ is the jet radius at $r_0$. We adopt an effective compression factor $\xi_{\rm jet} = \mathcal{O}(10)$, with $\xi_{\rm jet} = 10$ as our benchmark value, to account for the transverse compression induced by the overlapping flows.

This mutual compression produces a substantially shallower density profile than that of an isolated conical jet. In the absence of pressure interactions, mass conservation in free geometrical expansion dictates a proton number density scaling as $n_p'(r) \propto r^{-2}$, whereas a perfectly cylindrical flow approaches $n_p'(r) \sim {\rm const}$. The Cosmic Collider interaction zone is expected to occupy an intermediate regime between these two limiting cases. We parameterize the proton density inside the compressed interaction core as $n_p'(r)\propto r^{-a}$, adopting $a=0.5$ as a representative fiducial benchmark for a compressed, non-cylindrical flow. The robustness of the model is systematically tested in the Main Text by varying the compression index $a$ over the broad, physically motivated range $0 \le a \le 1.2$. This interval encompasses nearly cylindrical configurations ($a \sim 0$) as well as moderately expanding compressed flows, while excluding the presently unfavored freely expanding conical regime ($a \simeq 2$). Still, as a stress test, we also explicitly verified the extreme limit of uncompressed, free conical expansion ($a = 2$). Because the total hadronic emissivity scales as $\propto n_{p,1}^{(1)} n_{p,2}^{(2)} R_{pp}^2$, achieving equivalent multi-messenger yields under $a = 2$ simply requires a modest rescaling by a factor of $\sim 2.5$ in either the baseline proton densities $n_{p,i,0}^{(i)}$ ($i \in \{1,2\}$) or the core radius $R_{pp}(r_0)$, demonstrating the strong structural robustness of the Cosmic Collider mechanism against magnetohydrodynamic uncertainties.

The persistence of the compressed interaction region over distances $r_f \sim \mathcal{O}(10) \, \rm pc$ is further supported by relativistic hydrodynamics. Because the two jets remain nearly parallel, the shocked layer is continuously advected downstream at relativistic speed rather than forming a stationary turbulent interface. Relativistic bulk motion increases the growth length of Kelvin-Helmholtz modes, while rapid downstream advection limits the time available for nonlinear disruption~\cite{Hardee2007,Perucho2012,Bodo2013}. Although Kelvin-Helmholtz, current-driven, and other shear instabilities may locally develop within the interaction layer~\cite{Hardee2007,Bodo2013,Perucho2014}, they are not expected to disrupt the overall compressed interaction core on the scales relevant for the present model. We adopt $r_f = 10 \, \rm pc$ as our reference benchmark distance.

Importantly, the Cosmic Collider mechanism does not rely on perfectly laminar flows. The present treatment neglects the additional particle mixing produced by hydrodynamic instabilities. Local turbulent motions and oblique shocks can generate additional velocity dispersion, increasing the fraction of particle pairs with large relative angles and thereby enhancing the CM energy $\sqrt{s}$ available for proton-proton ($pp$) collisions. Such effects would preferentially increase the fraction of interactions occurring at larger CM energies, potentially enhancing the production of secondary particles rather than suppressing it. While a quantitative treatment of shock geometry and non-linear instability evolution requires dedicated three-dimensional relativistic magnetohydrodynamic simulations, neglecting these effects renders our estimates conservative.

The description adopted here should therefore be regarded as an effective large-scale parameterization of the compressed interaction core. The benchmark configuration with $a=0.5$, $\xi_{\rm jet}=10$, and $r_f = 10 \,\rm pc$ captures the enhanced particle density and spatial localization expected from mutual jet compression, while the comprehensive parameter study presented in the Main Text demonstrates that the core predictions of the Cosmic Collider framework remain robust across a broad range of hydrodynamic compression efficiencies.

\subsection{Cosmic Collider Regimes}
\label{sec:cosmic_collider_regimes}

For the bulk Lorentz factors considered in this work ($\Gamma_b \sim 10-20$), denoting the semi-aperture angles of the two parent jets as $\theta_{j,1}$ and $\theta_{j,2}$, their expected values are $\theta_{j,1} \simeq \theta_{j,2} \sim 1/\Gamma_b \sim 3^\circ - 6^\circ$. The effective semi-aperture angle of the resulting structured final jet $\theta_{j,f}$ instead describes the post-interaction morphology and does not constrain the initial overlap condition between the two parent jets.

As discussed in Sec.~\ref{sec:jet_spin_and_misalignment}, gas-rich SMBHBs naturally evolve toward a nearly aligned spin-orbit configuration through the combined action of circumbinary gas torques and viscous alignment within individual accretion flows~\cite{BardeenPetterson1975,Bogdanovic2007,Dotti2010}. Since relativistic jets are launched approximately along the SMBH spin axes, the two parent jets are likewise expected to remain nearly co-aligned during a substantial fraction of the hard-binary phase (see Sec.~\ref{sec:SMBHB_orbit}). Therefore, degree-scale relative jet inclinations are astrophysically favored, whereas large mutual inclinations require a significant residual spin-orbit misalignment.

The geometrical efficiency of the Cosmic Collider mechanism is determined by the competition between two physical effects governed by the jet-jet misalignment angle $\theta_{\rm mis}$. Larger values of $\theta_{\rm mis}$ yield higher CM energies for individual $pp$ collisions, at the expense of a reduced spatial extent of the overlap region between the two relativistic jets, as measured by $r_f$. The resulting contraction of the interaction volume can compensate for, or even dominate over, the gain in collision energy. The interplay between alignment and CM energy naturally defines three distinct Cosmic Collider regimes:

\begin{itemize}

\item \textbf{Extended-Overlap Regime ($\theta_{\rm mis} \rightarrow 0^\circ$):} In this nearly co-aligned configuration, the two relativistic jets overlap over the largest possible distance. The mutual hydrodynamic compression discussed in Sec.~\ref{sec:jet_jet_compression} maintains an extended interaction core, maximizing both the interaction volume and the $pp$ optical depth $\tau_{pp}$ (see Main Text). In this limiting geometry, the CM energy of individual collisions is reduced compared to misaligned configurations; however, the vast reservoir of available hadronic interactions and the quasi-isotropic proton distribution in the jet frames (see Main Text and Sec.~\ref{sec:jet_spin_and_misalignment}) can compensate for the lower CM energy.

\item \textbf{Efficient Cosmic Collider Regime ($\theta_{\rm mis} \sim 1^\circ - 5^\circ$):} For small but finite degree-scale misalignments, the two parent jets remain sufficiently overlapping to preserve the extended compressed interaction region up to $r_f \sim \mathcal{O}(10) \, \mathrm{pc}$, while the relative inclination enhances the available CM energy for $pp$ collisions. This regime provides the optimal physical balance between CM energy of individual collisions and interaction volume. The angular range adopted throughout the Main Text should be regarded as a representative realization of this physically favored regime rather than a rigid threshold.

\item \textbf{Crossing Regime ($\theta_{\rm mis} \gg \theta_{j,1}, \theta_{j,2}$):} For large mutual inclinations exceeding the semi-aperture angles of the parent jets, the two flows intersect only over a limited volume. Individual $pp$ collisions can occur at substantially higher CM energies, but the reduced overlap length strongly suppresses the total number of interactions and prevents the formation of the extended compressed interaction core considered in this work (see Sec.~\ref{sec:jet_jet_compression}).

\end{itemize}

These regimes should not be interpreted as equally probable geometric configurations. The alignment processes operating in gas-rich SMBHBs preferentially populate the nearly co-aligned region of the parameter space, whereas highly inclined configurations are expected to be less common. Consequently, the relevant quantity entering the demographic estimate in Sec.~\ref{sec:demographics} is not the fraction of systems lying within an arbitrary angular interval under an isotropic assumption, but rather the physically selected fraction of dual-jet SMBHBs capable of sustaining an extended overlap region, denoted by $f_{\rm align}$.

The precise value of $f_{\rm align}$ depends on the population distribution of residual spin and jet misalignments, as well as on the details of jet launching and propagation. Determining this distribution requires dedicated population-synthesis modeling and three-dimensional relativistic magnetohydrodynamic simulations. Here, we treat $f_{\rm align}$ as an effective demographic parameter that encapsulates these underlying astrophysical ingredients.

\subsection{Cosmic Collider Demographics and Rarity Estimation}
\label{sec:demographics}

To evaluate the observational viability of the Cosmic Collider framework and establish whether such systems represent an exceptionally rare anomaly or an accessible class of multi-messenger sources, we perform a Fermi-style demographic estimate of the number of Cosmic Colliders detectable through their neutrino emission, which is unaffected by intergalactic attenuation and thus provides the most extended observational horizon (see Main Text). The primary objective is to demonstrate that the framework naturally accounts for a small, non-zero population of detectable sources without requiring artificial fine-tuning of SMBH demographics or jet launching parameters. The total number of detectable Cosmic Colliders $N_{\rm CC}$ can be factored as
\begin{equation}
N_{\rm CC} = n_{\rm host} \, V_{\rm eff} \times f_{\rm AGN} \times f_{\rm RL} \times f_{\rm bin} \times f_{\rm dual} \times f_{\rm align} \times f_{\rm obs} \, ,
\label{eq:fermi_colliders}
\end{equation}
where $n_{\rm host}$ is the comoving density of suitable host galaxies, $V_{\rm eff}$ is an effective comoving volume that accounts for both relativistic beaming and the flux-limited horizon of the detector, and each remaining factor represents a physical probability or environmental fraction, as defined below.

\begin{enumerate}

\item {\bf Host Galaxy Density ($n_{\rm host} \sim 5 \times 10^{-4}\,\mathrm{Mpc}^{-3}$):} The spatial density of massive elliptical host galaxies ($M_\star \ge 10^{11}\,M_\odot$) capable of hosting $M = {\cal O}(10^8)\,M_\odot$ binary systems (see Sec.~\ref{sec:SMBHB_orbit}) is $n_{\rm host} = {\cal O}(10^{-4} - 10^{-3})\,\mathrm{Mpc}^{-3}$~\cite{Conselice2016, Marchesini2009}. We adopt $n_{\rm host} = 5 \times 10^{-4} \,\mathrm{Mpc}^{-3}$.

\item {\bf Beaming-Weighted Effective Volume ($V_{\rm eff}$):} The observed flux of a Cosmic Collider depends on the viewing angle $\theta_{\rm obs}$ between the jet axis and the line of sight through the Doppler factor ${\cal D}_D(\theta_{\rm obs}) = [\Gamma_b(1-\beta_b\cos\theta_{\rm obs})]^{-1}$, scaling as ${\cal D}_D^3/d_L^2$ for a stationary emission region [see Eq.~(5) of the Main Text]. Sources observed closer to the jet axis are therefore detectable out to larger distances. Normalizing to the benchmark SED of Fig.~4 of the Main Text (${\cal D}_D = \Gamma_b$), which exceeds the sensitivity of a given detector by a margin $M_{\rm ref}$ at $z_{\rm ref} = 0.2$, the horizon redshift $z_{\rm hor}(\theta_{\rm obs})$ is the distance at which the flux, scaled as ${\cal D}_D^3/d_L^2$ from this reference point, drops to the detector sensitivity, i.e.,
\begin{equation}
d_L[z_{\rm hor}(\theta_{\rm obs})] = d_L(z_{\rm ref})\, M_{\rm ref}^{1/2} \left[\frac{{\cal D}_D(\theta_{\rm obs})}{\Gamma_b}\right]^{3/2}~,
\end{equation}
which we conservatively cap at $z_{\rm hor} \le 1$. Here we neglect the redshift-induced energy shift of the nearly flat neutrino SED and adopt the margin at the most favorable point of the sensitivity curve, consistently with the order-of-magnitude nature of the estimate. Since randomly oriented systems have viewing angles distributed as $\frac{1}{2}\sin\theta_{\rm obs}\,{\rm d}\theta_{\rm obs}$ ($0 \le \theta_{\rm obs} \le \pi$), and both the jet and counter-jet pairs form an interaction region (as implicitly assumed in the standard beaming fraction below), the effective volume reads
\begin{equation}
V_{\rm eff} = \int_0^{\pi/2} {\rm d}\theta_{\rm obs}\, \sin\theta_{\rm obs}\, V_{\rm com}[z_{\rm hor}(\theta_{\rm obs})]~,
\label{eq:Veff}
\end{equation}
with $V_{\rm com}(z) = (4\pi/3)\,d_C^3(z)$ the comoving volume, where $d_C(z) = d_L(z)/(1+z)$ is the comoving distance in a flat Planck $\Lambda$CDM cosmology~\cite{Planck2018}, and $z_{\rm hor}$ is obtained by numerically inverting the $d_L(z)$ relation. This expression generalizes the standard product $f_{\rm beam}\,V_{\rm com}(z_{\rm hor})$, with $f_{\rm beam} = 1-\cos\theta_j \simeq 1/(2\Gamma_b^2)$, which is recovered for a single horizon inside the beaming cone and no detection outside it, and naturally weights the brighter on-axis sources. For IceCube-Gen2~\cite{IceCubeGen2} ($M_{\rm ref} \simeq 10$), the horizon ranges from $z_{\rm hor} \simeq 0.53$ at the beaming cone boundary ($\theta_{\rm obs} \simeq 1/\Gamma_b$) to the cap $z_{\rm hor} = 1$ for nearly on-axis sources, yielding $V_{\rm eff} \simeq 2.9 \times 10^8\,\mathrm{Mpc}^3$ ($V_{\rm eff} \simeq 4.6 \times 10^7\,\mathrm{Mpc}^3$ for the current IceCube sensitivity).

\item {\bf Active AGN Fraction ($f_{\rm AGN} \sim 10^{-1}$):} Only a subpopulation of massive host galaxies harbors SMBHs actively accreting matter from their circum-nuclear environment. While the overall galaxy population exhibits an average active accretion duty cycle of $f_{\rm AGN} \sim 10^{-2}$, we specifically target massive ellipticals hosting $M = {\cal O}(10^8)\,M_\odot$ SMBHB systems. For these high-mass systems, the active fraction is systematically higher, reaching $f_{\rm AGN} \sim 10^{-1}$~\cite{HeckmanBest2014}.

\item {\bf Relativistic Jet Fraction ($f_{\rm RL}\sim10^{-1}$):} Only a subset of active high-mass SMBHs launch powerful relativistic jets. The radio-loud fraction of massive SMBHs provides a robust empirical estimate for this probability, on the order of $f_{\rm RL}\sim 10^{-1}$~\cite{UrryPadovani1995,Ivezic2002}.

\item {\bf Sub-Parsec Binary Fraction ($f_{\rm bin} \sim 5 \times 10^{-2}$):} Following hierarchical galaxy mergers, the dynamical fraction of massive nuclear SMBHs residing in sub-parsec binary systems in the hard-binary regime [$d_{\rm bin} = {\cal O}(0.1) \, \mathrm{pc}$] is conservatively estimated to be $f_{\rm bin} \sim 5 \times 10^{-2}$~\cite{Begelman1980, Volonteri2003}. We emphasize that $f_{\rm bin}$ reflects purely the {\it dynamical} presence of a sub-parsec binary, governed by the long residence time associated with the final-parsec bottleneck (see Sec.~\ref{sec:SMBHB_orbit}). We explicitly do not adopt the much smaller binary quasar fraction ($\lesssim 0.3\%$, e.g., Ref.~\cite{Volonteri2003}), which requires simultaneous high-luminosity accretion onto both SMBHs; in our factorized framework, simultaneous accretion and jet activation are accounted for separately via $f_{\rm dual}$ to avoid double-counting.

\item {\bf Dual Jet Activation Fraction ($f_{\rm dual} \sim 10^{-1}$):} The fraction of time during which both SMBHs in the sub-parsec binary are simultaneously active and launching relativistic outflows is parameterized as $f_{\rm dual} \sim 10^{-1}$~\cite{VanWassenhove2012,Steinborn2016}. While gas accretion in circumbinary disks is highly asymmetric, hydrodynamic simulations of major mergers indicate that dual fueling occurs over a non-negligible fraction of the orbital residence time. This factor accounts for the overlapping duty cycles of the two AGN, ensuring that both engines are concurrently fueled during the hard-binary bottleneck. This choice keeps the combined probability of hosting an active dual system ($f_{\rm bin} \times f_{\rm dual}$) consistent with the statistical upper limits derived for binary quasar populations in hierarchical structure formation models~\cite{Volonteri2003}.

\item {\bf Jet Alignment Fraction ($f_{\rm align} \sim 5 \times 10^{-1}$):} As discussed in Sec.~\ref{sec:jet_spin_and_misalignment} and Sec.~\ref{sec:cosmic_collider_regimes}, relativistic jets in sub-parsec binaries do not possess randomly oriented vectors. Instead, circumbinary disk torques and Bardeen-Petterson spin alignment actively damp initial orbital misalignments, preferentially driving the twin jets toward small relative inclinations~\cite{Bogdanovic2007, Dotti2010}. While gas-driven alignment mechanisms heavily populate low-inclination states in the SMBHB systems under consideration, we conservatively estimate the fraction of dual-jet systems populating the optimal degree-scale misalignment window ($\theta_{\rm mis} \sim 1^\circ - 5^\circ$, corresponding to the Efficient Cosmic Collider Regime discussed in Sec.~\ref{sec:cosmic_collider_regimes}) as $f_{\rm align} \sim 5 \times 10^{-1}$. Because $\theta_{\rm mis}$ is comparable to or smaller than the jet beaming angles ($\theta_{\rm mis} \lesssim \theta_{j,1}, \theta_{j,2}$), the relativistic boost of the secondary gamma-ray and neutrino products remains directed along our line of sight, preserving observable multi-messenger fluxes within the same beaming envelope.

\item {\bf Observational Coverage Factor ($f_{\rm obs} \sim 5 \times 10^{-1}$):} Since the flux-limited horizon is already included in $V_{\rm eff}$, the fraction $f_{\rm obs}$ only accounts for sky coverage, instrument duty cycles, and Earth shadowing effects in deep-underground or under-ice neutrino observatories~\cite{IceCube,IceCubeGen2}.

\end{enumerate}

\noindent Evaluating Eq.~(\ref{eq:fermi_colliders}) with the baseline values above yields $N_{\rm CC} \simeq 1.8$ for IceCube-Gen2, i.e., ${\cal O}(1)$ detectable Cosmic Colliders, with several factors (e.g., $f_{\rm bin}$, $f_{\rm align}$) chosen conservatively. Adopting less conservative, yet plausible, astrophysical values, or source parameters toward the upper end of the explored ranges (which increase the margin $M_{\rm ref}$), can raise this estimate to $N_{\rm CC} \sim {\cal O}(10)$. For the current IceCube sensitivity ($M_{\rm ref} \simeq 1.6$, with horizon $z_{\rm hor} \simeq 0.25$ at the beaming cone boundary), the same calculation yields $N_{\rm CC} \simeq 0.3$, consistent with the present absence of identified Cosmic Colliders. Since the horizon is defined by the $90\%$ C.L. sensitivity rather than by the discovery potential, this also implies a non-negligible probability ($\sim 25\%$) that a Cosmic Collider already contributes to current IceCube data as a sub-threshold excess, a possibility that could be tested with dedicated stacking analyses of binary blazar candidates. Given the order-of-magnitude uncertainties affecting the individual factors, these numbers should be read as a plausibility argument rather than a precise demographic prediction: their key outcome is that $N_{\rm CC} \gtrsim 1$ emerges for next-generation detectors without any fine-tuning of the underlying astrophysical ingredients.

Cosmic Colliders are therefore sufficiently rare to have evaded current blind surveys, consistently with the expected number of systems within the horizon of present neutrino observatories, while not being so rare that their detection would require an implausible fine-tuning of SMBH and SMBHB demographics or of jet parameters and geometry. This makes the detection of the nearest Cosmic Colliders a realistic target for next-generation observatories.

\section{Competitive Hadronic Cooling Regimes across Blazar Classes}
\label{sec:supp_cooling_regimes}

To establish where and how the different radiative and hadronic processes operate within the dual-jet system, we evaluate the competitive cooling dynamics for relativistic protons. Building upon the jet profiles and target photon fields defined in Sec.~\ref{sec:target_photon_fields}, we evaluate the efficiency of the Cosmic Collider mechanism by continuously tracing the transition from pure BL Lacs to photon-rich FSRQs through intermediate regimes.

\subsection{Multiscale Spatial Geometry}
\label{sec:multiscale_geometry}

To accurately model the proton loss rates without introducing {\it ad hoc} radial profile functions, we adopt a physically grounded multiscale geometry based on mutual hydrodynamic compression (Sec.~\ref{sec:jet_jet_compression}). To parameterize the transition across blazar classes within a unified AGN framework, we explore bulk Lorentz factors $\Gamma_b \in [10, 20]$, adopting $\Gamma_b = 15$ as reference baseline, and fix the accretion disk luminosity to a representative intermediate range $L_{\rm disc} = {\cal O}(10^{44} - 10^{45}) \, \rm erg \, s^{-1}$ with benchmark value $L_{\rm disc} = 5 \times 10^{44} \, \rm erg \, s^{-1}$. Within this setup, variations across the blazar sequence~\cite{BlazarSeq} from BL Lac-like to FSRQ-like regimes are naturally encoded through a single external reprocessing parameter: the BLR covering factor $\eta_{\rm BLR}$ with $\eta_{\rm DT} = 5 \, \eta_{\rm BLR}$, as defined in Sec.~\ref{sec:target_photon_fields}.

\medskip

\noindent {\it Local Quasi-Neutrality in the Jet.} Macro-scale charge neutrality is a fundamental requirement in astrophysical plasmas. Any macroscopic charge separation on scales larger than the Debye length $\lambda_D$ would induce catastrophic radial electric fields ${\bf E}_R$, destroying the coherence of the flow. Consequently, recalling the total integrated comoving proton number density $n_{p,0}'$ (in $\text{cm}^{-3}$) defined in the Main Text and the corresponding total integrated comoving electron number density $n_{e,0}'$ defined after Eq.~(\ref{eq:jet_profiles_ne}), the plasma strictly satisfies $n_{e,0}' \simeq n_{p,0}'$. Since protons and electrons follow a similar spatial dependence [see Eq.~(1) of the Main Text and Eq.~(\ref{eq:jet_profiles_ne}), respectively], this ensures zero net charge density across the entire jet cross-section of radius $R_{\rm jet}$, as well as along the jet propagation axis $r$.

\medskip

\noindent {\it Spatial Hierarchy of Emission Regions ($R_{\rm jet} \gg R_{pp} \gg R_{e\text{-sync}}$).} Rather than assuming a uniform single-zone emitter, our framework naturally establishes a three-scale spatial hierarchy ($R_{\rm jet} \gg R_{pp} \gg R_{e\text{-sync}}$) governed by jet-jet interaction, jet hydrodynamics, and radiative cooling timescales. As detailed in Sec.~\ref{sec:jet_jet_compression}, the transverse localization across the overlap region is parameterized by the hydrodynamic compression factor $\xi_{\rm jet} \equiv R_{\rm jet}/R_{pp} = {\cal O}(10)$. Using our baseline benchmark value $\xi_{\rm jet} = 10$, the full geometric hierarchy is defined as follows:

\begin{itemize}
\item { \bf Global Jet Envelope ($R_{\rm jet}$):} At the jet-jet interaction onset distance $r_0 = {\cal O}(0.1-1)\,{\rm pc}$ (see Main Text), the unperturbed jet boundary exhibits a typical effective semi-aperture angle $\theta_j^{\rm eff} \sim 0.1\,\text{rad}$~\cite{Pushkarev2009} averaged over the initial acceleration phase, yielding a global jet radius $R_{\rm jet}(r_0) \simeq r_0 \tan(\theta_j^{\rm eff})= {\cal O}(0.01-0.1)\,{\rm pc}$. Note that $\theta_j^{\rm eff} \neq \theta_j \equiv 1/\Gamma_b$, as $\theta_j^{\rm eff}$ explicitly accounts for the spatial evolution and saturation of $\Gamma_b(r)$ during the extended magnetohydrodynamic acceleration phase~\cite{VlahakisKoenigl2004, Komissarov2007}. As benchmark values we adopt $r_0 = 0.5\,{\rm pc} \simeq 1.5 \times 10^{18}\, {\rm cm}$ and $R_{\rm jet}(r_0) = 0.05\,\text{pc} \simeq 1.5 \times 10^{17}\,\text{cm}$.

\item {\bf Hadronic Interaction Core ($R_{pp}$):} Compressed by the mutual hydrodynamic pressure described in Sec.~\ref{sec:jet_jet_compression}, the interacting proton populations are concentrated into a central core of radius $R_{pp}(r_0) = R_{\rm jet}(r_0)/\xi_{\rm jet} = {\cal O}(10^{16}) \, {\rm cm}$. For our benchmark values, this yields $R_{pp}(r_0) \simeq 1.5 \times 10^{16} \, {\rm cm} \ll R_{\rm jet}(r_0)$. We have verified that our physical results remain robust against variations of $R_{pp}$ across $\xi_{\rm jet} \in [5, 20]$. To parameterize the comoving proton spectral density profile $n_p^\prime(E_p^\prime,r)$, we employ Eq.~(1) of the Main Text (dropping jet index subscripts and employing primed quantities for comoving variables, as outlined in Sec.~\ref{sec:supp_agn_framework}). In the core region delimited by $R_{pp}(r_0)$, the physical parameters entering Eq.~(1) of the Main Text are constrained within realistic astrophysical intervals: total integrated comoving proton number density $n_{p,0}' = {\cal O}(10^3 - 10^4) \, {\rm cm}^{-3}$, spectral index $\alpha_p \sim 2$, minimum and maximum Lorentz factors $\gamma_p^{\rm min} \sim 1$ and $\gamma_p^{\rm max} = {\cal O}(10^8 - 10^{11})$, respectively, and compression index $a \in [0, 1.2]$ (see Sec.~\ref{sec:jet_jet_compression}). As reference benchmark values, we adopt $n_{p,0}' = 10^4 \, {\rm cm}^{-3}$, $\alpha_p = 2.2$, $\gamma_p^{\rm min} = 1$, $\gamma_p^{\rm max} = 10^{10}$, and $a = 0.5$. The physical admissibility of this baseline proton density -- which carries the bulk of the jet kinetic energy -- is explicitly verified via the total power evaluation below. The benchmark value $\gamma_p^{\rm max} = 10^{10}$ is consistent with the Hillas confinement criterion~\cite{protonAcc1}, $E_p^{\prime\,\rm max} \lesssim e B' R$. For $B'_0 = {\cal O}(0.1)\,{\rm G}$ [see Eq.~(\ref{eq:jet_profiles_B}) and below], the jet-scale field at $r_0$ yields $e B'_0 R_{\rm jet}(r_0) = {\cal O}(10^{10})\,{\rm GeV}$, while proton acceleration takes place at $r < r_0$, upstream of the jet-jet interaction, where the magnetic field is dominated by its poloidal component ($B' \propto r^{-2}$) and the product $B'R$ is correspondingly larger~\cite{VlahakisKoenigl2004, Komissarov2007}. Within the compressed core, projectiles with $E_p' \gtrsim e B'_0 R_{pp} = {\cal O}(10^{9})\,{\rm GeV}$ have Larmor radii exceeding $R_{pp}$ and sample the core only over a fraction of their path. This affects only the highest-energy decade of the projectile distribution, while the bulk of the target population remains confined, and the robustness of our conclusions is confirmed by the results obtained for $\gamma_p^{\rm max} = 10^8$ (see Sec.~\ref{sec:extended_results}).

\item {\bf Leptonic Synchrotron Emission Zone ($R_{e\text{-sync}}$):} Because ultra-relativistic electrons undergo rapid synchrotron cooling on timescales far shorter than the dynamical crossing time ($t_{e\text{-sync}} \ll t_{\rm dyn}$), non-thermal leptonic emission is strictly confined to compact sub-structures within the shocked layer~\cite{BlandfordKoenigl1979, Ghisellini2005}. We evaluate the radius $R_{e\text{-sync}}$ of the electron-synchrotron emission zone via the causality constraint imposed by the characteristic sub-daily variability of blazar leptonic continua ($\Delta t_{\rm var} \lesssim  1 \, \rm h$)~\cite{Aharonian2007, Ackermann2016}, setting $R_{e\text{-sync}}(r_0) \sim c \Delta t_{\rm var} {\cal D}_D = {\cal O}(10^{14}-10^{15})\,\text{cm}$, with Doppler factor ${\cal D}_D \sim \Gamma_b$. As our reference benchmark, we adopt $R_{e\text{-sync}}(r_0) = 5 \times 10^{14}\,{\rm cm}$, naturally satisfying $R_{e\text{-sync}} \ll R_{pp} \ll R_{\rm jet}$. Ultra-fast flares on timescales of $\sim 10\,\text{min}$ would probe even smaller emission regions, reinforcing the conservative nature of our spatial hierarchy. The comoving electron spectral density profile $n_e^\prime(E_e^\prime,r)$ is parametrized via Eq.~(\ref{eq:jet_profiles_ne}). Assuming charge neutrality which sets $n_{e,0}' = n_{p,0}' = {\cal O}(10^3 - 10^4) \, {\rm cm}^{-3}$, the physical parameters entering Eq.~(\ref{eq:jet_profiles_ne}) within $R_{e\text{-sync}}(r_0)$ span standard intervals: spectral index $\alpha_e \sim 2$, minimum and maximum Lorentz factor $\gamma_e^{\rm min} \sim 1$ and $\gamma_e^{\rm max} = {\cal O}(10^6 - 10^9)$, respectively, and $a \in [0, 1.2]$ (see Sec.~\ref{sec:jet_jet_compression}). As reference benchmark values, we adopt $n_{e,0}' = n_{p,0}' = 10^4\, {\rm cm}^{-3}$, $\alpha_e = 2.2$, $\gamma_e^{\rm min} = 1$, $\gamma_e^{\rm max} = 10^{7}$, and $a = 0.5$. To evaluate the synchrotron radiative cooling and the resulting spectra, the comoving magnetic field profile $B'(r)$ is parameterized via Eq.~(\ref{eq:jet_profiles_B}), taking a characteristic value of $B'_0 = {\cal O}( 0.1)\,\text{G}$ at $r_0$, for which we adopt $B'_0 = 0.1\,\text{G}$ as our baseline reference. While this leptonic component drives the synchrotron emission, its integrated energy density remains strictly sub-dominant with respect to the hadronic component.

\end{itemize}
This geometric scaling yields an effective transverse area ratio $A_{pp}/A_{e\text{-sync}} = {\cal O}(10^2 - 10^3)$, naturally decoupling the extended hadronic interaction core from the highly compact leptonic emission zone.

\medskip

\noindent {\it Kinetic Jet Power.} To evaluate the global energetic viability of the framework, we estimate the total kinetic jet power in the laboratory frame as
\begin{equation}
\label{Ltot}
L_{\rm jet} = \pi R_{\rm jet}^{2} \Gamma_b^2 \beta c \left(U_p^\prime + U_e^\prime + U_B^\prime\right)~,
\end{equation}
where $U_p^\prime$, $U_e^\prime$, and $U_B^\prime = B^{\prime2}/(8\pi)$ denote the comoving energy densities of protons, electrons, and magnetic field, respectively, and $\beta = v/c \simeq 1$. The particle (proton $p$ or electron $e$) energy densities $U_{x=p,e}^\prime(r)$ are evaluated as
\begin{equation}
\label{up}
U_{x=p,e}^\prime(r) = 4\pi \int_{E_x^{\prime\rm min}}^{E_x^{\prime\rm max}} E_x^\prime n_x^\prime(E_x^\prime, r) \,{\rm d}E_x^\prime~,
\end{equation}
where $E_x^{\prime\rm min,max} = m_x c^2 \gamma_x^{\rm min,max}$, and the factor $4\pi$ arises from the integration over all solid angles, recalling that $n_{x=p,e}'$ is the particle isotropic comoving spectral number density per unit solid angle [see Eq.~(1) of the Main Text and Eq.~(\ref{eq:jet_profiles_ne}), respectively]. Given the benchmark parameters defined above -- and noting $m_p \simeq 1836 \, m_e$ -- the non-thermal energy budget is heavily dominated by hadrons with $U_p^\prime \gg U_e^\prime$. Furthermore, evaluating $U_B^\prime$ with the baseline $B_0^\prime = 0.1 \, \rm G$ defined above yields $U_p^\prime \gg U_B^\prime$, confirming that the jet is kinetically dominated at these scales. The mutual hydrodynamic compression described in Sec.~\ref{sec:jet_jet_compression} strongly concentrates the flow within the innermost core region, allowing us to factorize the kinetic power as
\begin{equation}
\label{LtotVsCore}
L_{\rm jet} = K \, L_{\rm jet, core} \simeq K \left( \pi R_{pp}^{2} \Gamma_b^2 \beta c \, U_p^\prime \right)~,
\end{equation}
where we have employed $U_p^\prime \gg U_e^\prime, U_B^\prime$, with $L_{\rm jet, core}$ isolating the power contained inside the core radius $R_{pp} = R_{\rm jet}/\xi_{\rm jet}$, and $K \sim 1.5 - 2$ accounting for the low power distribution outside the core in highly collimated relativistic flows~\cite{Komissarov2007}. 

Scanning over our physical parameter ranges yields $L_{\rm jet} = {\cal O}(10^{46} - 10^{48})\,\mathrm{erg\,s^{-1}}$, with our reference benchmark yielding $L_{\rm jet} = {\cal O}(10^{47})\,\mathrm{erg\,s^{-1}}$. Since this power must be supplied at the jet base, it should be compared with the Eddington accretion power of the central engine, $\dot{M}_{\rm Edd}c^2 \sim 10^{47}\,(M_{1,2}/10^8\,M_\odot)\,\mathrm{erg\,s^{-1}}$. For SMBH masses $M_{1,2} = {\cal O}(10^8 - 10^9)\,M_\odot$, the required jet power is thus at most comparable to $\dot{M}_{\rm Edd}c^2$, within the reach of efficient, magnetically arrested jet launching and spanning from typical to the powerful observed blazars~\cite{Ghisellini2014}.

\medskip

\noindent {\it Inner Boundary Self-Consistency ($r \ge r_0$).}
We define $r_0$ as the physical inner boundary of the active Cosmic Collider region, where jet-jet interaction effectively starts and the hadronic interaction zone is established. While electron-synchrotron emission may also be produced at smaller radii ($r<r_0$), we restrict the calculation of the hadronic and leptonic cooling channels ($pp$, $p\gamma$, and $p\text{-sync}$) to the confined region $r_0 \le r \le r_f$, where $r_f$ defines the outer boundary of the active Cosmic Collider region. This separation is justified by the different dynamical and radiative conditions on the two sides of $r_0$:

\begin{enumerate}

\item {\bf Relativistic escape and Doppler de-boosting:}
Photons produced at $r<r_0$ and propagating outward travel at $c$, faster than the bulk plasma moving at $\beta c$, and therefore progressively outrun the emitting shells. At $r>r_0$, the surviving upstream radiation is consequently strongly forward-directed with respect to the jet motion. In the comoving frame of the downstream plasma, nearly co-propagating photons are redshifted. This reduces both their comoving photon energies and, consequently, their effectiveness as target photons for $p\gamma$ interactions.

\item {\bf Geometric dilution and local radiation dominance:} As upstream electron-synchrotron photons propagate from $r<r_0$ to larger distances, their radiation energy density decreases due to geometric dilution. In contrast, within the confined region $r_0 \le r \le r_f$, electron-synchrotron photons are continuously produced locally by the non-thermal electron population. The local synchrotron field is therefore expected to provide the dominant internal non-thermal photon target for $p\gamma$ interactions, while the residual contribution from electron-synchrotron photons produced at $r<r_0$ is assumed to be subdominant. The local synchrotron field is considered together with the CMB, and potential external BLR and dusty-torus radiation fields.

\end{enumerate}

\medskip

\begin{table}[h]
\centering
\begin{threeparttable}
\setlength{\tabcolsep}{12pt} 
\caption{Reference physical parameters adopted for the proton cooling and efficiency calculations at the onset scale $r_0 = 0.5\,\text{pc}$.}
\label{tab:model_params}
\begin{tabular}{@{}lll@{}}
\hline\hline
\noalign{\vspace{2pt}}
Parameter & Symbol & Benchmark Value \\[3pt]
\hline
\noalign{\vspace{2pt}}
Bulk Lorentz Factor & $\Gamma_b$ & $15$ \\
Accretion Disk Luminosity & $L_{\rm disc}$ & $5 \times 10^{44} \, \rm erg \, s^{-1}$ \\
Interaction Onset Distance & $r_0$ & $0.5 \, \rm pc$ ($1.5 \times 10^{18} \, \rm cm$) \\
Final Interaction Distance & $r_f$ & $10 \, \rm pc$  ($3.1 \times 10^{19} \, \rm cm$) \\
Global Jet Radius at $r_0$ & $R_{\rm jet}(r_0)$ & $1.5 \times 10^{17} \, \rm cm$ \\
Hydrodynamic Compression Factor & $\xi_{\rm jet}$ & 10 \\
Hadronic Core Radius at $r_0$ & $R_{pp}(r_0)$ & $1.5 \times 10^{16} \, \rm cm$ \\
Synchrotron Region Radius at $r_0$ & $R_{e \text{-sync}}(r_0)$ & $5 \times 10^{14} \, \rm cm$ \\
Proton/Electron Total Number Density & $n_{p,0}' = n_{e,0}'$ & $10^4 \,  \rm cm^{-3}$ \\
Proton/Electron Spectral Index & $\alpha_p = \alpha_e$ & $2.2$ \\
Proton Minimum/Maximum Lorentz Factor & $\gamma_p^{\rm min}$, $\gamma_p^{\rm max}$ & $1$, $10^{10}$ \\
Electron Minimum/Maximum Lorentz Factor & $\gamma_e^{\rm min}$, $\gamma_e^{\rm max}$ & $1$, $10^7$ \\
Compression Index & $a$ & $0.5$ \\
Comoving Magnetic Field Strength at $r_0$ & $B'_0$ & $0.1 \, \rm G$ \\
BLR / DT Covering Factors & $\eta_{\rm BLR}, \eta_{\rm DT}$ & $[0, 0.1], \, 5\eta_{\rm BLR}$ \\
Dusty Torus Temperature & $T_{\rm DT}$ & $1000 \, \rm K$ \\
External Anisotropy Factors & $f_{\rm BLR}, f_{\rm DT}$ & $1, 1$ \\[3pt]
\hline\hline
\end{tabular}
\end{threeparttable}
\end{table}

\noindent
Consequently, to systematically compare the dominant interaction regimes and energy-loss channels across the parameter space, we compute the comoving cooling timescales $t_{\chi}'$ and relative integrated energy-loss fractions $f_\chi$, where $\chi \in \{pp, p\gamma, p\text{-sync}\}$, via Eqs.~(\ref{eq:t_pp_explicit})--(\ref{eq:t_psyn}) and~(\ref{eq:f_integrated}), adopting the baseline model parameters summarized in Table~\ref{tab:model_params} and used in the Main Text.

\subsection{Proton Cooling Phase Diagram in the Energy-Distance Plane}
\label{sec:proton_cooling_phase_diagram}

The relative dominance among the three proton energy-loss channels -- proton-proton collisions ($pp$), photohadronic production ($p\gamma$), and proton-synchrotron radiation ($p\text{-sync}$) -- is mapped in Fig.~\ref{fig:tcool} as a phase diagram in the comoving proton energy versus distance plane ($E_p', r$).

\begin{figure}[h]
\centering
\includegraphics[width=.47\textwidth]{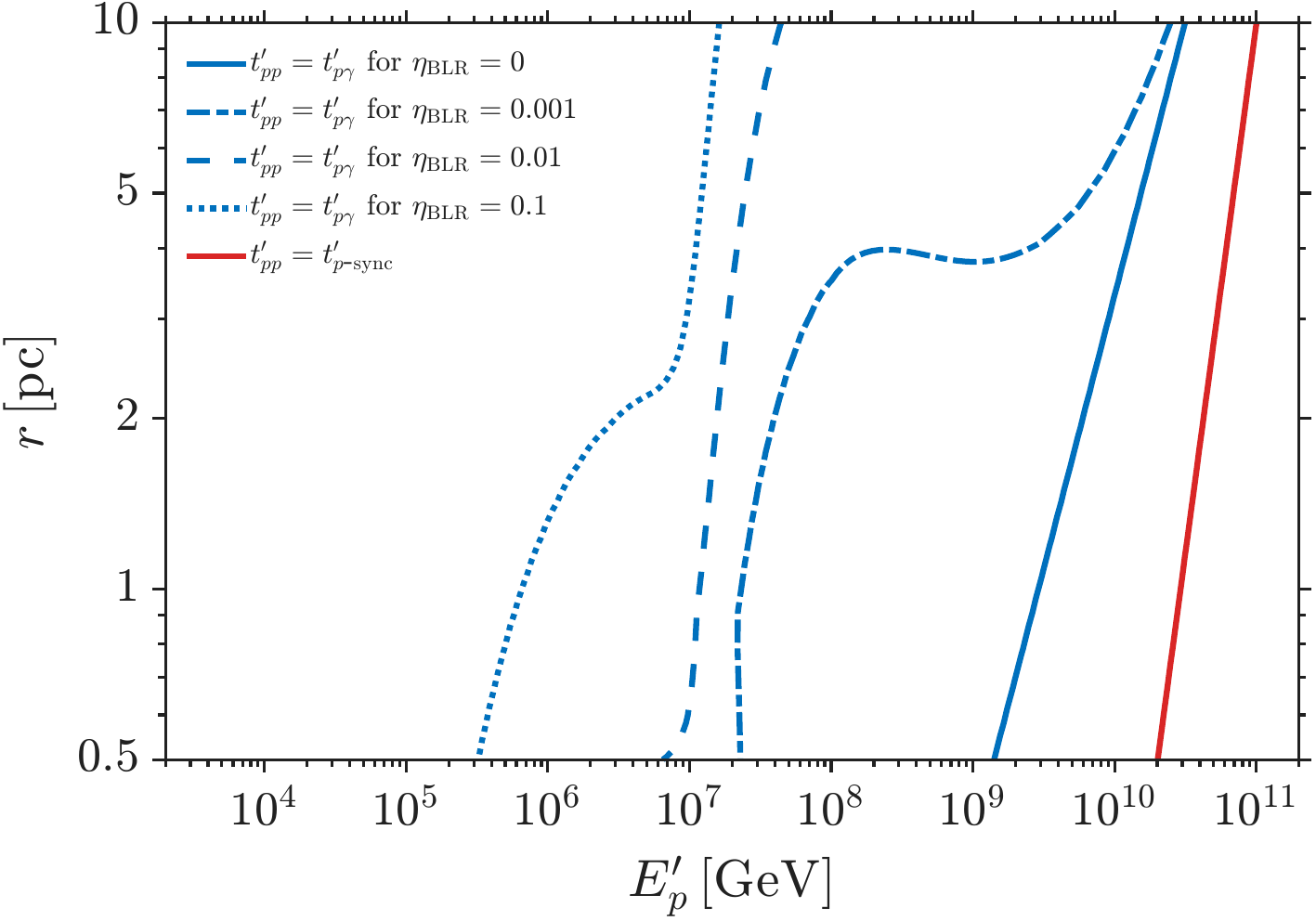}
\caption{Equilibrium contours in the comoving proton energy-distance plane ($E_p', r$) marking the transition boundaries between dominant proton cooling channels. Blue curves trace where the hadronic $pp$ cooling timescale equals the photohadronic $p\gamma$ timescale ($t_{pp}' = t_{p\gamma}'$) for increasing values of the BLR covering factor, ranging from $\eta_{\rm BLR} = 0$ (BL Lac regime) to $\eta_{\rm BLR} = 0.1$ (efficient FSRQ regime), with $\eta_{\rm DT} = 5 \,  \eta_{\rm BLR}$. The red curve marks the boundary where hadronic $pp$ cooling balances proton-synchrotron ($p\text{-sync}$) losses ($t_{pp}' = t_{p\text{-sync}}'$).}
\label{fig:tcool}
\end{figure}

The blue curves in Fig.~\ref{fig:tcool} mark the equilibrium boundaries where $t_{pp}' = t_{p\gamma}'$ for different environmental photon densities, varied systematically via $\eta_{\rm BLR} \in [0, 0.1]$. To the left of these curves, $pp$ collisions represent the fastest cooling mechanism ($t_{pp}' < t_{p\gamma}'$), whereas to the right, photohadronic interactions take over. 

In the pure BL Lac regime ($\eta_{\rm BLR} = 0$), external BLR and DT photon fields are absent, and $p\gamma$ interactions occur solely off internal electron-synchrotron photons and the CMB. Because internal electron-synchrotron photon densities decay along the jet axis more steeply than the proton densities $n_p' \propto r^{-a}$, the $t_{pp}' = t_{p\gamma}'$ equilibrium boundary shifts toward higher proton energies at larger distances. Consequently, $pp$ collisions dominate the cooling budget across the entire spatial extent of the interaction zone up to proton energies $E_p' \sim (10^9 - 10^{10}) \, \rm GeV$.

As external photon fields are introduced ($\eta_{\rm BLR} > 0$), the dense quasi-thermal photon environment characteristic of FSRQs provides an abundant target for photopion production off relativistic protons. This transition from the BL Lac regime ($\eta_{\rm BLR} = 0$) to powerful FSRQs ($\eta_{\rm BLR} = 0.1$) is smooth and continuous: as $\eta_{\rm BLR}$ increases, the $t_{pp}' = t_{p\gamma}'$ equilibrium curve shifts systematically toward lower proton energies, reaching $E_p' \sim 3 \times 10^5\, \rm GeV$ near $r_0 = 0.5 \, \rm pc$ for $\eta_{\rm BLR} = 0.1$, and progressively shrinking the domain where $pp$ collisions act as the primary cooling channel.  

The red curve in Fig.~\ref{fig:tcool} delineates the boundary where $t_{pp}' = t_{p\text{-sync}}'$. Because the proton-synchrotron cooling rate scales linearly with energy [$t_{p\text{-sync}}^{\prime -1} \propto E_p' B^{\prime 2}$, see Eq.~(\ref{eq:t_psyn})], $p\text{-sync}$ losses become faster than $pp$ collisions only at extremely high proton energies, $E_p' \gtrsim 2 \times 10^{10} \, \rm GeV$ near the onset distance $r_0$, where the magnetic field is highest [$B'(r_0)=B'_0 = 0.1 \, \rm G$], increasing to $E_p' \sim 10^{11}\,\rm GeV$ at $r_f$.

\subsection{Integrated Energy-Loss Fractions and Robustness of the Cosmic Collider}
\label{sec:proton_loss_efficiencies}

While the cooling timescales in Fig.~\ref{fig:tcool} determine which process is locally fastest, the overall impact on the proton population depends on the spatially integrated energy-loss fractions $f_\chi(E_p')$, with $\chi \in \{pp, p\gamma, p\text{-sync}\}$, defined in Eq.~(\ref{eq:f_integrated}). Figure~\ref{fig:fchi} displays $f_{pp}$, $f_{p\gamma}$, and $f_{p\text{-sync}}$ integrated over $r \in  [r_0, r_f]$ for four representative blazar regimes.

\begin{figure}[t]
\centering
\includegraphics[width=.47\textwidth]{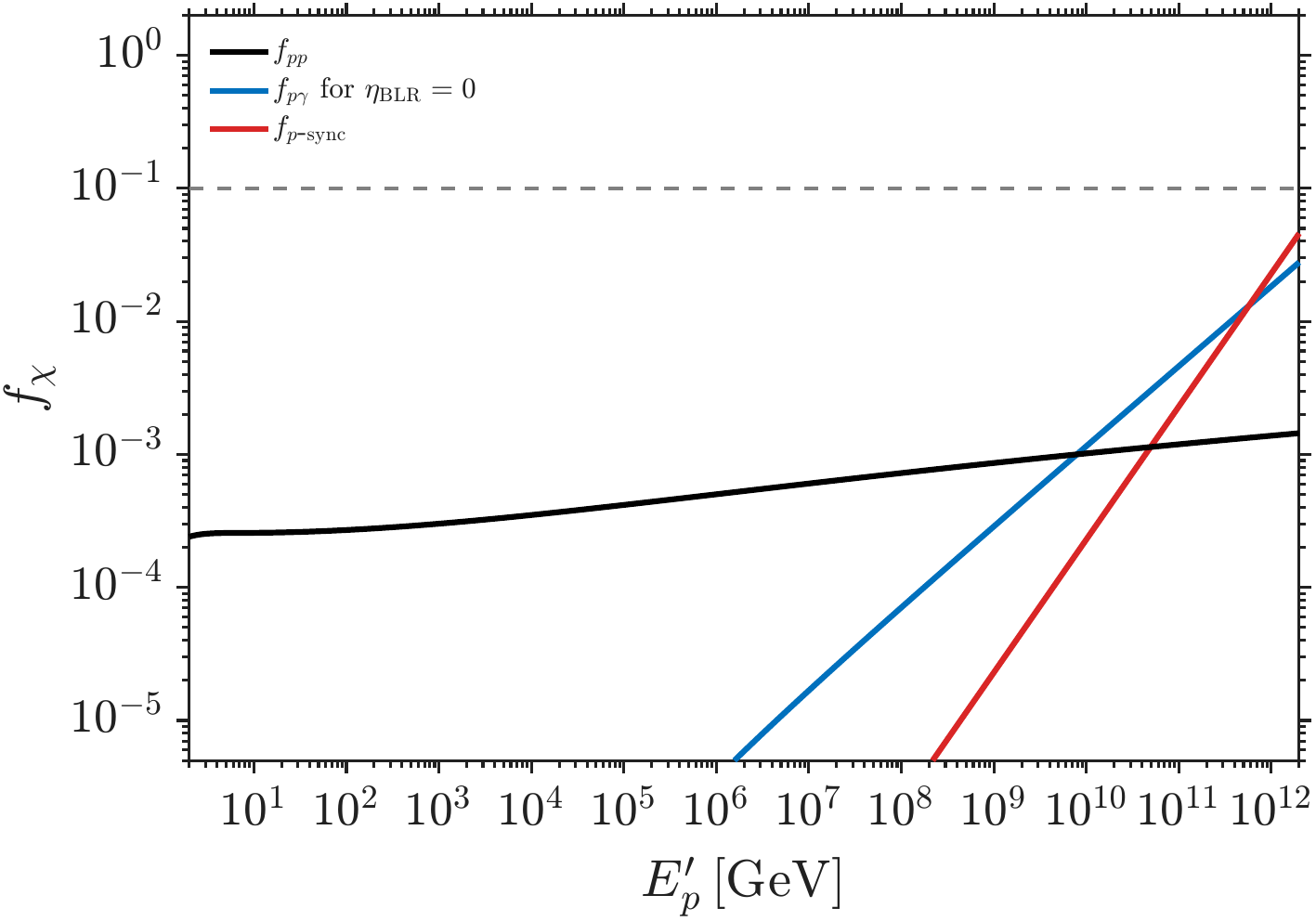}\hspace{0.4cm}
\includegraphics[width=.47\textwidth]{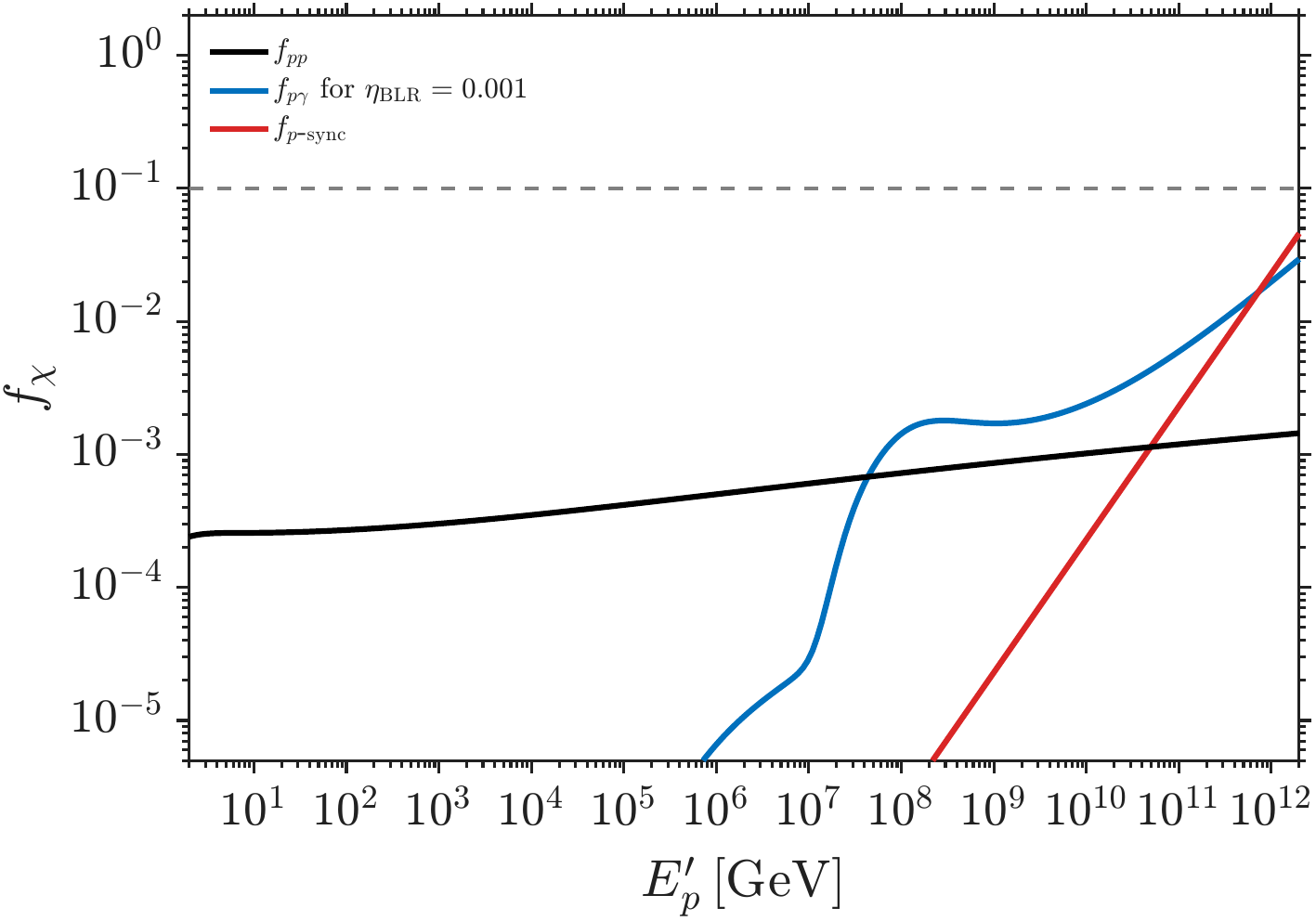}

\vspace{0.3cm}

\includegraphics[width=.47\textwidth]{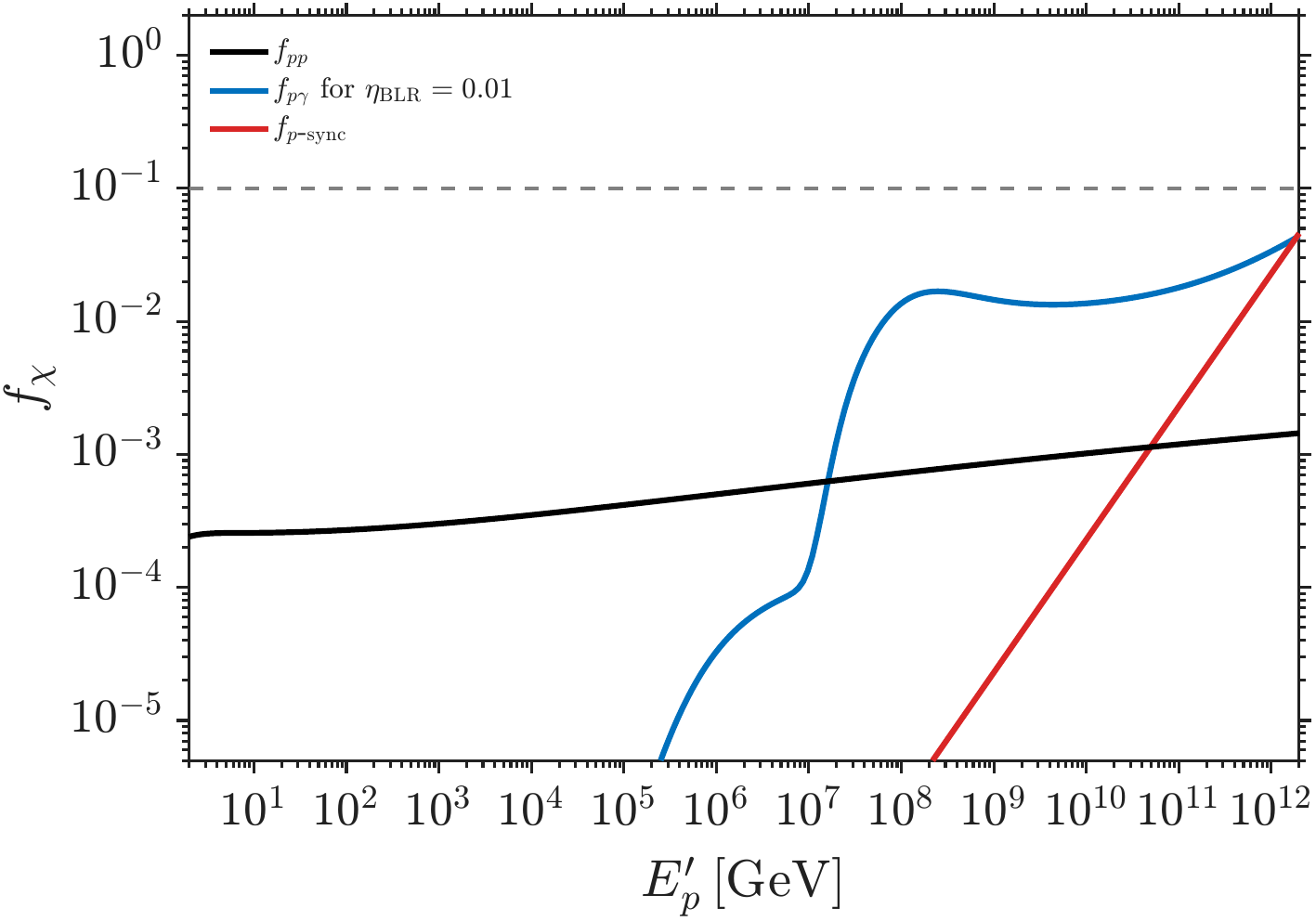}\hspace{0.4cm}
\includegraphics[width=.47\textwidth]{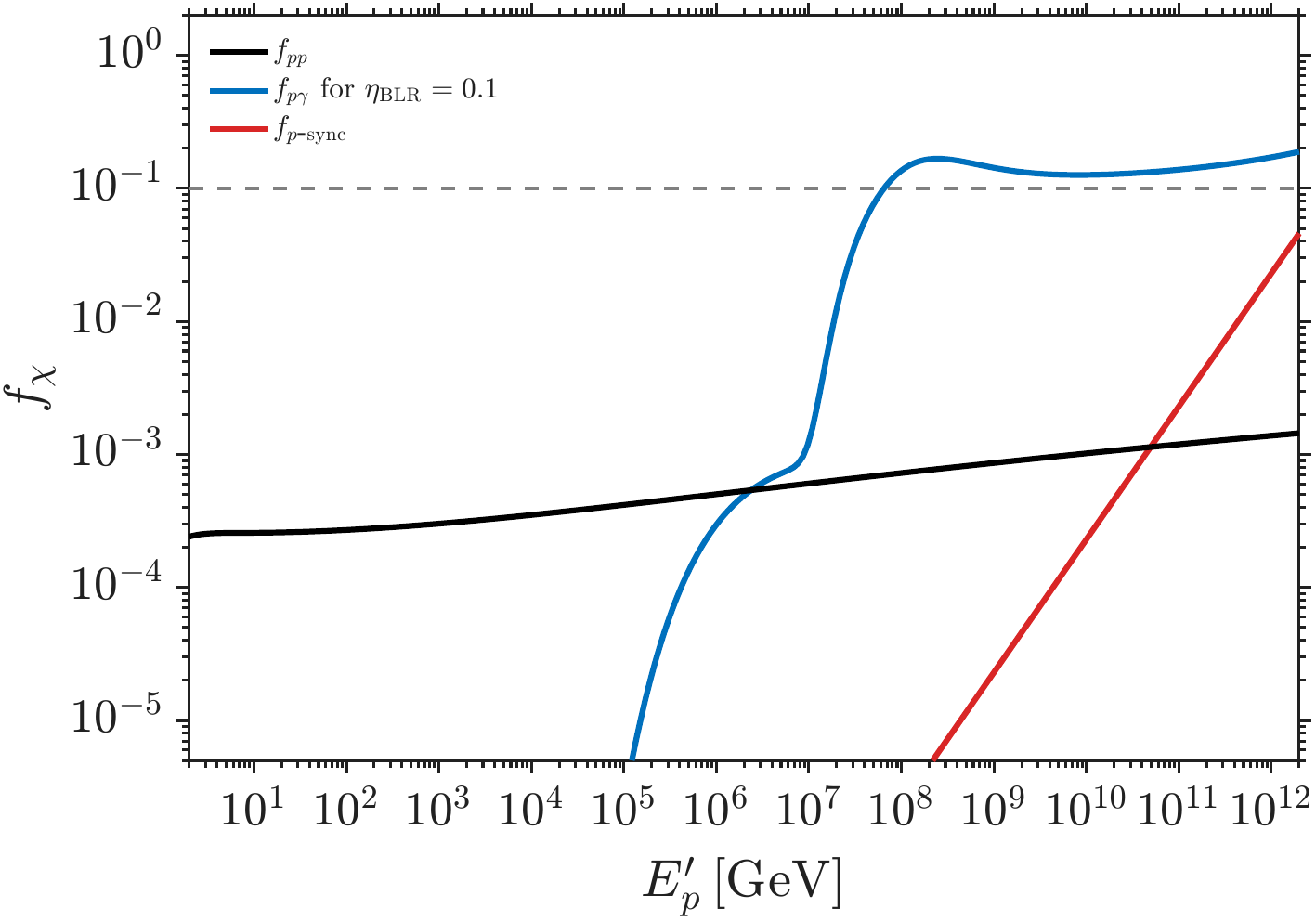}
\caption{Integrated energy-loss fractions $f_\chi$ ($\chi \in \{pp, p\gamma, p\text{-sync}\}$) as a function of the comoving proton energy $E_p'$, evaluated over the interaction region $r \in [r_0, r_f]$ for four representative values of the BLR covering factor: $\eta_{\rm BLR} = 0$ (top-left, BL Lac regime), $\eta_{\rm BLR} = 0.001$ (top-right), $\eta_{\rm BLR} = 0.01$ (bottom-left), and $\eta_{\rm BLR} = 0.1$ (bottom-right, efficient FSRQ regime). Curves denote $f_{pp}$ (black), $f_{p\gamma}$ (blue), and $f_{p\text{-sync}}$ (red). The horizontal dashed line highlights the reference efficiency level $f_\chi = 0.1$.}
\label{fig:fchi}
\end{figure}

A key physical insight revealed by Fig.~\ref{fig:fchi} concerns the operational independence of the Cosmic Collider. The horizontal dashed line marks an efficiency threshold of $f_\chi = 0.1$, corresponding to a $10\%$ fractional energy loss for a proton traversing the interaction zone. As long as $f_\chi \lesssim 0.1$, the process $\chi \in \{pp, p\gamma, p\text{-sync}\}$ extracts only a minor fraction of the primary proton energy (see Sec.~\ref{sec:fChi}). Even where $p\gamma$ cooling is locally faster than $pp$ ($t_{p\gamma}' < t_{pp}'$), photohadronic losses do not significantly deplete the proton population as long as $f_{p\gamma} \lesssim 0.1$, leaving the $pp$ collision rate and secondary particle production effectively unperturbed.

In all the subpanels of Fig.~\ref{fig:fchi}, both $f_{pp}$ and $f_{p\text{-sync}}$ remain identical, while $f_{p\gamma}$ grows as $\eta_{\rm BLR}$ increases from $\eta_{\rm BLR}=0$ (BL Lac limit, top-left), through intermediate states $\eta_{\rm BLR}=0.001 - 0.01$ (top-right and bottom-left), up to $\eta_{\rm BLR}=0.1$ (efficient FSRQ regime, bottom-right). Across all panels and throughout the explored energy range ($2 \, {\rm GeV} \le E_p' \le 2 \times 10^{12} \, {\rm GeV}$), $f_{pp}$ maintains a stable level $f_{pp} \sim 2.5 \times 10^{-4} - 1.5 \times 10^{-3}$ [corresponding to an optical depth $\tau_{pp} \sim 5 \times 10^{-4} - 3 \times 10^{-3}$, see Eq.~(\ref{eq:f_vs_tau_pp})], while $f_{p\text{-sync}} \propto E_p'$ [see Eqs.~(\ref{eq:t_psyn}) and~(\ref{eq:f_integrated})] exceeds $f_{pp}$ only at $E_p' \gtrsim 5 \times 10^{10} \, \rm GeV$ and remains below $0.1$ across the entire energy range. In contrast, the scaling of $f_{p\gamma}$ with $\eta_{\rm BLR}$ delineates three physical regimes:

\begin{itemize}

\item {\bf BL Lac Regime ($\eta_{\rm BLR} = 0$, top-left):} Hadronic $pp$ collisions dominate the energy loss ($f_{pp} > f_{p\gamma}, f_{p\text{-sync}}$) up to $E_p' \sim 10^{10} \, \rm GeV$, while photohadronic losses remain negligible ($f_{p\gamma} < 0.1$) across the entire energy range. The $pp$ mechanism feeding the Cosmic Collider thus operates at maximum efficiency in BL Lac objects without significant photohadronic competition. Moreover, with secondaries produced almost exclusively by $pp$ collisions, BL Lac binaries provide the cleanest laboratories for probing UHE hadronic physics.

\item {\bf Intermediate Regime ($\eta_{\rm BLR} = 0.001 - 0.01$, top-right and bottom-left):} As external photon fields are gradually introduced, $f_{p\gamma}$ increases at intermediate energies, $E_p' \sim (10^6 - 10^{9}) \, \rm GeV$, due to interactions off BLR and dusty torus photons, so that $f_{pp} > f_{p\gamma}$ holds up to $E_p' \sim (2 - 5) \times 10^7 \, \rm GeV$. However, $f_{p\gamma}$ remains below the $0.1$ threshold across the entire energy range, allowing the $pp$-driven Cosmic Collider to operate efficiently up to the highest energies.
    
\item {\bf Efficient FSRQ Regime ($\eta_{\rm BLR} = 0.1$, bottom-right):} In powerful FSRQs, intense external photon fields from the BLR and dusty torus boost $f_{p\gamma}$, causing it to exceed $f_{pp}$ for $E_p' \gtrsim 2 \times 10^6 \, \rm GeV$. However, $f_{p\gamma}$ remains below $0.1$ up to $E_p' \sim 5 \times 10^7 \, \rm GeV$, above which it saturates at $f_{p\gamma} \sim 0.15 - 0.2$. As a result, $pp$ interactions proceed without substantial energy depletion even in luminous FSRQs up to these energies, ensuring that the Cosmic Collider remains fully functional well beyond PeV scales.

\end{itemize}

In summary, the Cosmic Collider mechanism driven by $pp$ collisions in dual-jetted SMBHBs is intrinsically robust, even under the conservative assumption of isotropic external photon fields ($f_{\rm BLR} = f_{\rm DT} = 1$, see Sec.~\ref{sec:target_photon_fields}). In BL Lac objects, $pp$ interactions dictate the proton energy-loss budget up to $E_p' \sim 10^{10} \, \rm GeV$, with no channel exceeding $f_\chi = 0.1$, allowing the Cosmic Collider to probe particle physics in the UHE regime. In FSRQs, although $p\gamma$ losses dominate for $E_p' \gtrsim (10^6 - 10^8) \, \rm GeV$, the condition $f_{p\gamma} < 0.1$ guarantees that hadronic $pp$ collisions feeding the Cosmic Collider remain fully operational and efficient up to near-UHE scales.

\section{Extended Results}
\label{sec:extended_results}

In this section, we provide extended parameter studies illustrating the dependence of the $pp$ optical depth $\tau_{pp}$, the differential collider luminosity, and the resulting emitted secondary spectra (gamma rays and neutrinos) on the primary proton injection parameters, specifically the spectral indices $\alpha_{p,i}$ and maximum Lorentz factors $\gamma_{p,i}^{\mathrm{max}}$, with $i \in \{1, 2\}$ denoting the proton populations in the two colliding jets (see Main Text for notation). We vary $\alpha_{p,i} \in [2.1, 2.4]$ and $\gamma_{p,i}^{\mathrm{max}} \in [10^8, 10^{11}]$ so as to span the physically plausible range expected from standard relativistic shock acceleration and magnetic reconnection mechanisms in blazar jets~\cite{Drury1983, BlandfordEichler1987, SironiSpitkovsky2014,Marcowith2016}. Unless specified otherwise, all other Cosmic Collider parameters are held fixed at their baseline benchmark values, as reported in Table~\ref{tab:model_params}.

\subsection{Proton-Proton Optical Depth}
\label{sec:extended_pp_optical_depth}

\begin{figure}[t]
\centering
\includegraphics[width=.47\textwidth]{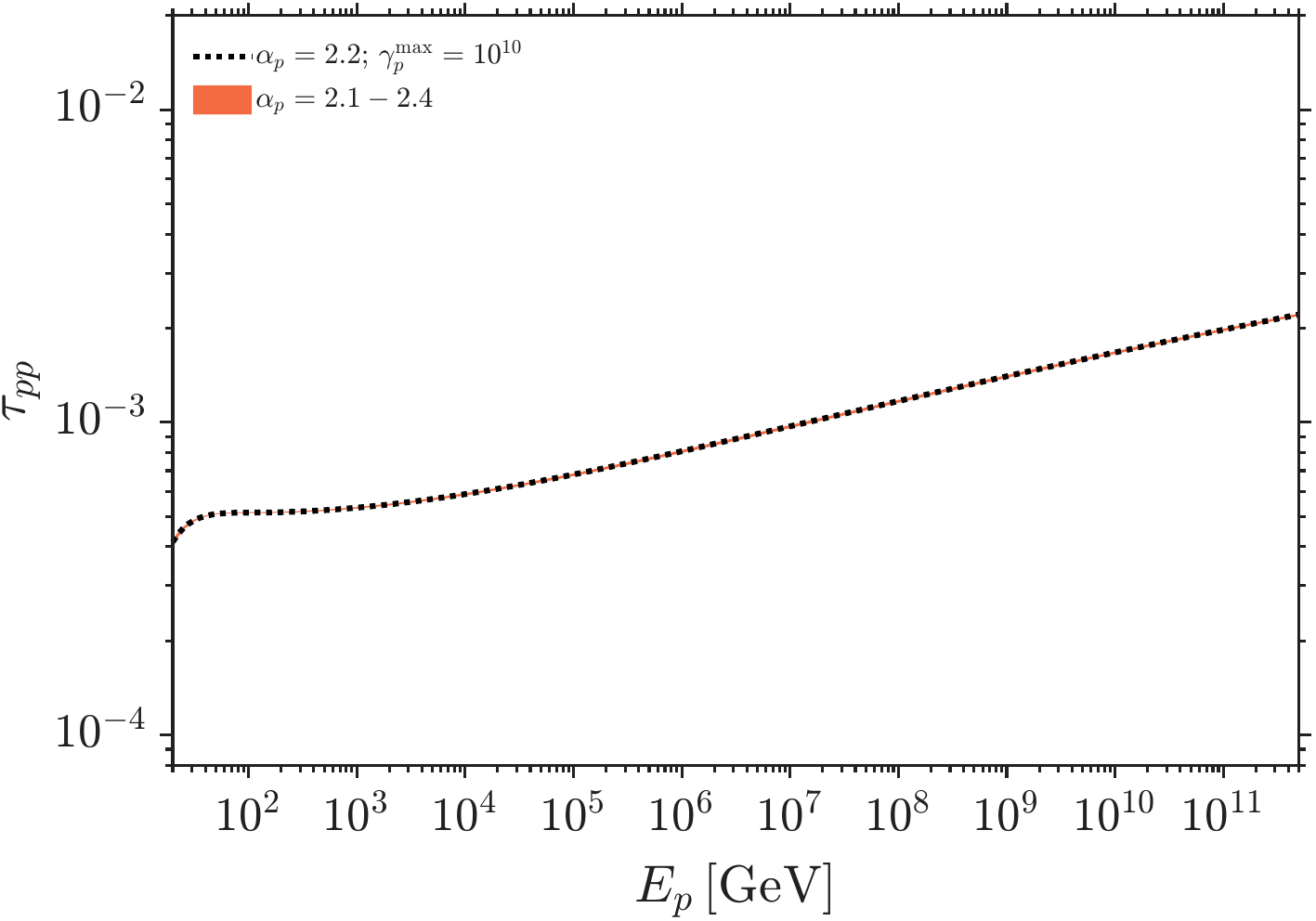}\hspace{0.4cm}
\includegraphics[width=.47\textwidth]{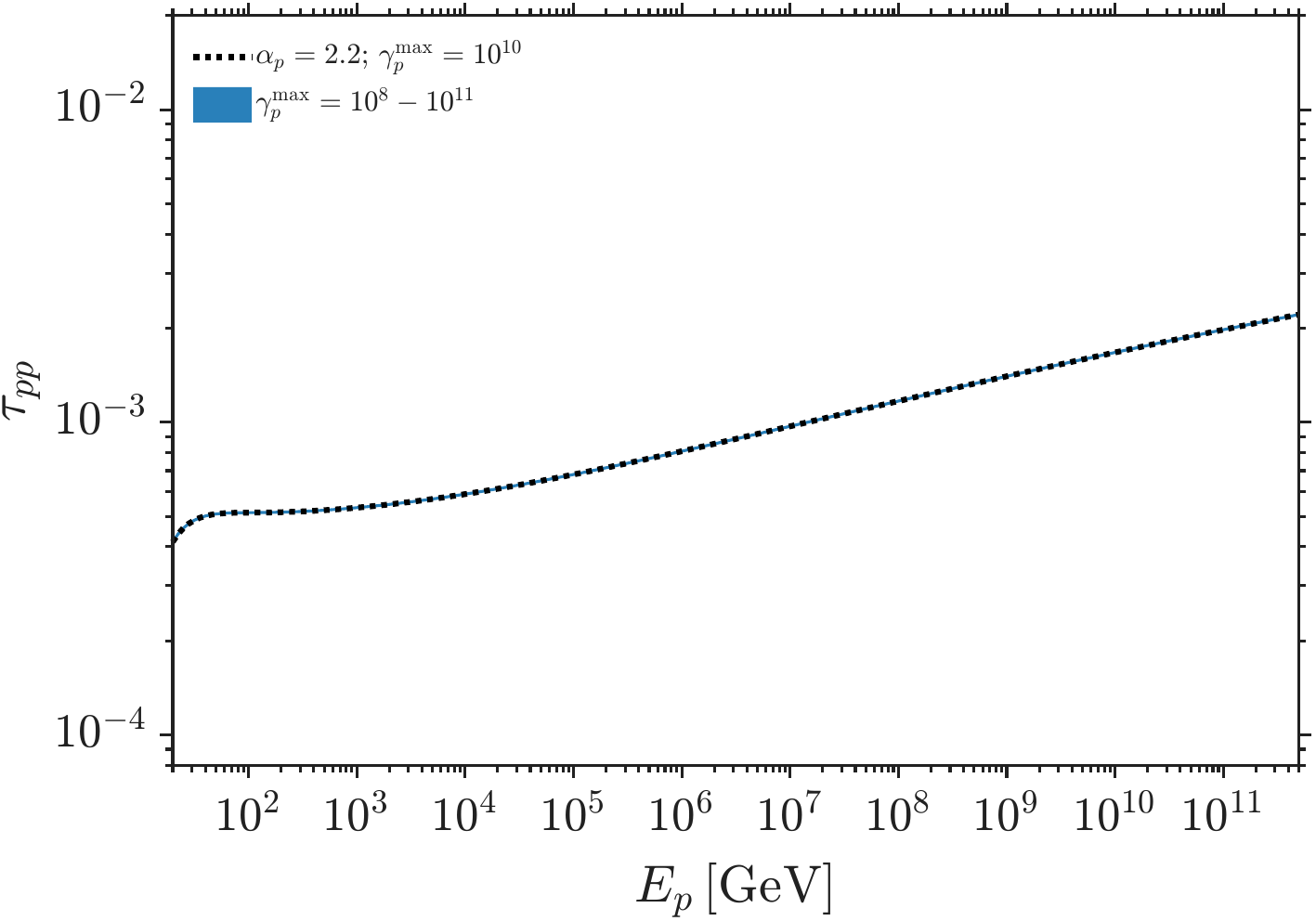}
\caption{Proton-proton optical depth $\tau_{pp}$ as a function of the host-galaxy frame proton energy $E_p$. The dotted black line represents our benchmark model, obtained using the same parameter values ($n_{p,0}'$, $r_0$, $r_f$, $a$, $\theta_{\rm mis}$, and $\Gamma_b$) as in the Main Text. The shaded regions show the sensitivity of $\tau_{pp}$ to individual variations of the two additional parameters: the proton spectral index $\alpha_p$ (left panel) and the maximum proton Lorentz factor $\gamma_p^{\rm max}$ (right panel).}
\label{fig:tauPPsm}
\end{figure}

To assess the impact of hadronic parameters on the beam interaction efficiency, we examine the sensitivity of the proton-proton optical depth $\tau_{pp}(E_p)$ to variations in the proton spectrum. For this analysis, quantities are evaluated in the laboratory frame (anchored to the host galaxy environment, thereby independent of redshift $z$). Furthermore, since we track a single test proton traversing the target proton background, we adopt the simplified notation $E_p$ for its host-galaxy frame proton energy, $\alpha_p$ for its spectral index, and $\gamma_p^{\rm max}$ for its maximum Lorentz factor (omitting jet identification subscripts, as anticipated in Sec.~\ref{sec:supp_agn_framework}).

The baseline hadronic optical depth is computed using the reference set of parameters established in the Main Text: total integrated comoving proton number density $n_{p,0}' = 10^4 \,{\rm cm}^{-3}$, onset interaction distance $r_0 = 0.5 \, \rm pc$, final interaction boundary $r_f = 10 \, \rm pc$, compression index $a = 0.5$, misalignment angle $\theta_{\rm mis} = 3^\circ$, and bulk Lorentz factor $\Gamma_b = 15$. The baseline proton spectrum adopts a spectral index $\alpha_p = 2.2$ and a maximum proton Lorentz factor $\gamma_p^{\rm max} = 10^{10}$.

As illustrated in Fig.~\ref{fig:tauPPsm}, the resulting optical depth $\tau_{pp}(E_p)$ exhibits remarkable stability against variations in both $\alpha_p$ and $\gamma_p^{\rm max}$:

\begin{itemize}
\item {\bf Proton spectral index $\alpha_p$:} The variation of $\alpha_p$ in the physically motivated range $[2.1, 2.4]$ leaves $\tau_{pp}(E_p)$ virtually unchanged, as shown in the left panel of Fig.~\ref{fig:tauPPsm}. A harder spectrum ($\alpha_p = 2.1$) slightly increases the proportion of high-energy target protons, whereas a softer spectrum ($\alpha_p = 2.4$) slightly enhances the target density near the low-energy threshold. However, since $\tau_{pp}$ integrates over the full target proton distribution, whose total density is fixed by $n_{p,0}'$, these local spectral shifts largely offset each other, producing a negligible net effect on the optical depth.

\item {\bf Maximum proton energy $\gamma_p^{\rm max}$:} The optical depth displays a complete insensitivity to the high-energy cutoff of the target distribution, as shown in the right panel of Fig.~\ref{fig:tauPPsm}. Physically, $\tau_{pp}(E_p)$ measures the interaction probability of a test proton probing the integrated target proton spectral number density $n_p'$. Because the power-law spectrum ${\rm d}n_p'/{\rm d}E_p' \propto {E_p'}^{-\alpha_p}$ is steep [see Eq.~(1) of the Main Text], the total target density is dominated by low-energy protons. Extending or truncating the target distribution in the range $\gamma_p^{\rm max} \in [10^8, 10^{11}]$ has a negligible effect on the total target proton density available for $pp$ collisions.
\end{itemize}

Consequently, $\tau_{pp}$ serves as an indicator of the hadronic interaction regime, confirming that the optical depth $\tau_{pp}$ derived in the Main Text is robust with respect to parameter value variations inside physically reasonable ranges and demonstrating the feasibility of the Cosmic Collider mechanism.

\subsection{Collider Luminosity Across Different Scenarios}
\label{sec:collider_luminosity_scenarios}

To evaluate the energy-dependent interaction rate of proton-proton collisions and the resulting secondary production within the overlapping jet region, we compute the differential collider luminosity $\sqrt{s}L_{\sqrt{s}} \equiv s\,\mathrm{d}N_\gamma/\mathrm{d}\sqrt{s}$ as a function of the CM energy $\sqrt{s}$ (see Main Text). Since acceleration mechanisms and environmental conditions may differ between the two SMBH accretion systems, we systematically explore symmetric variations as well as asymmetric configurations across the parameter space $(\alpha_{p,i}, \gamma_{p,i}^{\mathrm{max}})$ for jets $i \in \{1,2\}$ (see notation in the Main Text). 

Specifically, we organize our investigation into three distinct physical scenarios: 
Fig.~\ref{fig:CollLumSimm} addresses symmetric parameter variations across both jets; 
Fig.~\ref{fig:CollLumLow} evaluates the robustness of the collider under asymmetric parameter degradation; and 
Fig.~\ref{fig:CollLumHigh} explores the maximum reachable efficiency in highly energetic asymmetric configurations. 
In all three figures, the symmetric benchmark scenario introduced in the Main Text ($\alpha_{p,1} = \alpha_{p,2} = 2.2$, $\gamma_{p,1}^{\mathrm{max}} = \gamma_{p,2}^{\mathrm{max}} = 10^{10}$) is displayed as a blue dotted curve for direct reference, while the different explored scenarios are represented by the solid blue curves.

\begin{figure}[t]
\centering
\includegraphics[width=.47\textwidth]{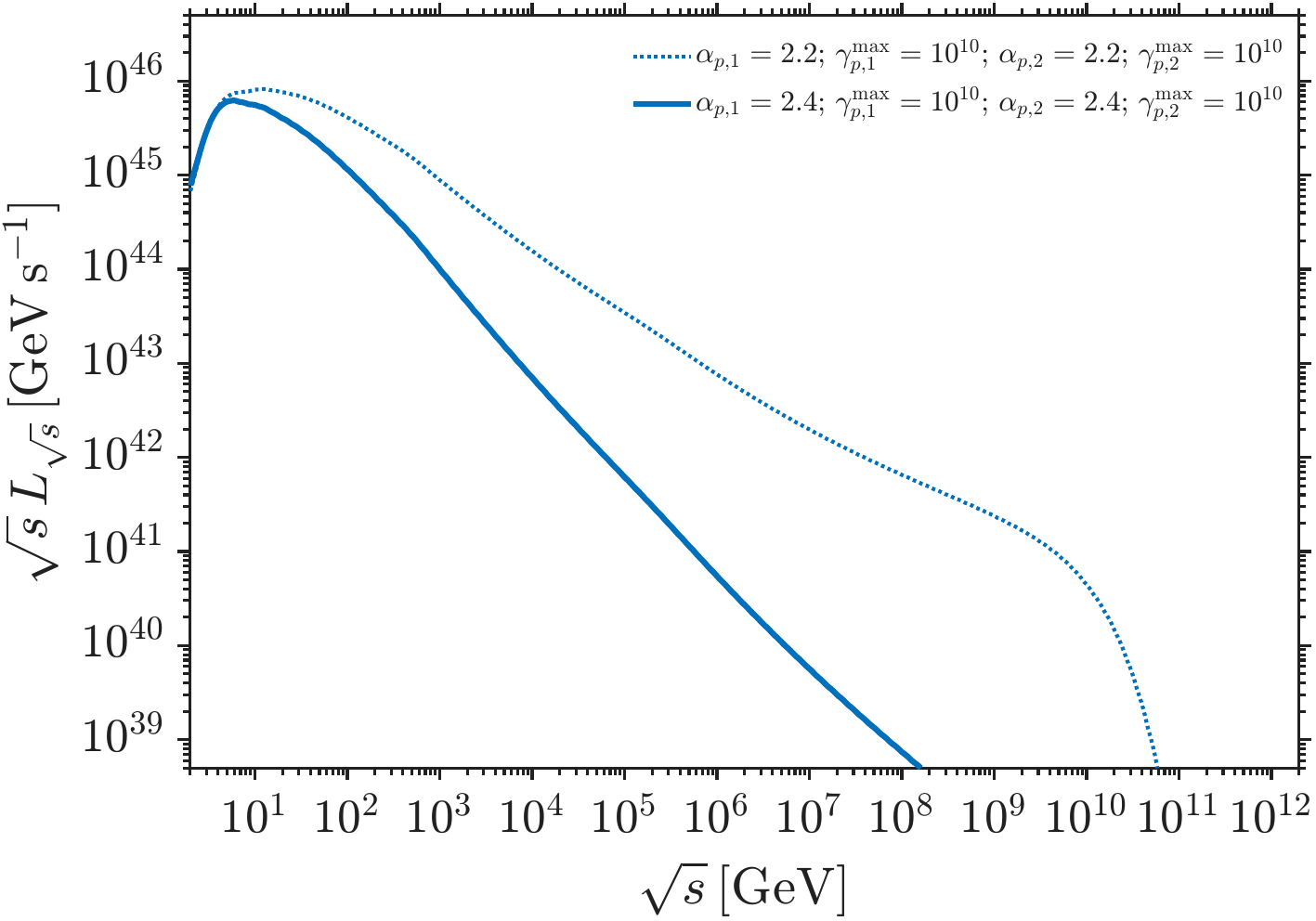}\hspace{0.4cm}
\includegraphics[width=.47\textwidth]{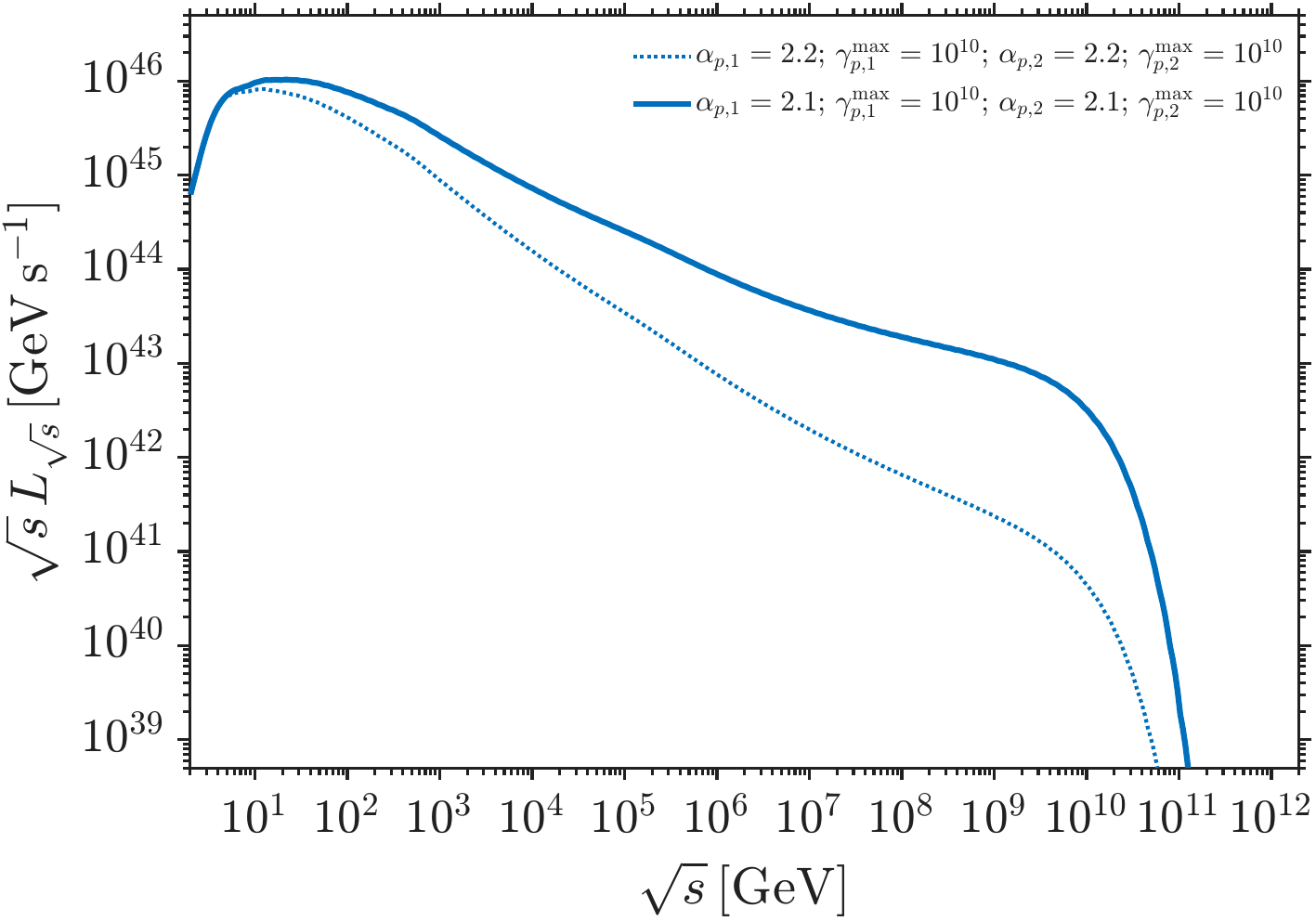}

\vspace{0.3cm}

\includegraphics[width=.47\textwidth]{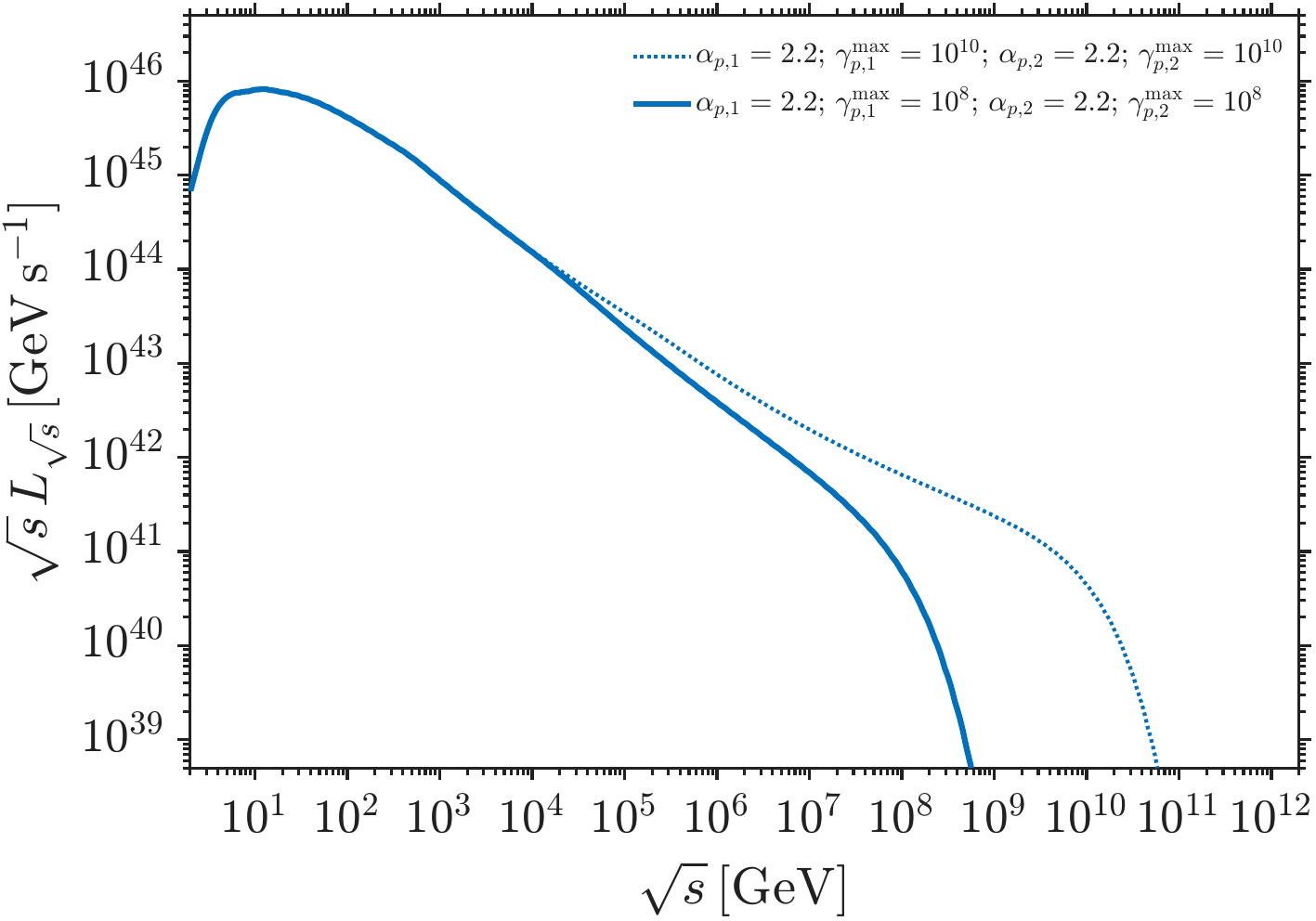}\hspace{0.4cm}
\includegraphics[width=.47\textwidth]{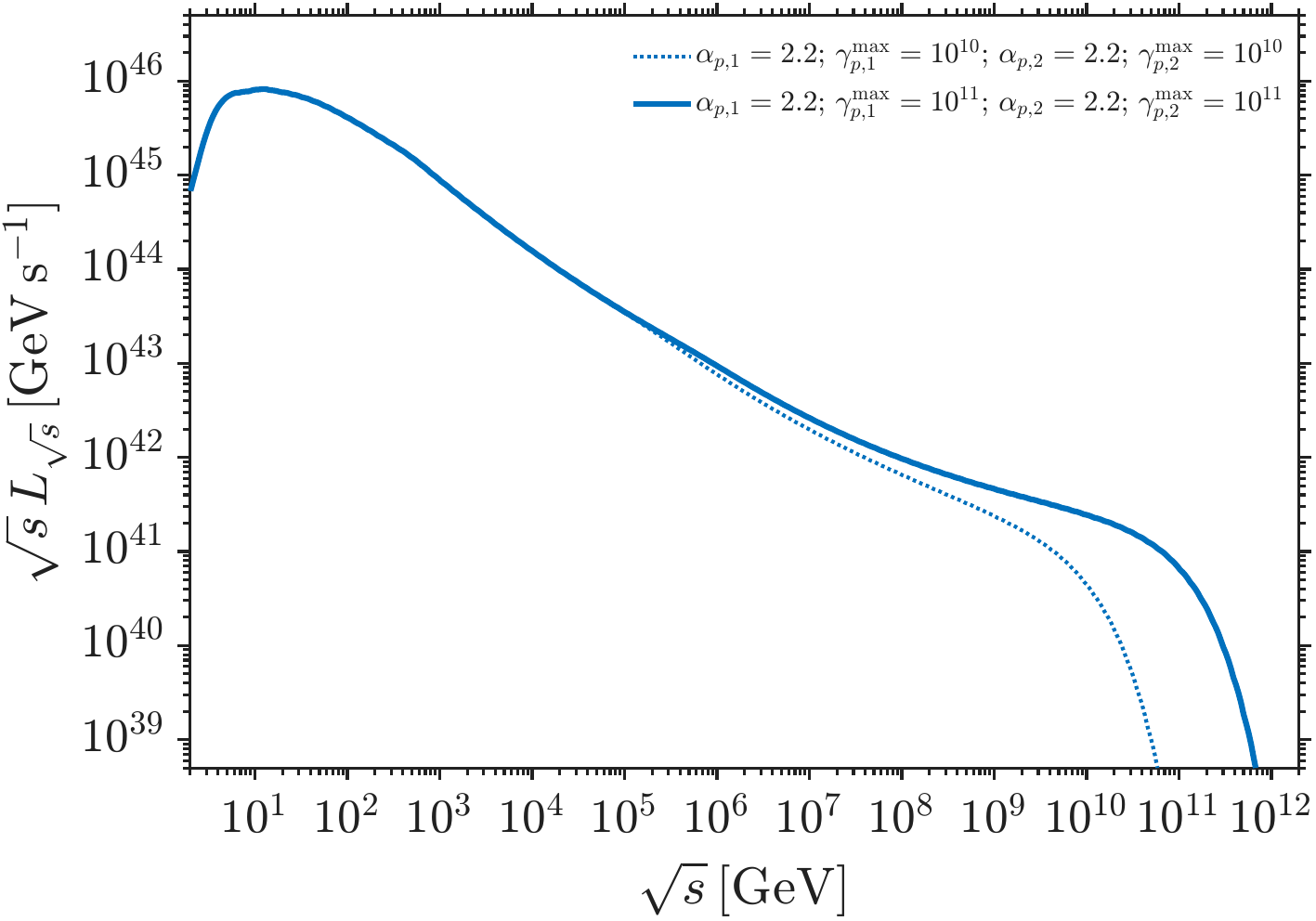}
\caption{Differential collider luminosity $\sqrt{s}L_{\sqrt{s}}$ as a function of the CM energy $\sqrt{s}$ for symmetric variations in both jets ($\alpha_{p,1} = \alpha_{p,2}$, $\gamma_{p,1}^{\mathrm{max}} = \gamma_{p,2}^{\mathrm{max}}$, solid blue curves). Top left: softer proton index ($\alpha_{p,i} = 2.4$). Top right: harder proton index ($\alpha_{p,i} = 2.1$). Bottom left: reduced cutoff ($\gamma_{p,i}^{\mathrm{max}} = 10^8$). Bottom right: extended cutoff ($\gamma_{p,i}^{\mathrm{max}} = 10^{11}$). The benchmark model ($\alpha_{p,i}=2.2, \gamma_{p,i}^{\mathrm{max}}=10^{10}$) is shown as a blue dotted line in all panels.}
\label{fig:CollLumSimm}
\end{figure}

\begin{figure}[t]
\centering
\includegraphics[width=.47\textwidth]{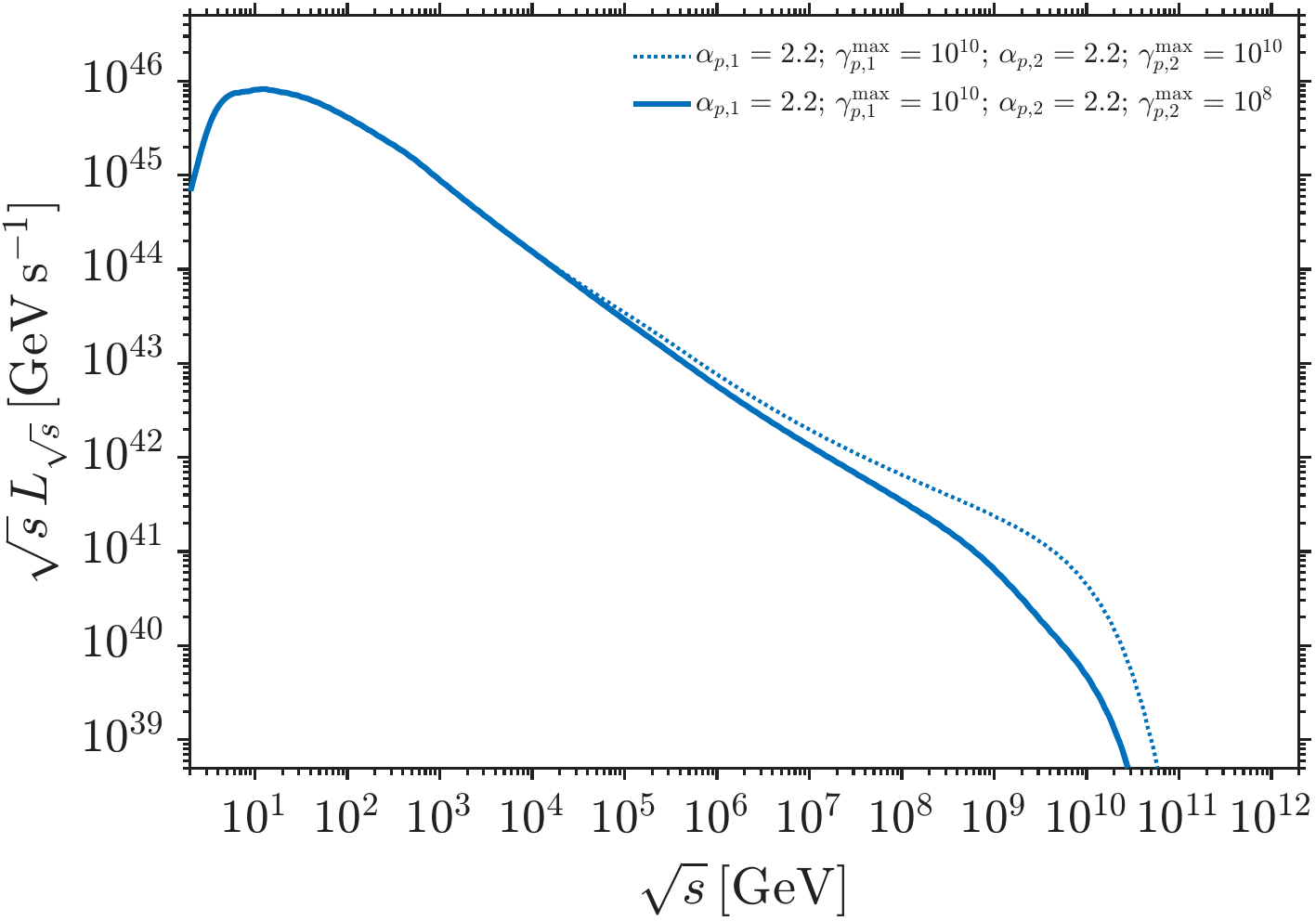}\hspace{0.4cm}
\includegraphics[width=.47\textwidth]{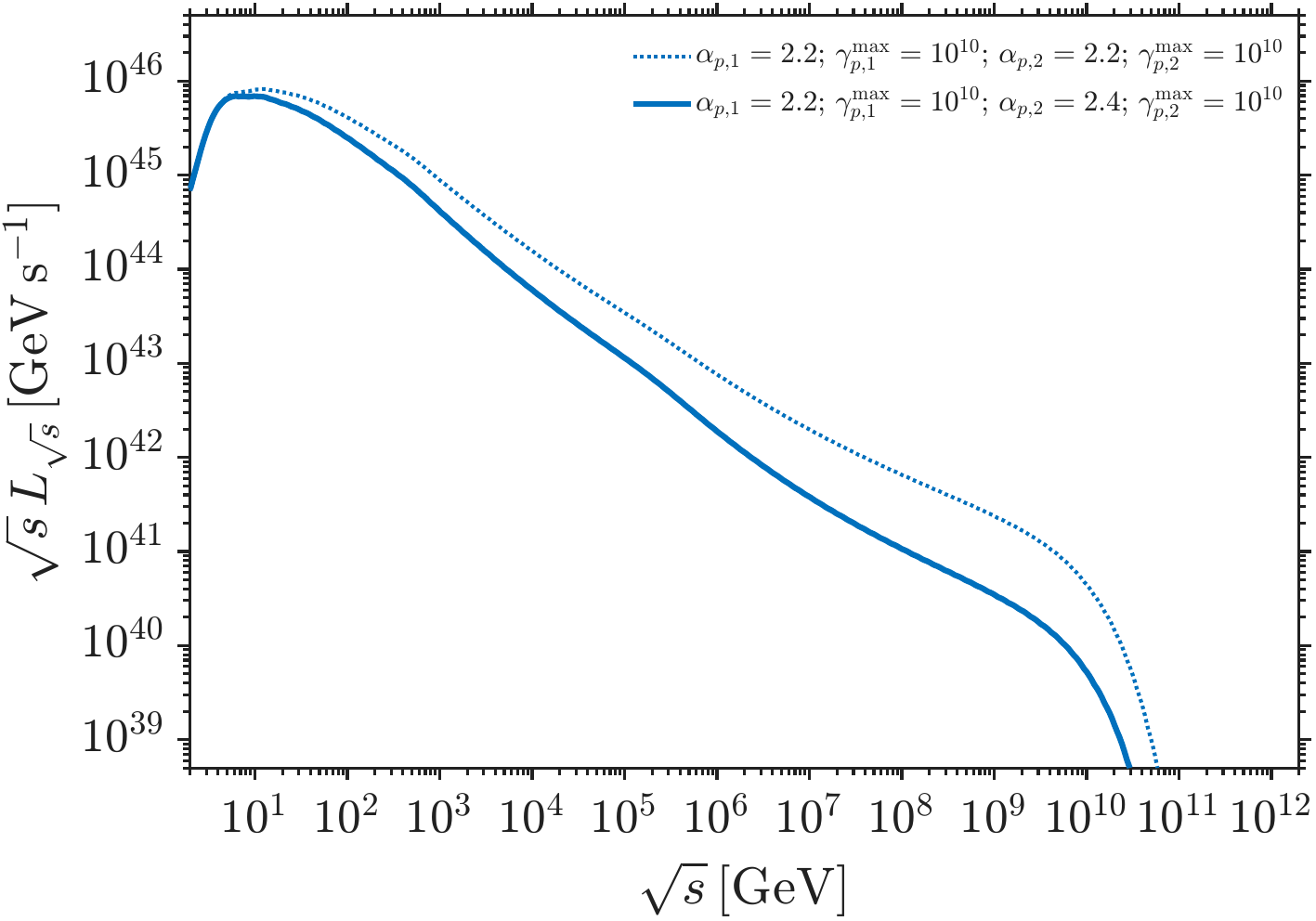}

\vspace{0.3cm}

\includegraphics[width=.47\textwidth]{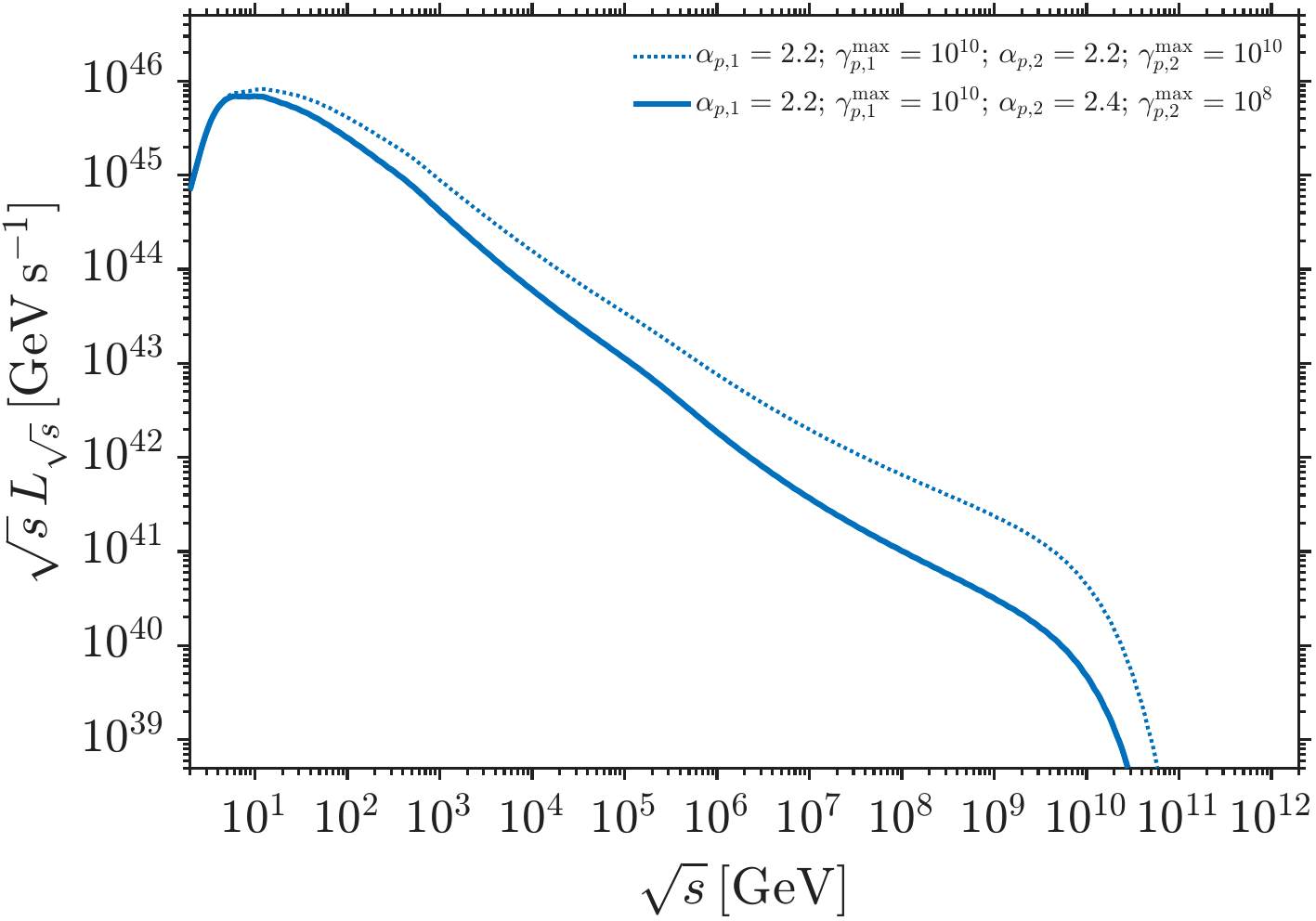}\hspace{0.4cm}
\includegraphics[width=.47\textwidth]{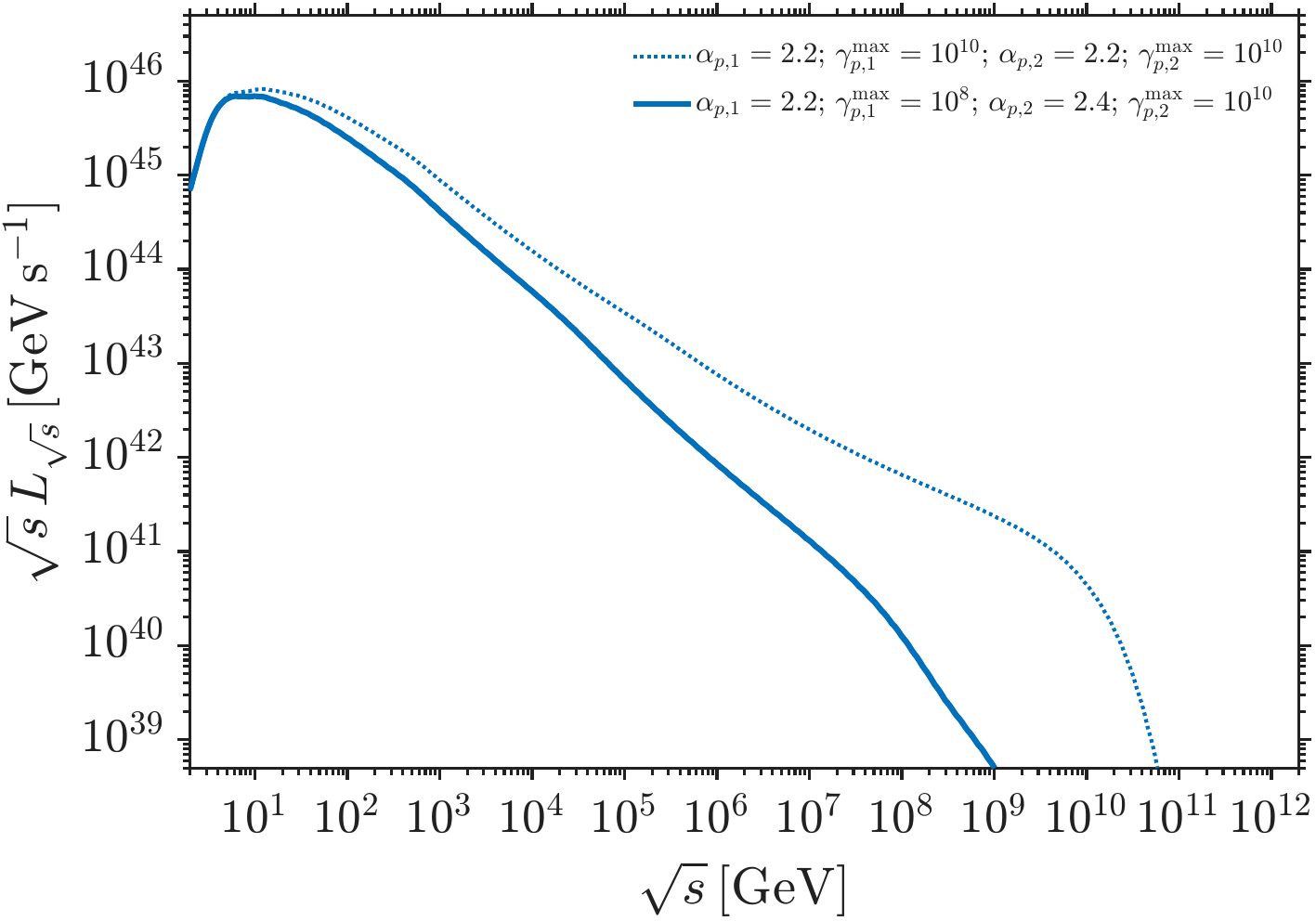}
\caption{Differential collider luminosity $\sqrt{s}L_{\sqrt{s}}$ as a function of the CM energy $\sqrt{s}$ for conservative/sub-benchmark asymmetric configurations (solid blue curves). Top left: reduced cutoff
in jet 2 ($\gamma_{p,2}^{\mathrm{max}} = 10^8$). Top right: softer spectrum in jet 2  ($\alpha_{p,2} = 2.4$). Bottom left: softer index and reduced cutoff in jet
2 ($\alpha_{p,2} = 2.4, \gamma_{p,2}^{\mathrm{max}} = 10^8$). Bottom right: softer index in jet 2 with reduced cutoff in jet 1 ($\alpha_{p,2}=2.4$, $\gamma_{p,1}^{\mathrm{max}} = 10^8$). The benchmark model ($\alpha_{p,i}=2.2, \gamma_{p,i}^{\mathrm{max}}=10^{10}$) is shown as a blue dotted line in all panels.}
\label{fig:CollLumLow}
\end{figure}

\begin{figure}[t]
\centering
\includegraphics[width=.47\textwidth]{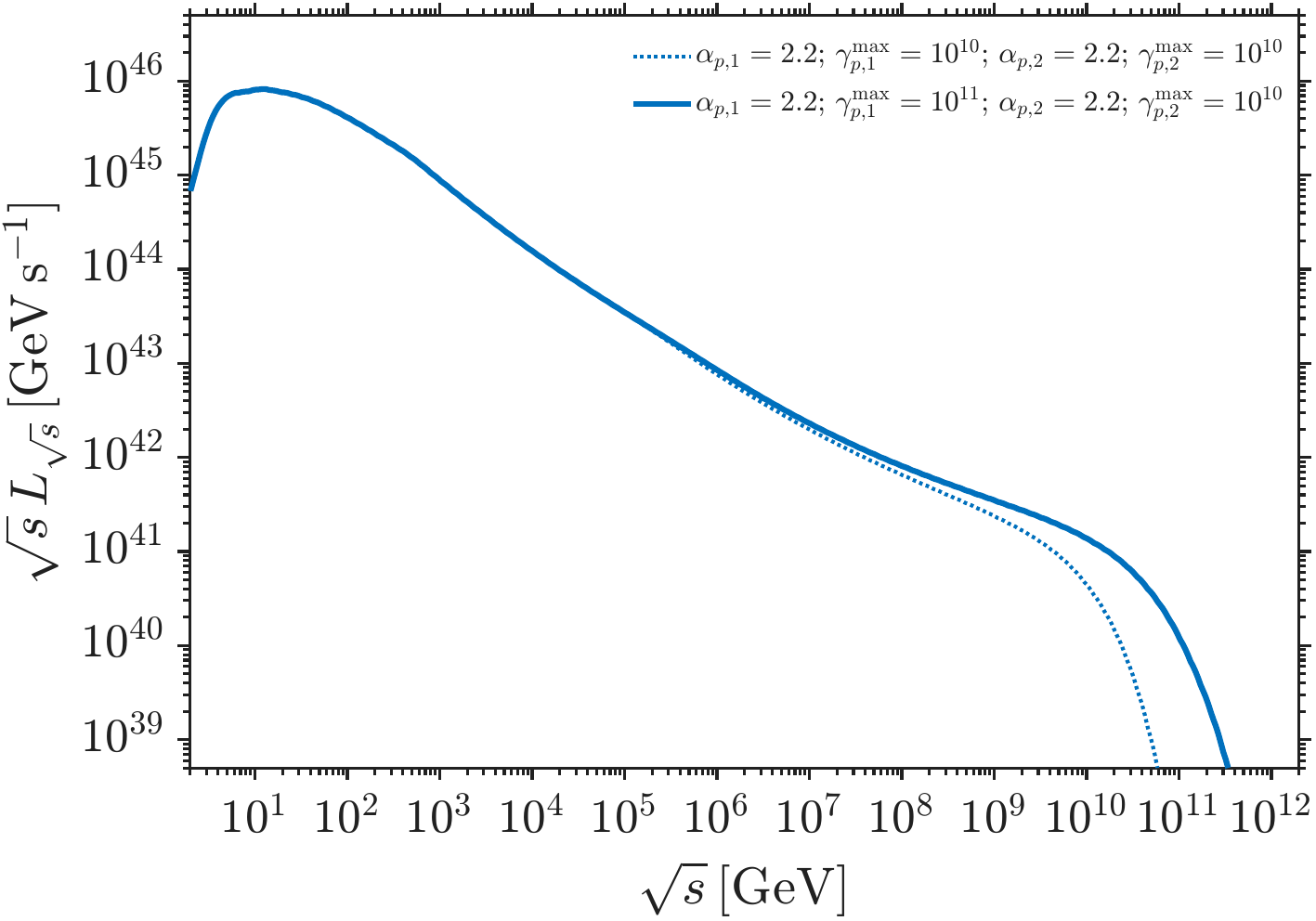}\hspace{0.4cm}
\includegraphics[width=.47\textwidth]{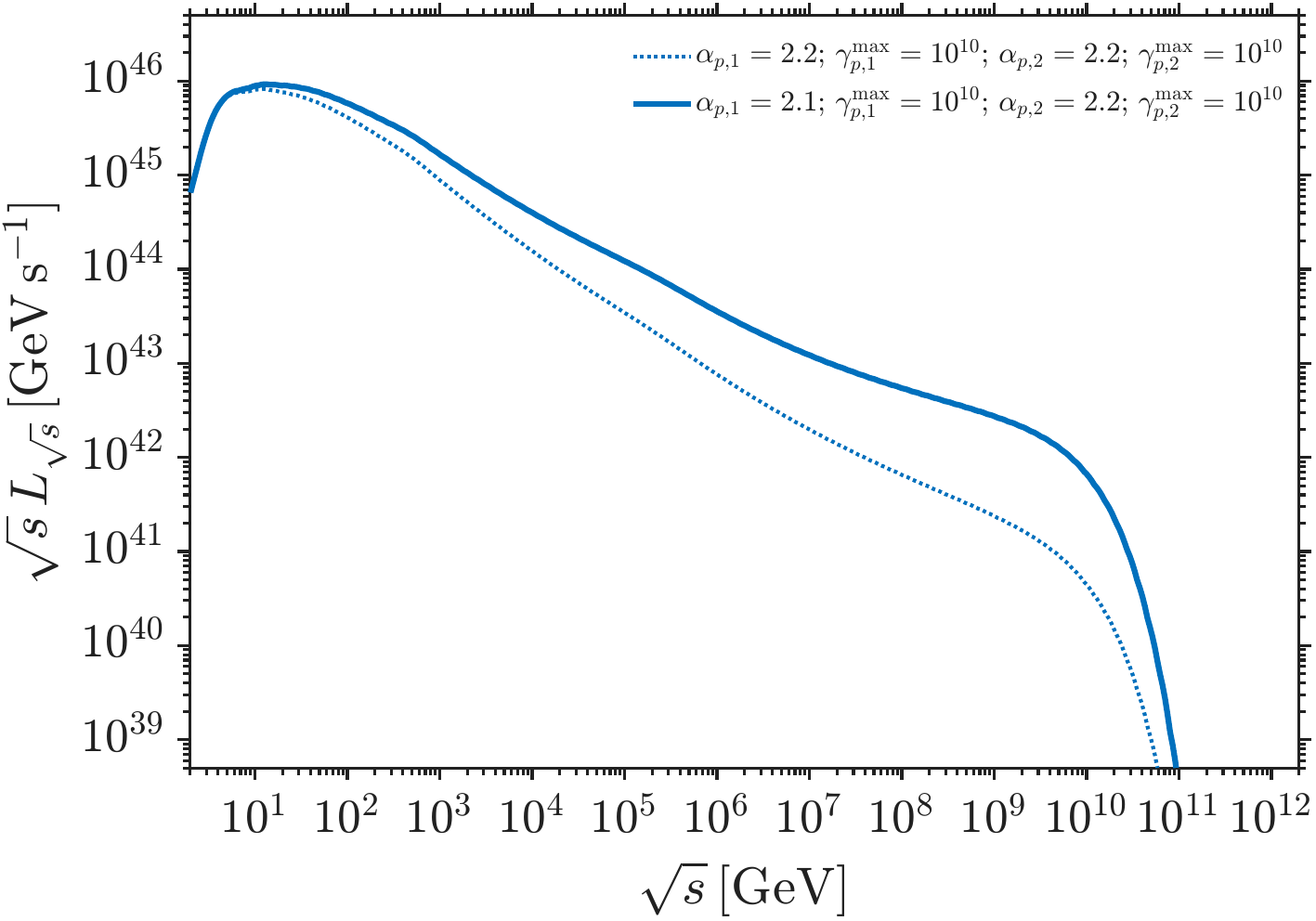}

\vspace{0.3cm}

\includegraphics[width=.47\textwidth]{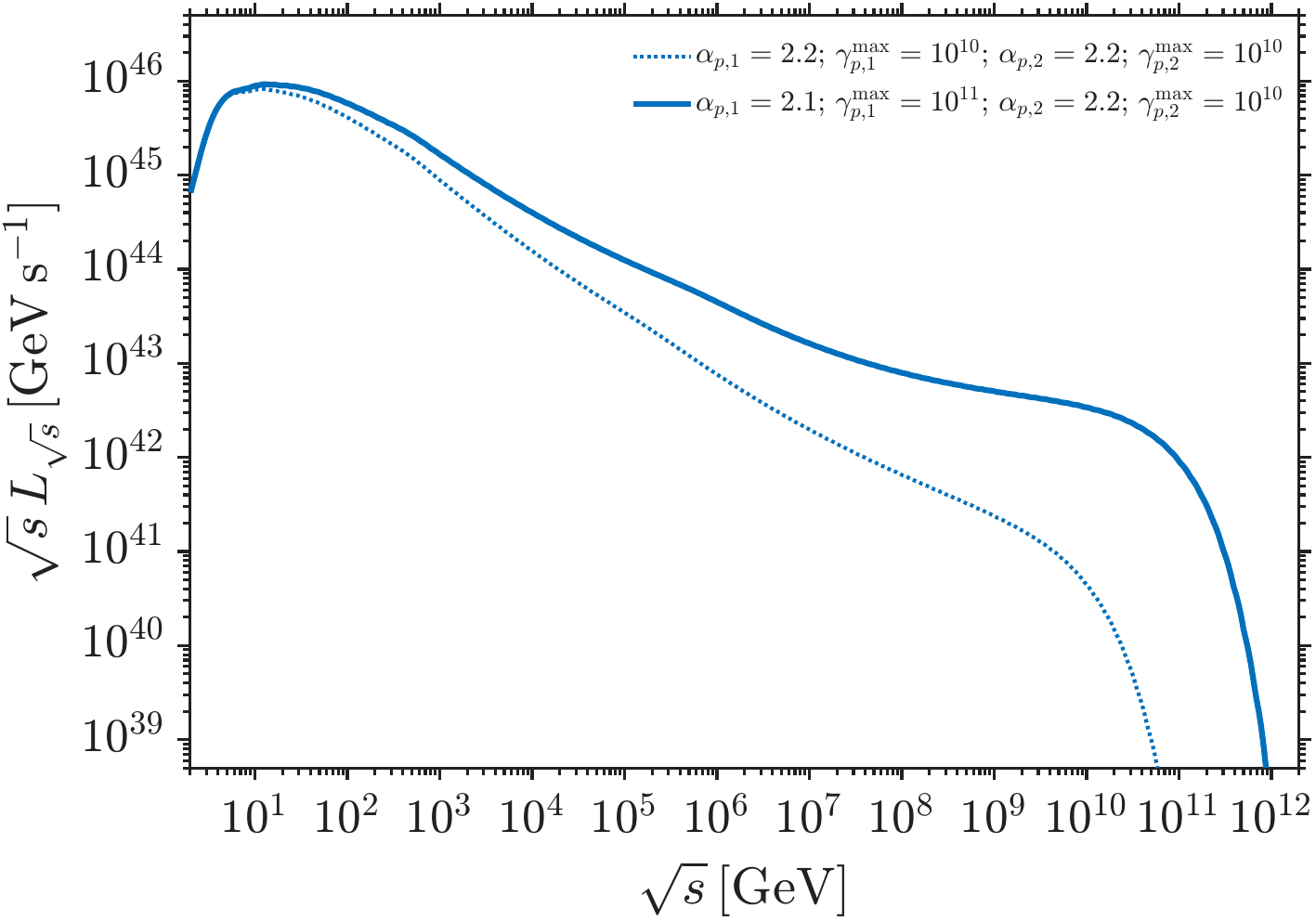}\hspace{0.4cm}
\includegraphics[width=.47\textwidth]{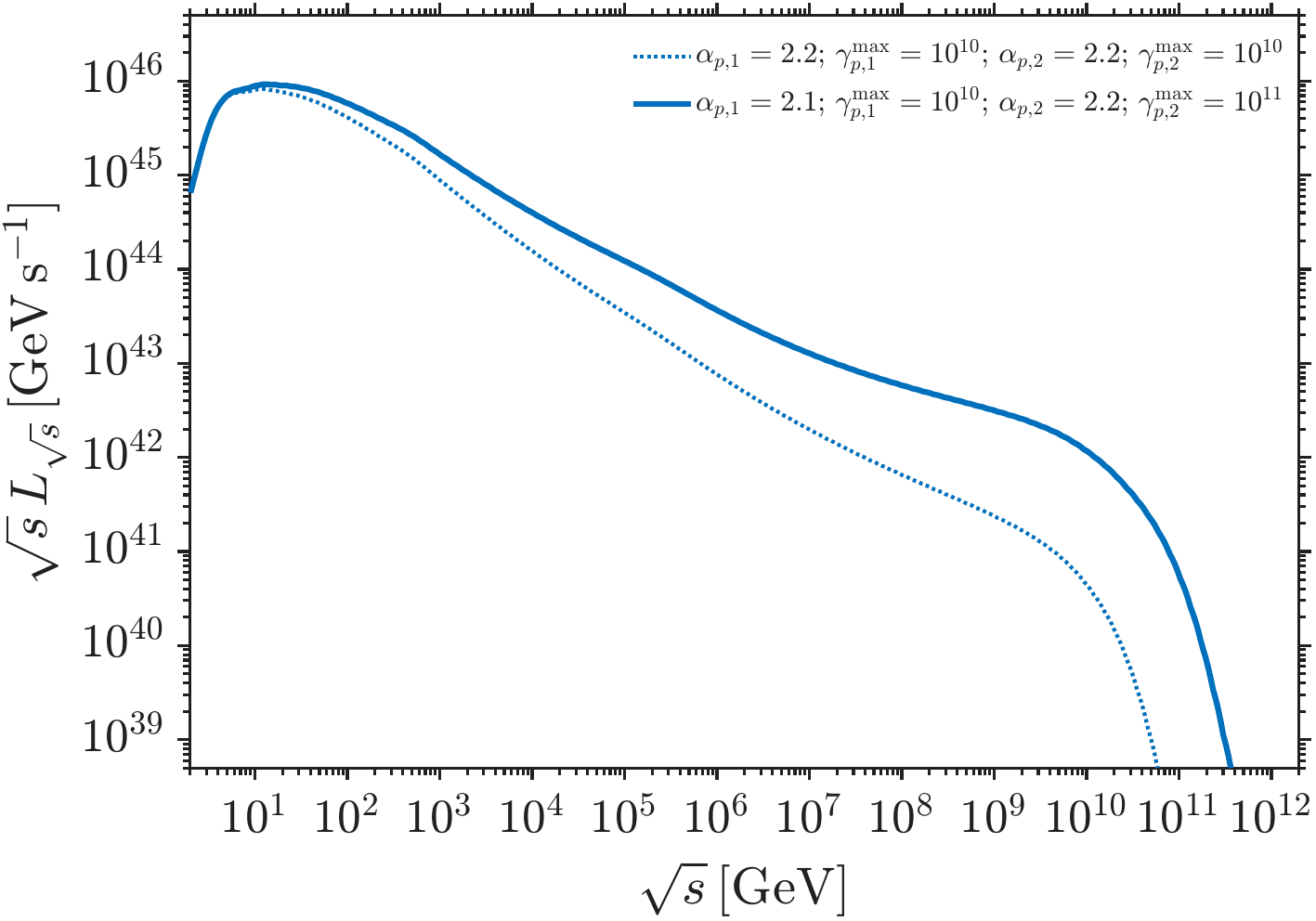}
\caption{Differential collider luminosity $\sqrt{s}L_{\sqrt{s}}$ as a function of the CM energy $\sqrt{s}$ for optimistic/super-benchmark asymmetric configurations (solid blue curves). Top left: extended cutoff in jet 1 ($\gamma_{p,1}^{\mathrm{max}} = 10^{11}$). Top right: harder spectrum in jet 1 ($\alpha_{p,1} = 2.1$). Bottom left: harder index and extended cutoff in jet 1 ($\alpha_{p,1} = 2.1, \gamma_{p,1}^{\mathrm{max}} = 10^{11}$). Bottom right: harder index in jet 1 with extended cutoff in jet 2 ($\alpha_{p,1} = 2.1, \gamma_{p,2}^{\mathrm{max}} = 10^{11}$). The benchmark model ($\alpha_{p,i}=2.2, \gamma_{p,i}^{\mathrm{max}}=10^{10}$) is shown as a blue dotted line in all panels.}
\label{fig:CollLumHigh}
\end{figure}

The behavior of the differential collider luminosity across the parameter space highlights three distinct main operational cases:

\begin{itemize}

\item {\bf Symmetric Variations (Fig.~\ref{fig:CollLumSimm}):} When both jets undergo identical modifications, the collider luminosity profile reflects the redistribution of the relativistic proton density of both jets. Softening the spectral indices ($\alpha_{p,1} = \alpha_{p,2} = 2.4$, top left) shifts the available power toward low CM energies, suppressing high-energy collisions above $\sqrt{s} \sim 10^5\,\mathrm{GeV}$. Conversely, harder proton spectra ($\alpha_{p,1} = \alpha_{p,2} = 2.1$, top right) enhance high-energy collisions, producing a prominent luminosity shoulder that remains elevated up to $\sqrt{s} \sim 10^{10}\,\mathrm{GeV}$. Adjusting the maximum Lorentz factors symmetrically directly sets the absolute kinematical threshold: setting $\gamma_{p,1}^{\mathrm{max}} = \gamma_{p,2}^{\mathrm{max}} = 10^8$ (bottom left) truncates the collider luminosity sharply at $\sqrt{s} \sim 10^8\,\mathrm{GeV}$, whereas extending it to $\gamma_{p,1}^{\mathrm{max}} = \gamma_{p,2}^{\mathrm{max}} = 10^{11}$ (bottom right) stretches the high-energy tail beyond $10^{11}\,\mathrm{GeV}$.

\item {\bf Robustness Under Asymmetric Degradation (Fig.~\ref{fig:CollLumLow}):} To assess the robustness of the hadronic collider in sub-optimal environments, we consider asymmetric setups where one or both jets exhibit steeper proton spectra and/or reduced cutoff energies. Degrading a single jet by reducing its maximum Lorentz factor to $\gamma_{p,2}^{\mathrm{max}} = 10^8$ (top left) only slightly lowers the maximum CM energy, while softening its spectral index to $\alpha_{p,2} = 2.4$ (top right) attenuates -- but not dramatically -- the differential luminosity at high energies. Combining a soft spectral index and a reduced cutoff in jet 2 (bottom left) yields a similarly moderate suppression. The strongest degradation occurs when the soft spectrum of jet 2 is paired with a reduced cutoff in jet 1 (bottom right): the luminosity at $\sqrt{s} \sim 10^8\,\mathrm{GeV}$ drops by about one order of magnitude relative to the benchmark. Even in this least favorable case, however, the Cosmic Collider still populates CM energies up to $\sqrt{s} \sim 10^9\,\mathrm{GeV}$, more than four orders of magnitude beyond terrestrial colliders, confirming the intrinsic robustness of the mechanism.

\item {\bf High-Energy Enhancements and Super-Benchmark Potential (Fig.~\ref{fig:CollLumHigh}):} When asymmetric variations involve superior acceleration efficiency in at least one jet, the collider luminosity exceeds the benchmark level. Improving a single jet by extending its maximum Lorentz factor to $\gamma_{p,1}^{\mathrm{max}} = 10^{11}$ (top left) increases the maximum CM energy, while hardening its index to $\alpha_{p,1} = 2.1$ (top right) boosts the differential luminosity at highest energies ($\sqrt{s} \gtrsim 10^6\,\mathrm{GeV}$). Combining a hard spectral index in jet 1 with extended cutoffs in either jet 1 or jet 2 (bottom panels) increases both the maximum reachable luminosity -- elevating the high-energy collision yield -- and the maximum CM energy. This shows that in favorable situations the proton collision rate at extremely high CM energies can become very efficient.

\end{itemize}

\subsection{Secondary Photon and Neutrino Luminosities}
\label{sec:asymmetric_luminosities}

Having established the behavior of the $pp$ interaction optical depth and the differential collider luminosity, we evaluate the resulting spectral distributions of emitted secondary particles -- specifically, broadband neutrinos ($E'_\nu L'_{E'_{\nu}}$) and intrinsic gamma rays ($E'_\gamma L'_{E'_{\gamma}}$). All quantities are calculated in the comoving frame as a function of the secondary particle energy $E'_{\gamma,\nu}$.

In direct analogy with the collider luminosity investigation, we systematically explore secondary production across symmetric parameter variations (Fig.~\ref{fig:EnerLumSimm}), sub-benchmark conservative asymmetric configurations (Fig.~\ref{fig:EnerLumLow}), and super-benchmark optimistic asymmetric configurations (Fig.~\ref{fig:EnerLumHigh}) for jets $i \in \{1,2\}$. In each figure, secondary photon and neutrino spectral luminosities are depicted by solid blue and red curves, respectively, while the baseline benchmark scenario ($\alpha_{p,i}=2.2, \gamma_{p,i}^{\mathrm{max}}=10^{10}$) is superimposed as dotted curves in all panels for direct reference.

\begin{figure}[t]
\centering
\includegraphics[width=.47\textwidth]{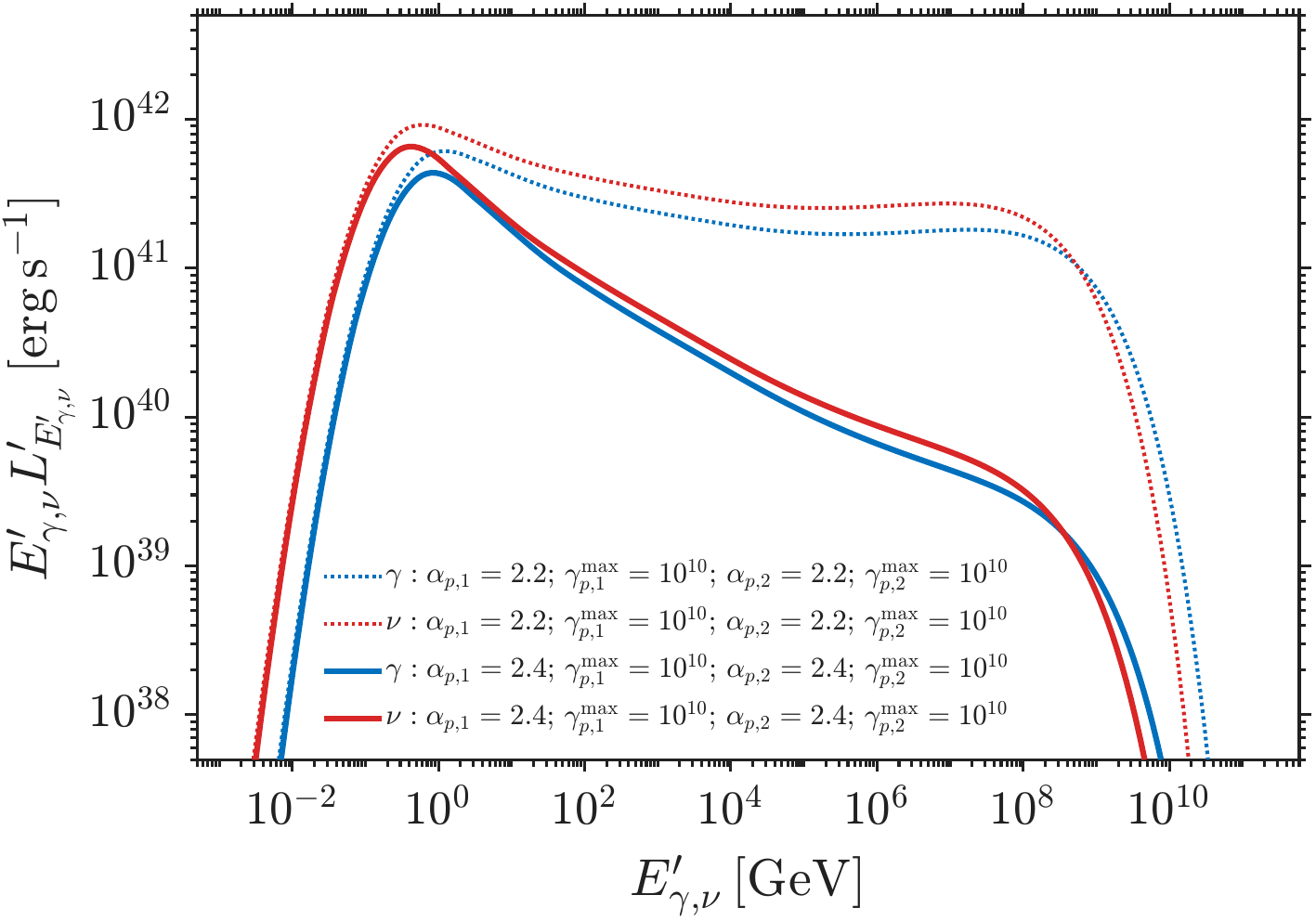}\hspace{0.4cm}
\includegraphics[width=.47\textwidth]{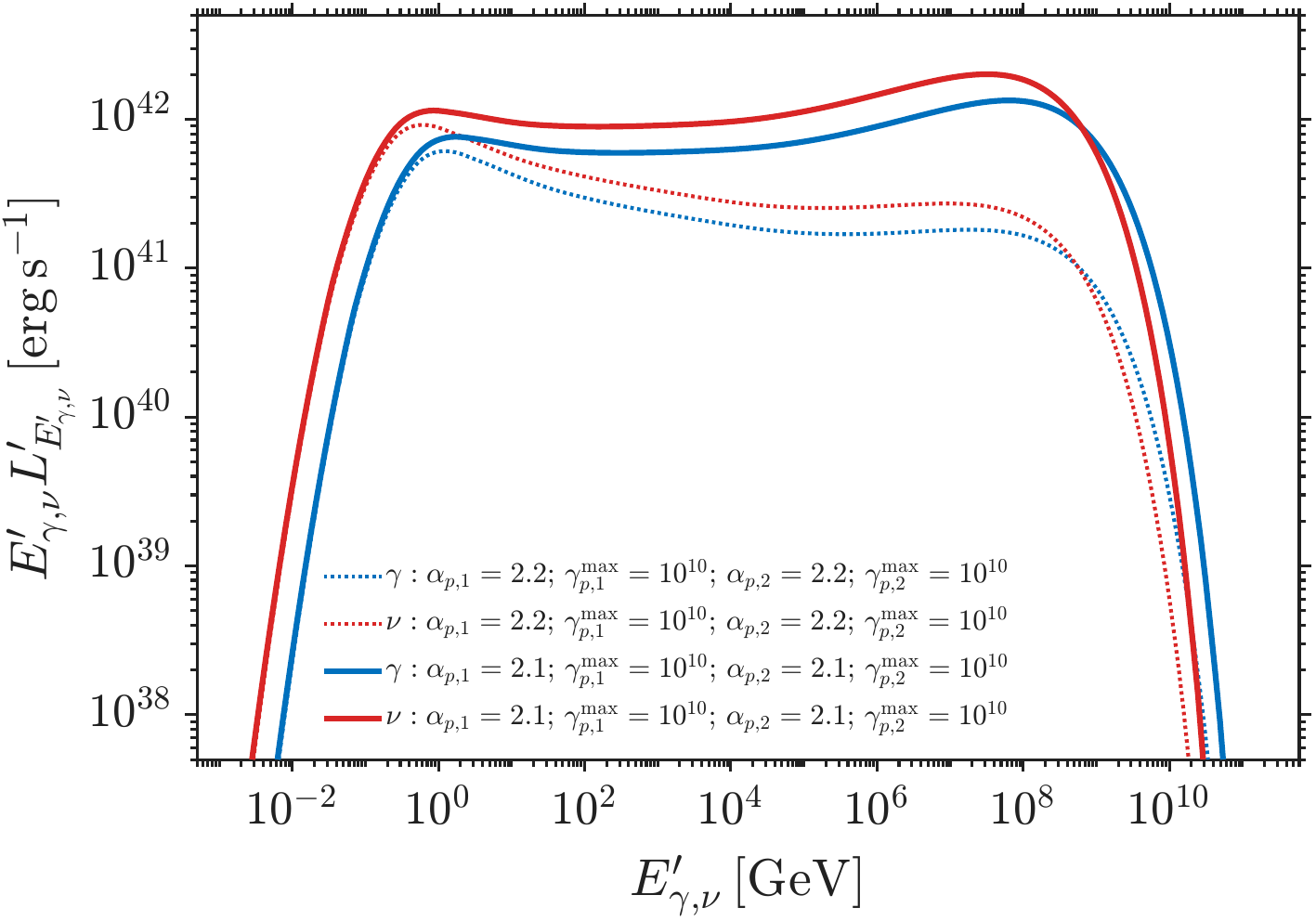}

\vspace{0.3cm}

\includegraphics[width=.47\textwidth]{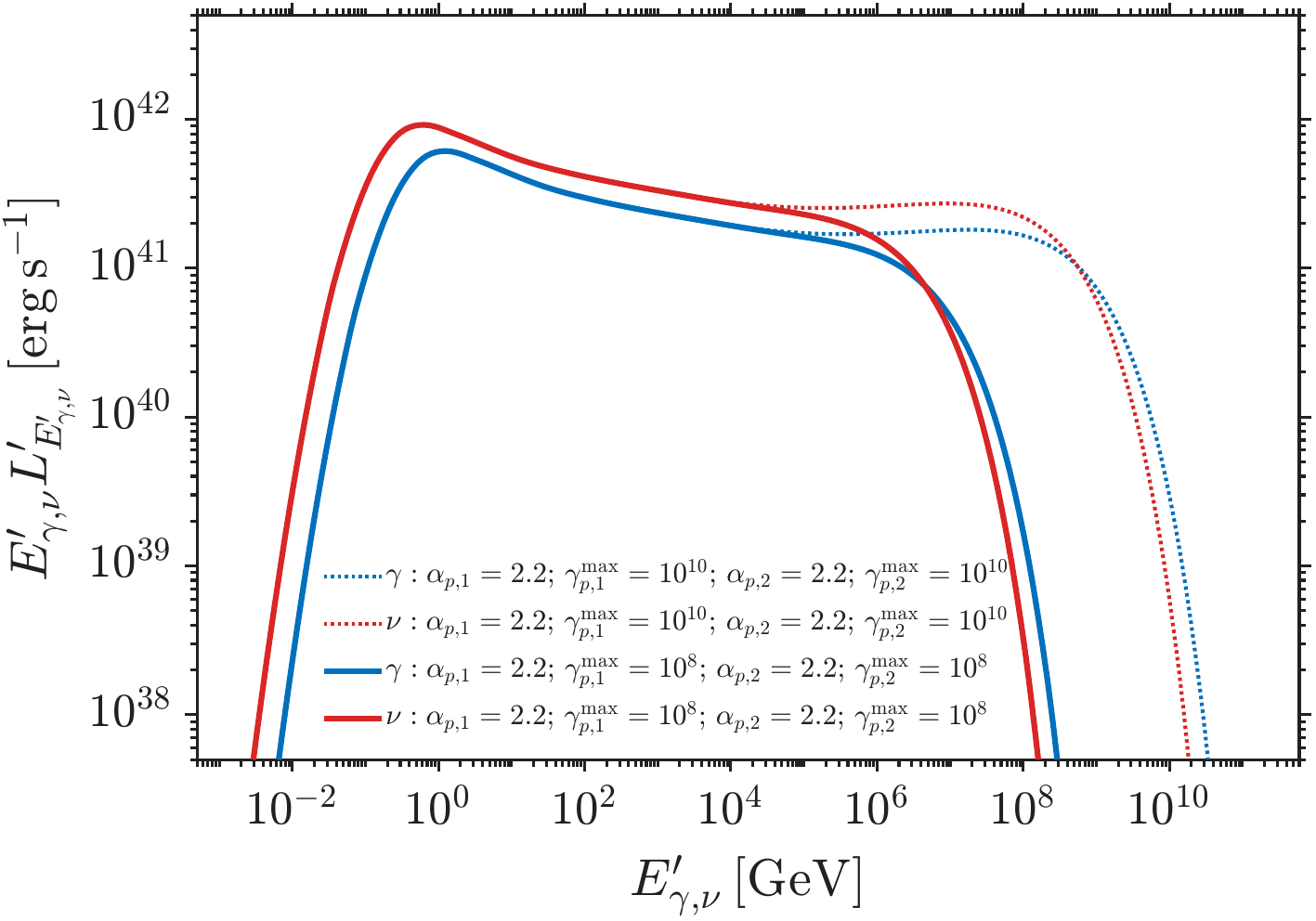}\hspace{0.4cm}
\includegraphics[width=.47\textwidth]{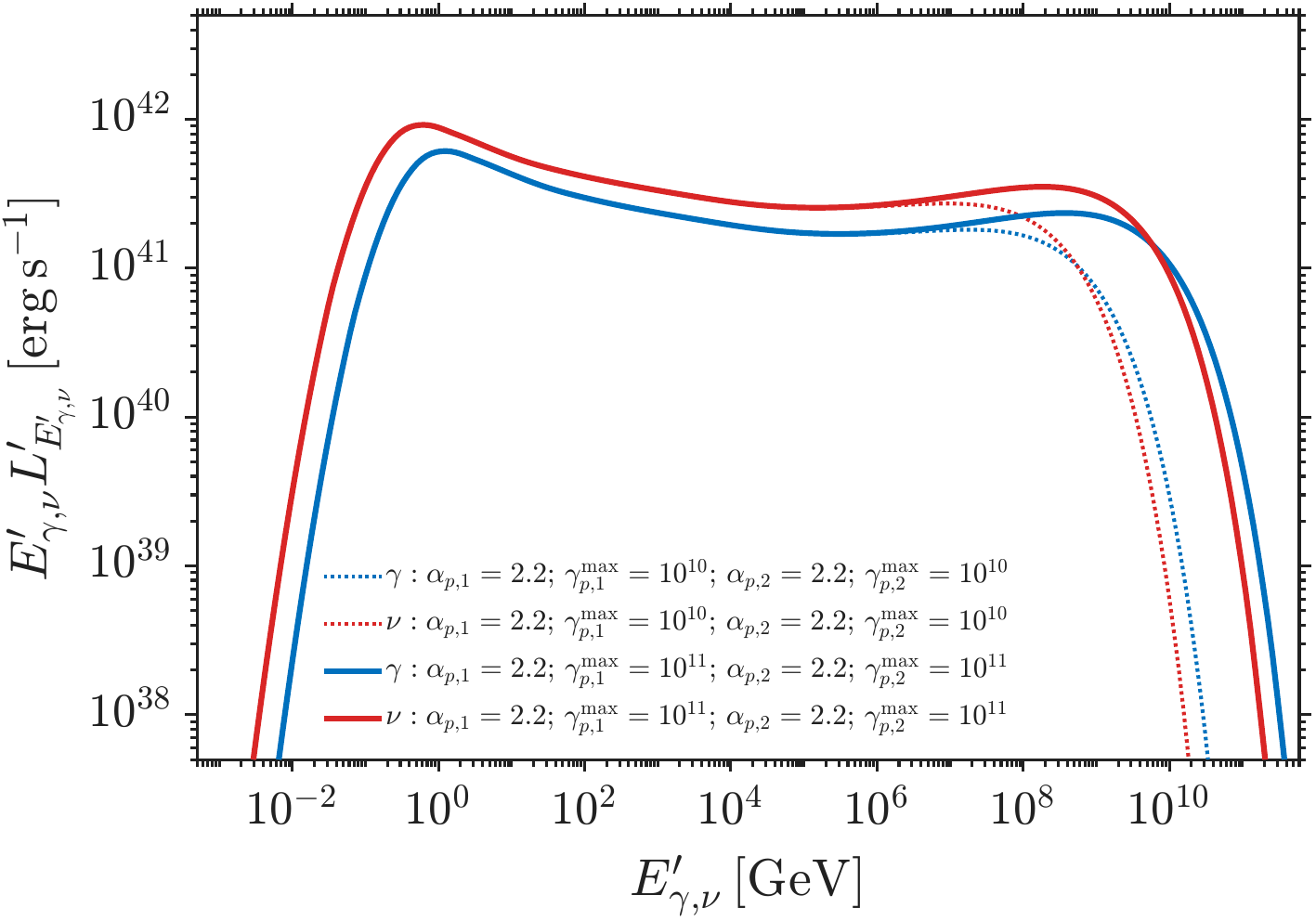}
\caption{Secondary photon ($E'_\gamma L'_{E'_{\gamma}}$, solid blue) and neutrino ($E'_\nu L'_{E'_{\nu}}$, solid red) spectral luminosities as a function of the energy $E'_{\gamma,\nu}$ for symmetric parameter variations in both jets ($\alpha_{p,1} = \alpha_{p,2}$, $\gamma_{p,1}^{\mathrm{max}} = \gamma_{p,2}^{\mathrm{max}}$). Top left: softer proton index ($\alpha_{p,i} = 2.4$). Top right: harder proton index ($\alpha_{p,i} = 2.1$). Bottom left: reduced cutoff ($\gamma_{p,i}^{\mathrm{max}} = 10^8$). Bottom right: extended cutoff ($\gamma_{p,i}^{\mathrm{max}} = 10^{11}$). Dotted lines represent the benchmark model ($\alpha_{p,i} = 2.2$, $\gamma_{p,i}^{\mathrm{max}} = 10^{10}$).}
\label{fig:EnerLumSimm}
\end{figure}

\begin{figure}[t]
\centering
\includegraphics[width=.47\textwidth]{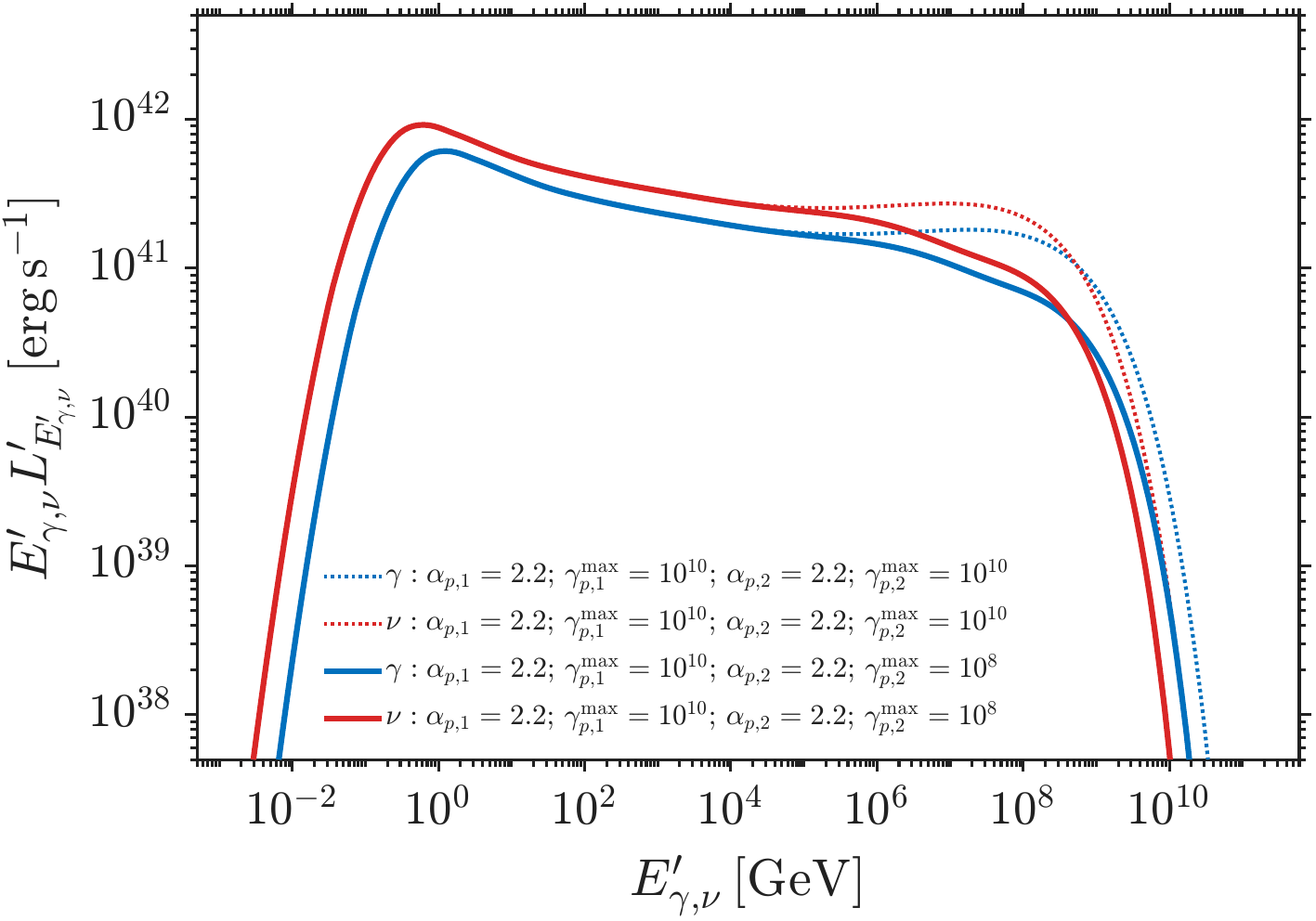}\hspace{0.4cm}
\includegraphics[width=.47\textwidth]{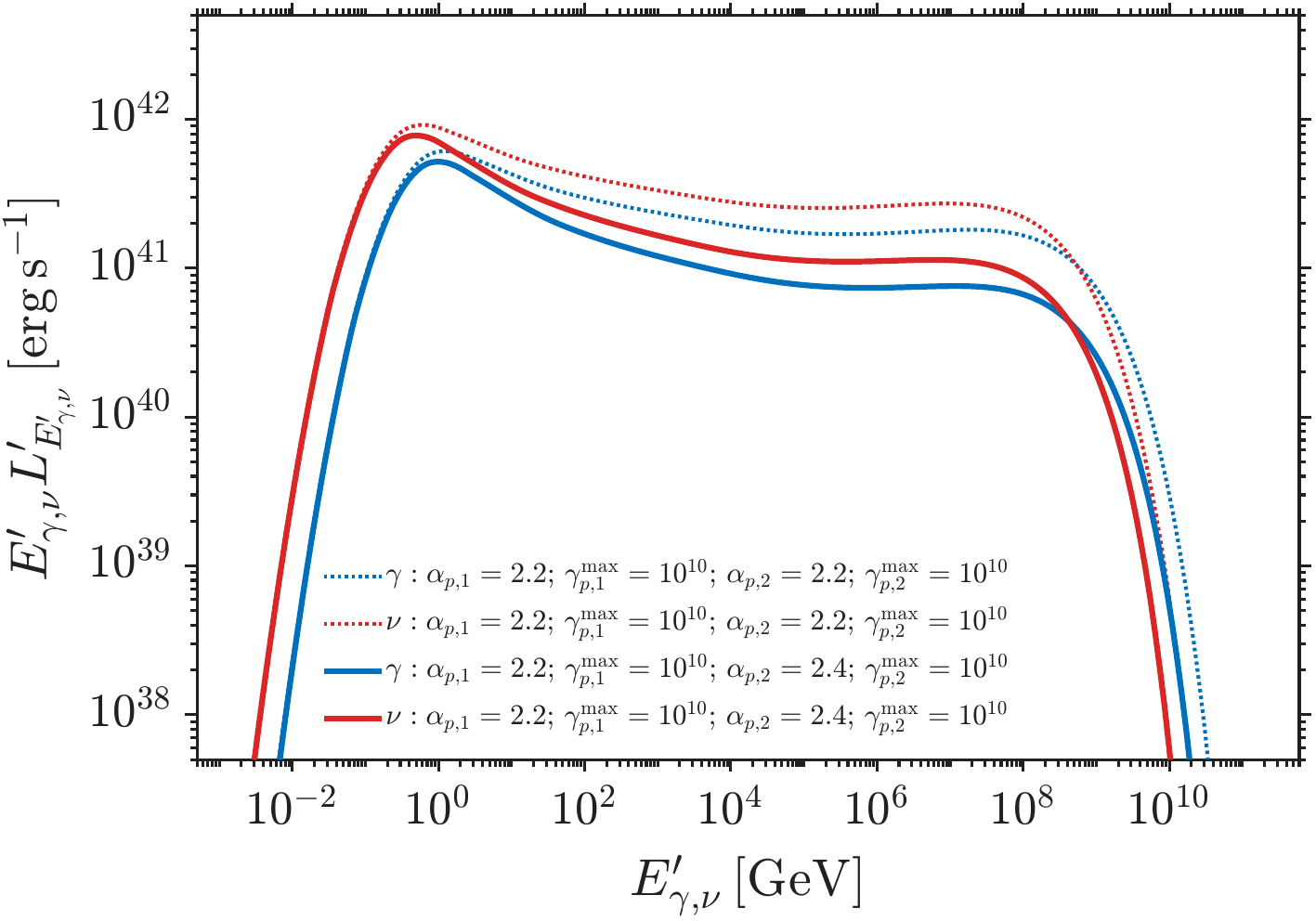}

\vspace{0.3cm}

\includegraphics[width=.47\textwidth]{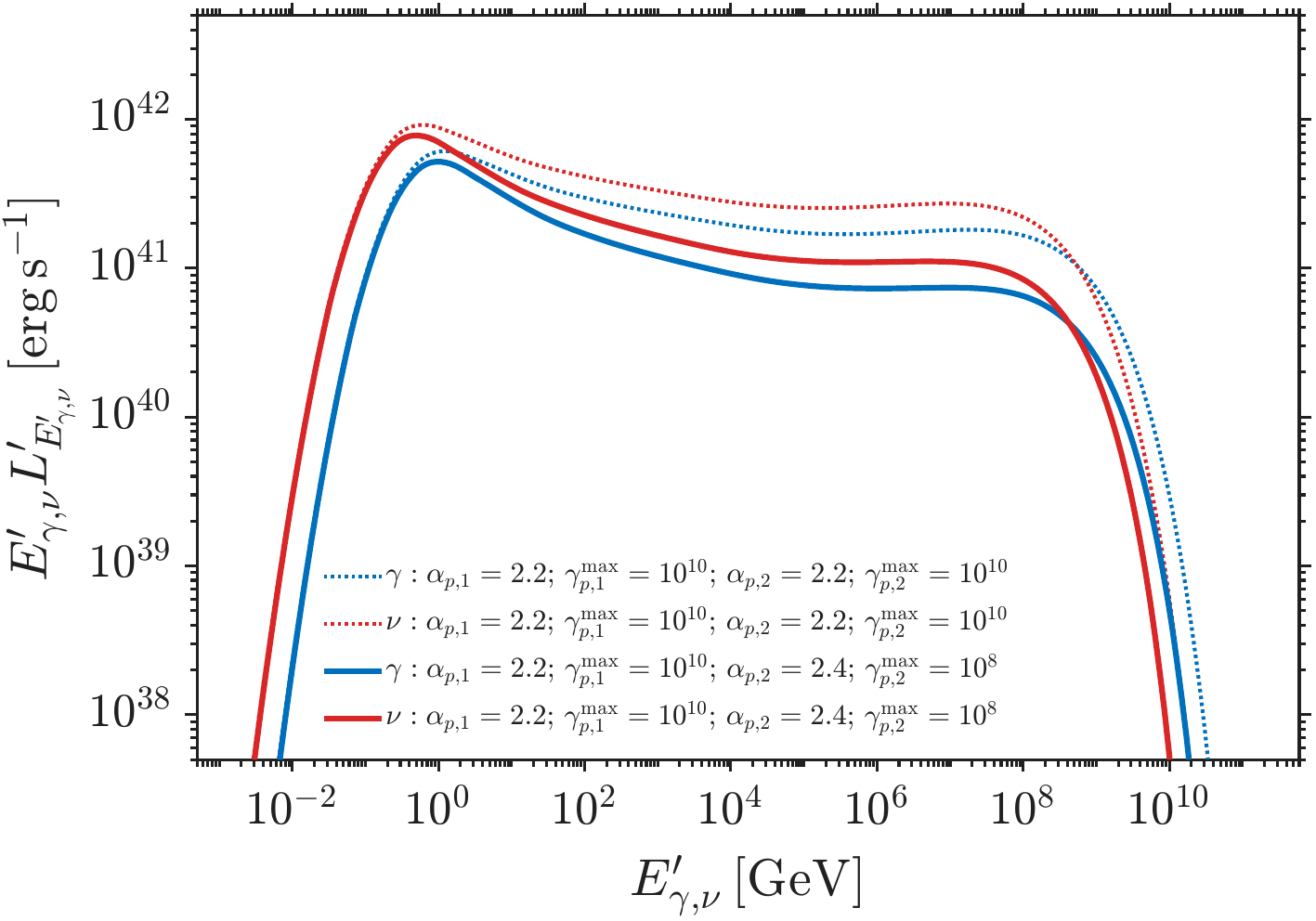}\hspace{0.4cm}
\includegraphics[width=.47\textwidth]{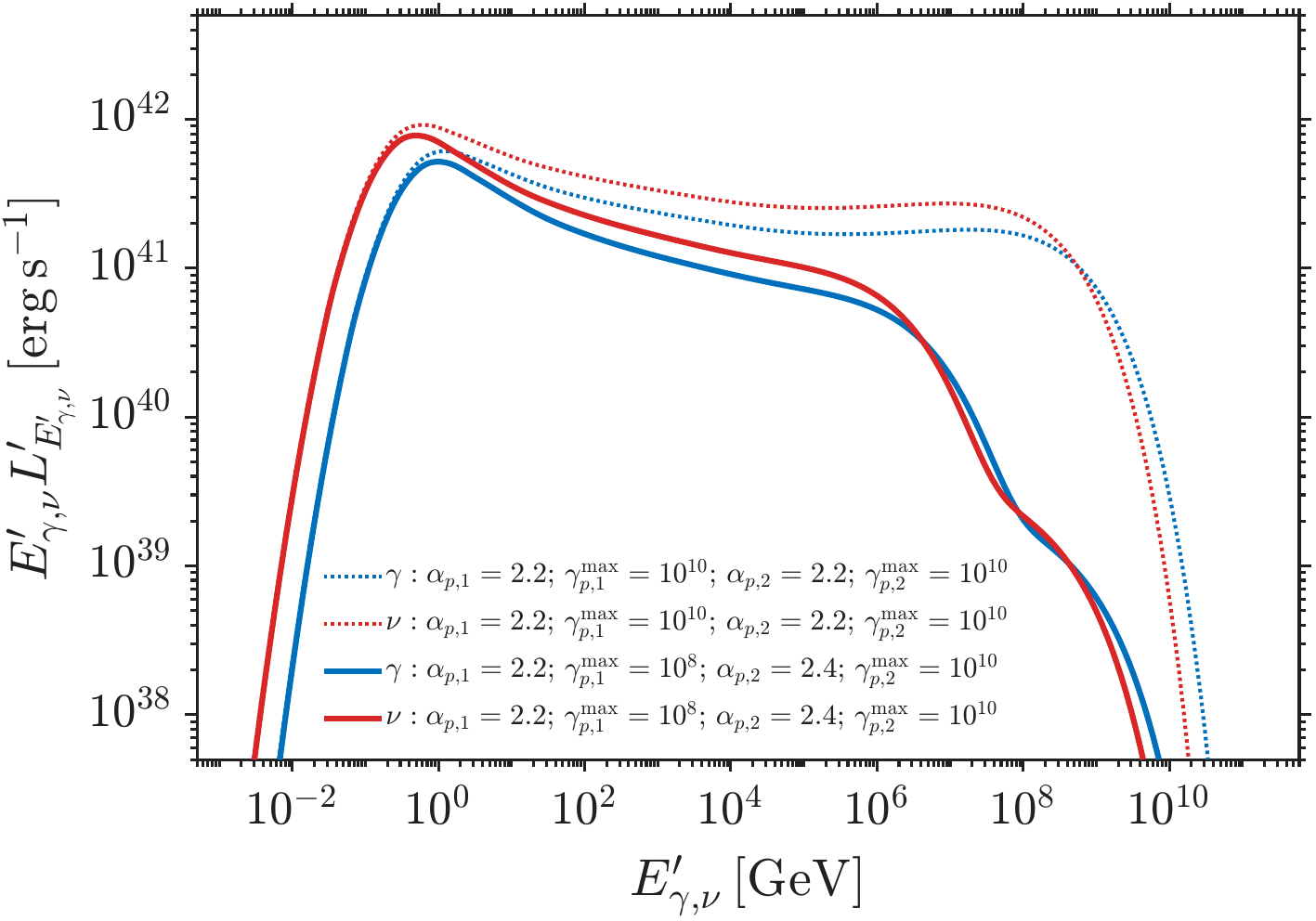}
\caption{Secondary photon ($E'_\gamma L'_{E'_{\gamma}}$, solid blue) and neutrino ($E'_\nu L'_{E'_{\nu}}$, solid red) spectral luminosities as a function of the energy $E'_{\gamma,\nu}$ for conservative/sub-benchmark asymmetric configurations. Top left: reduced cutoff in jet 2 ($\gamma_{p,2}^{\mathrm{max}} = 10^8$). Top right: softer spectrum in jet 2 ($\alpha_{p,2} = 2.4$). Bottom left: softer index and reduced cutoff in jet 2 ($\alpha_{p,2} = 2.4, \gamma_{p,2}^{\mathrm{max}} = 10^8$). Bottom right: softer index in jet 2 with reduced cutoff in jet 1 ($\alpha_{p,2} = 2.4, \gamma_{p,1}^{\mathrm{max}} = 10^{8}$). The benchmark model ($\alpha_{p,i}=2.2, \gamma_{p,i}^{\mathrm{max}}=10^{10}$) is shown as dotted curves in all panels.}
\label{fig:EnerLumLow}
\end{figure}

\begin{figure}[t]
\centering
\includegraphics[width=.47\textwidth]{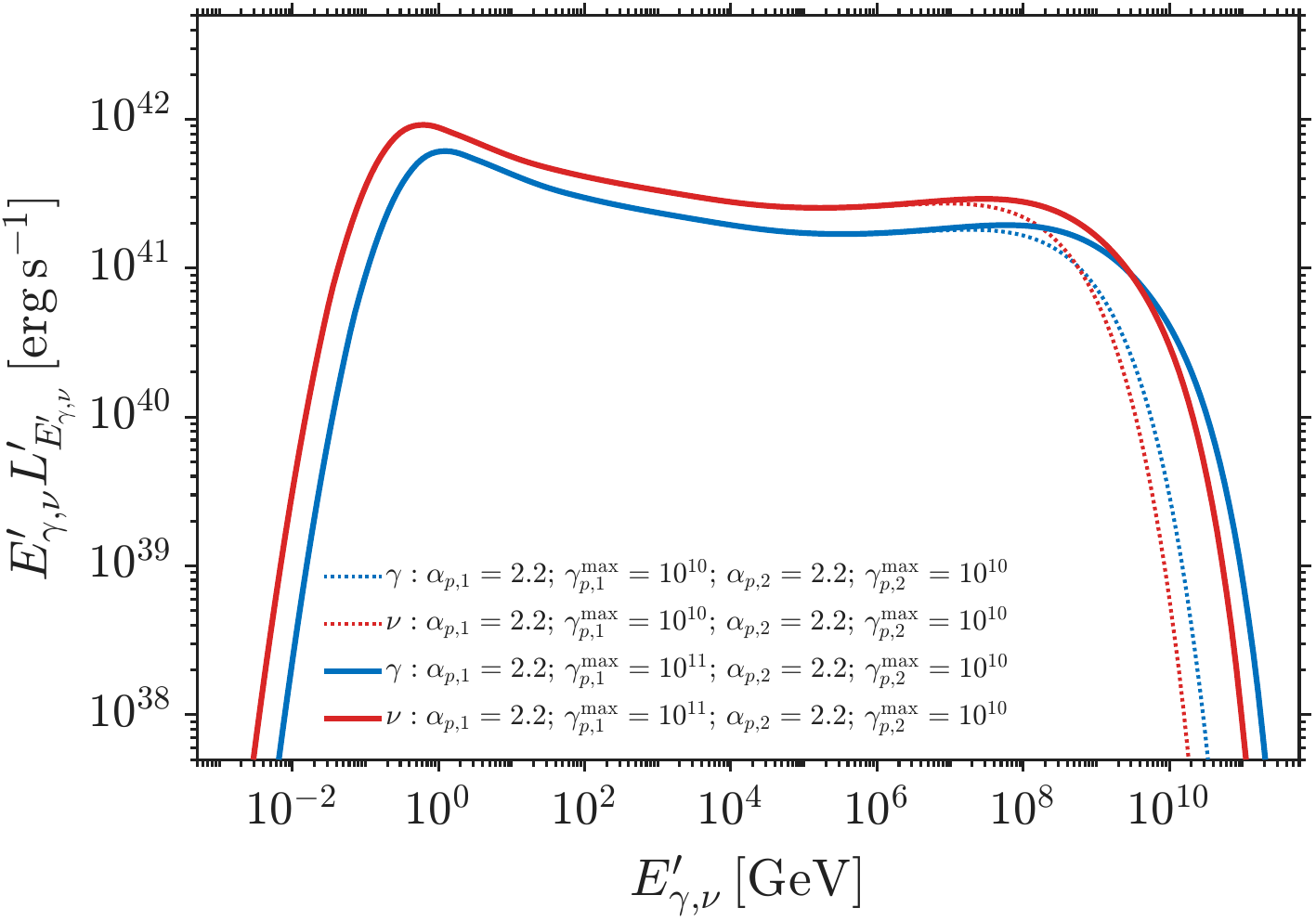}\hspace{0.4cm}
\includegraphics[width=.47\textwidth]{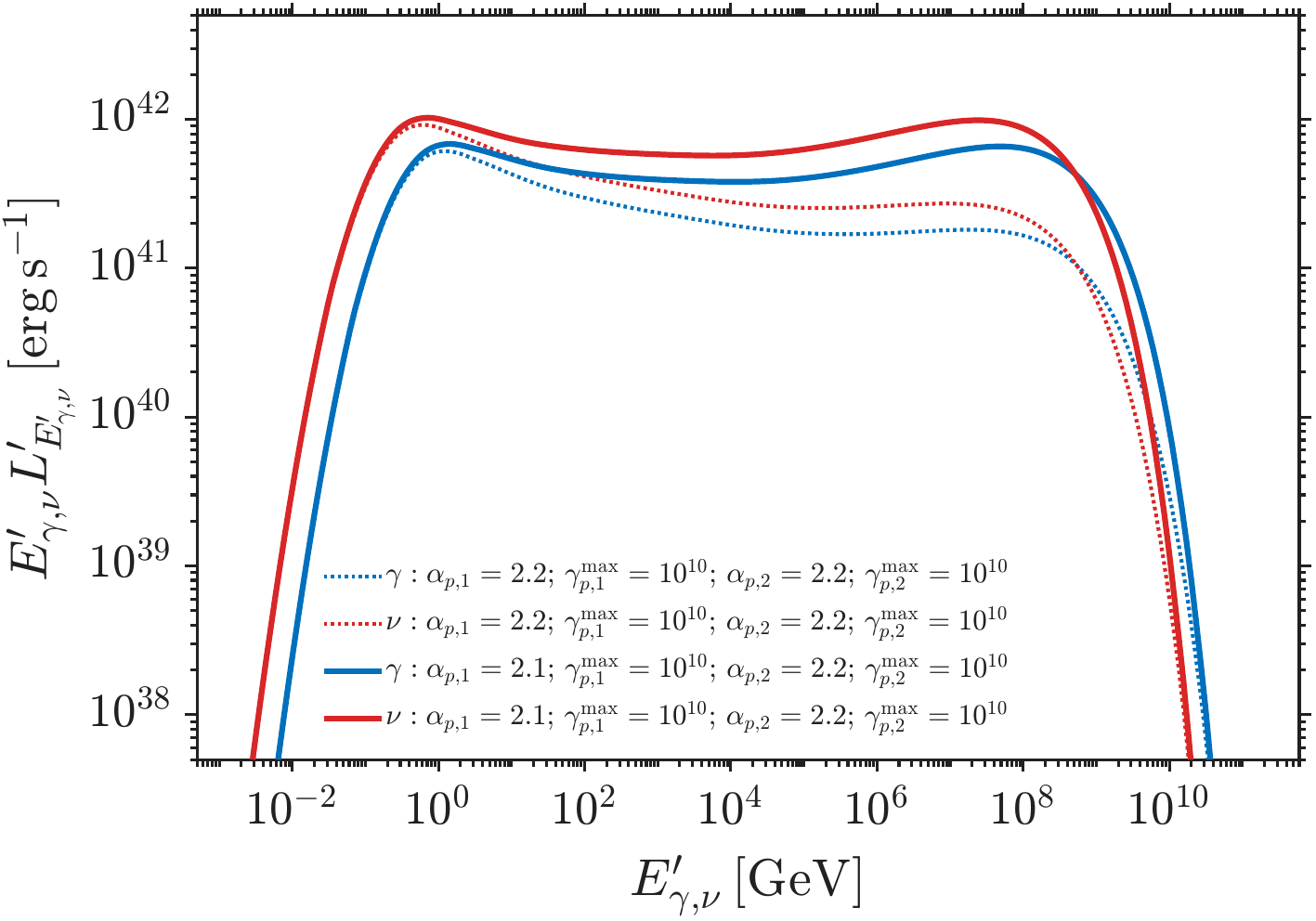}

\vspace{0.3cm}

\includegraphics[width=.47\textwidth]{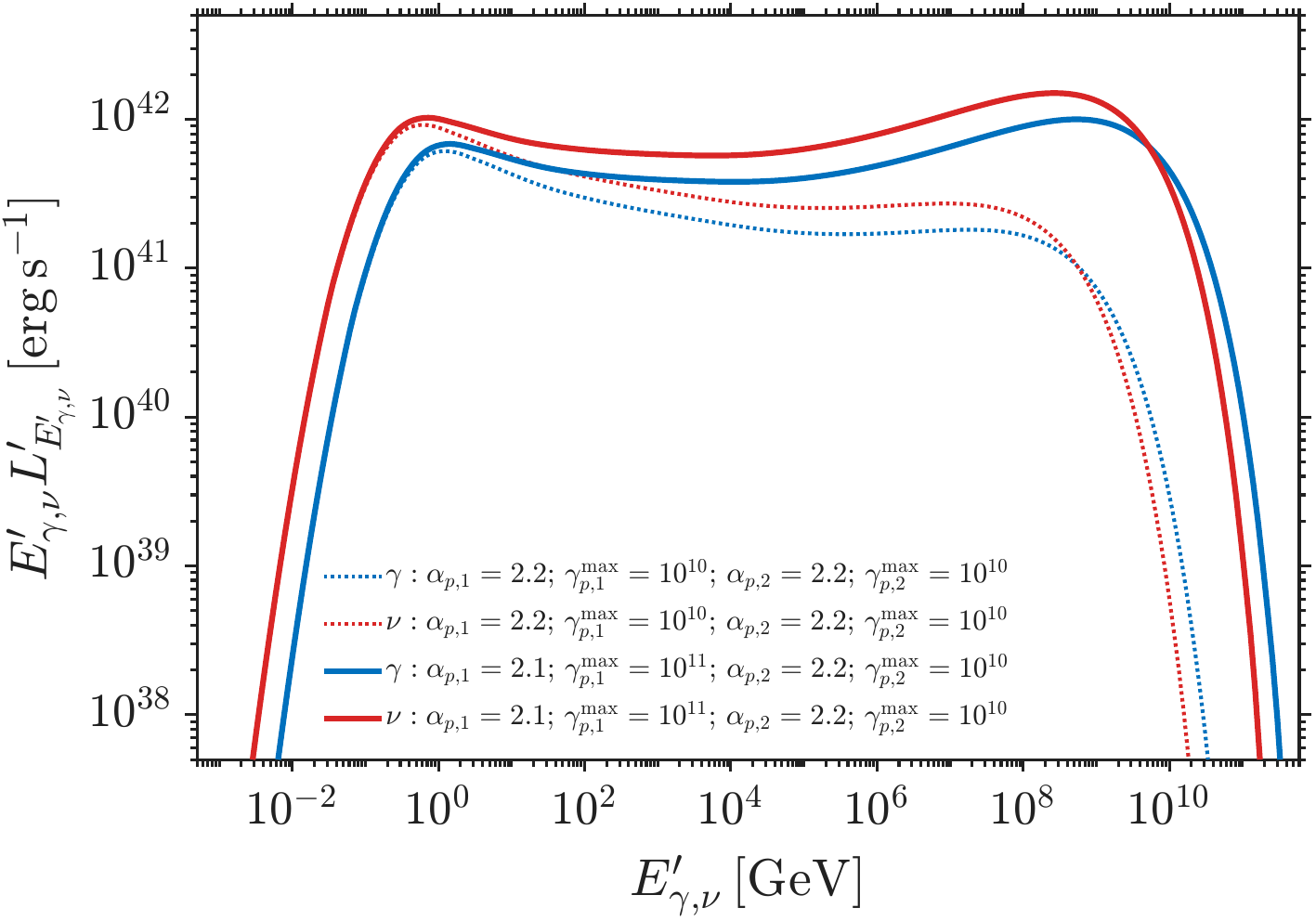}\hspace{0.4cm}
\includegraphics[width=.47\textwidth]{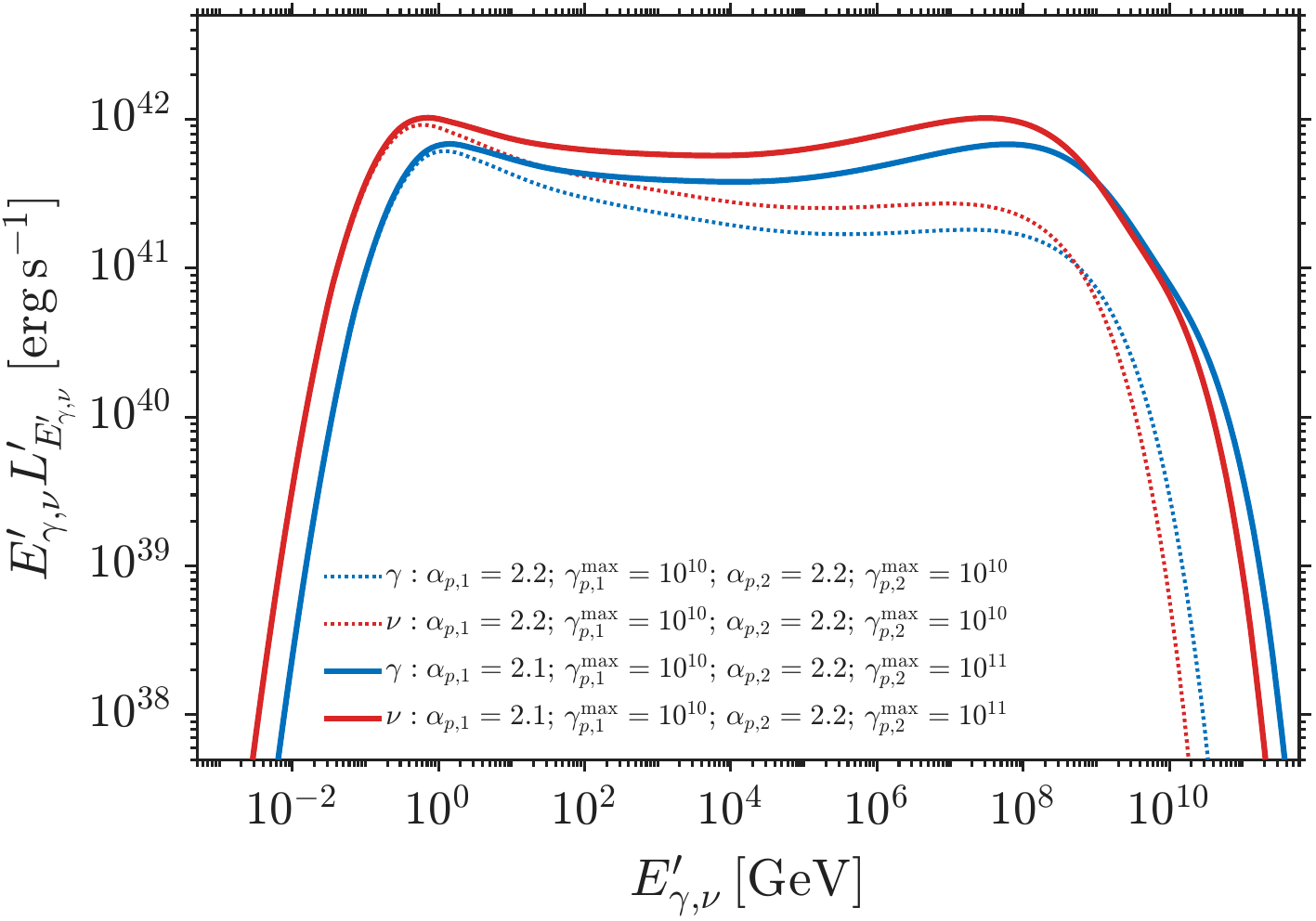}
\caption{Secondary photon ($E'_\gamma L'_{E'_{\gamma}}$, solid blue) and neutrino ($E'_\nu L'_{E'_{\nu}}$, solid red) spectral luminosities as a function of the energy $E'_{\gamma,\nu}$ for optimistic/super-benchmark asymmetric configurations. Top left: extended cutoff in jet 1 ($\gamma_{p,1}^{\mathrm{max}} = 10^{11}$). Top right: harder spectrum in jet 1 ($\alpha_{p,1} = 2.1$). Bottom left: harder index and extended cutoff in jet 1 ($\alpha_{p,1} = 2.1, \gamma_{p,1}^{\mathrm{max}} = 10^{11}$). Bottom right: harder index in jet 1 with extended cutoff in jet 2 ($\alpha_{p,1} = 2.1, \gamma_{p,2}^{\mathrm{max}} = 10^{11}$). The benchmark model ($\alpha_{p,i}=2.2, \gamma_{p,i}^{\mathrm{max}}=10^{10}$) is shown as dotted curves in all panels.}
\label{fig:EnerLumHigh}
\end{figure}

The main physical insights gained from analyzing the secondary spectra across the explored scenarios can be summarized as follows:

\begin{itemize}

\item {\bf Symmetric Variations (Fig.~\ref{fig:EnerLumSimm}):} When both jets undergo identical modifications, the secondary spectral profiles directly track the energy distribution of the underlying primary protons. Softening the proton spectral index ($\alpha_{p,i} = 2.4$, top left) redistributes power toward lower energies, producing a steeper secondary continuum that lies about one order of magnitude below the benchmark at $E'_{\gamma,\nu} \sim (10^6 - 10^8)\,\mathrm{GeV}$. Conversely, a harder index ($\alpha_{p,i} = 2.1$, top right) hardens the secondary emission, which rises up to a maximum around $E'_{\gamma,\nu} \sim 10^8\,\mathrm{GeV}$. This panel clearly illustrates a universal feature of hadronic interactions: owing to the logarithmic growth of the inelastic $pp$ cross section $\sigma_{pp}^{\rm inel}(s)$ with CM energy $s$ (Sec.~\ref{sec:cross_section_pp}), secondary spectra are systematically harder than the parent proton spectrum, causing $E'_{\gamma,\nu} L'_{E'_{\gamma,\nu}}$ to visibly rise toward high energies before the exponential cutoff. Altering the maximum Lorentz factor $\gamma_{p,i}^{\mathrm{max}}$ directly controls the high-energy cutoffs of the secondary spectra, which steepen above $E'_{\gamma,\nu} \sim 10^6\,\mathrm{GeV}$ and cut off at $E'_{\gamma,\nu} \sim (10^7 - 10^8)\,\mathrm{GeV}$ when $\gamma_{p,i}^{\mathrm{max}} = 10^8$ (bottom left), while extending up to $E'_{\gamma,\nu} \sim (10^{10} - 10^{11})\,\mathrm{GeV}$ in the case $\gamma_{p,i}^{\mathrm{max}} = 10^{11}$ (bottom right).

\item {\bf Asymmetric Degradation and Emergence of Spectral Features (Fig.~\ref{fig:EnerLumLow}):} Asymmetric configurations demonstrate both the overall resilience of the secondary output and the emergence of diagnostic spectral signatures driven by non-identical primary proton distributions. When jet 2 suffers a reduced maximum cutoff ($\gamma_{p,2}^{\mathrm{max}} = 10^8$, top left), the secondary emission undergoes a moderate drop above $E'_{\gamma,\nu} \sim 10^6\,\mathrm{GeV}$ due to the lack of high-energy target protons from jet 2. Softening the spectral index of jet 2 ($\alpha_{p,2} = 2.4$, top right) leads to a progressive, mild suppression of secondary luminosity toward higher energies, as fewer target protons are available for UHE $pp$ interactions. The visibility of these asymmetric effects depends on how spectral slope and energy cutoff are paired between the two jets. In the bottom-left panel ($\alpha_{p,1} = 2.2, \gamma_{p,1}^{\mathrm{max}} = 10^{10}$ vs $\alpha_{p,2} = 2.4, \gamma_{p,2}^{\mathrm{max}} = 10^8$), because the jet with the harder proton spectrum also possesses the higher cutoff, the jet 1 proton spectrum completely dictates the shape of secondary production across intermediate and UHE scales, effectively masking jet 2 proton spectral properties and yielding a smooth profile without distinct features. Inverting these parameters ($\alpha_{p,1} = 2.2, \gamma_{p,1}^{\mathrm{max}} = 10^8$ vs $\alpha_{p,2} = 2.4, \gamma_{p,2}^{\mathrm{max}} = 10^{10}$, bottom right) the secondary spectra reveal a prominent {\it smoothed step-like spectral break} around $E'_{\gamma,\nu} \sim (10^7 - 10^8) \,\mathrm{GeV}$. Physically, the harder proton population of jet 1 governs secondary production at intermediate energies but terminates at $\gamma_{p,1}^{\mathrm{max}} = 10^8$; above this threshold, the secondary emission is forced to track the softer primary proton population of jet 2 ($\alpha_{p,2} = 2.4$) extending up to $\gamma_{p,2}^{\mathrm{max}} = 10^{10}$, resulting in a characteristic spectrum with an intermediate knee followed by an ankle prior to the final cutoff.

\item {\bf Super-Benchmark Enhancements (Fig.~\ref{fig:EnerLumHigh}):} Optimistic asymmetric configurations ($\alpha_{p,i} = 2.1$ or $\gamma_{p,i}^{\mathrm{max}} = 10^{11}$) demonstrate how harder proton injections and higher energy cutoffs enhance secondary production across the UHE regime. Increasing the cutoff of jet 1 to $\gamma_{p,1}^{\mathrm{max}} = 10^{11}$ (top left) directly extends the secondary neutrino and photon spectra toward higher maximum energies. Adopting a harder spectral slope for jet 1 ($\alpha_{p,1} = 2.1$, top right) hardens the secondary emission profiles, significantly boosting fluxes at intermediate and high energies while retaining the baseline cutoff. Combining both optimistic parameters in jet 1 ($\alpha_{p,1} = 2.1, \gamma_{p,1}^{\mathrm{max}} = 10^{11}$, bottom left) produces a cumulative enhancement, yielding both a harder secondary spectrum and an extended energy cutoff. Finally, pairing the harder proton spectrum in jet 1 with the higher cutoff in jet 2 ($\alpha_{p,1} = 2.1, \gamma_{p,1}^{\mathrm{max}} = 10^{10}$ vs $\alpha_{p,2} = 2.2, \gamma_{p,2}^{\mathrm{max}} = 10^{11}$, bottom right) leads to an earlier, smoother onset of spectral softening above the maximum at $E'_{\gamma,\nu} \sim 10^8\,\mathrm{GeV}$, followed by a shoulder around $E'_{\gamma,\nu} \sim 10^{10}\,\mathrm{GeV}$. Since the configuration is conceptually similar to the one in Fig.~\ref{fig:EnerLumLow} (bottom right), this shoulder is an attenuated version of the step-like feature observed there: because the two proton cutoffs are much closer ($\gamma_{p,1}^{\mathrm{max}} = 10^{10}$ vs $\gamma_{p,2}^{\mathrm{max}} = 10^{11}$), the sharp step-like break is replaced by a significantly gentler and smoother secondary cutoff.

\end{itemize}

Since PeV-EeV gamma rays are subject to severe pair-production attenuation ($\gamma\gamma \to e^+ e^-$) on background photons (CMB, EBL, and radio, see Main Text) during cosmological propagation~\cite{dwekkrennrich,GalantiPiccinini2020}, while {\it neutrinos travel completely unattenuated} (see Main Text), such a composite step-like spectral feature in the UHE neutrino sky would provide a distinctive observational signature of the interaction of two distinct relativistic proton populations, as expected in the Cosmic Collider scenario.

\end{widetext}


\begin{thebibliography}{9}
\bibitem{footnoteSingleJet} Owing to the isotropy of the proton distributions in their rest frames, collisions among protons of the same jet (1--1 and 2--2) occur at large angles and are included. The 2--2 term is evaluated in the jet 2 frame, where both distributions are isotropic; since $\Gamma_{\rm rel} \simeq 1.3$, the two jet frames nearly coincide, and treating its secondaries as those of jet 1 affects the observed flux by $\lesssim 10\%$.
\bibitem{footnoteGammaRel} For symmetric jets, $\Gamma_{\rm rel} = \Gamma_b^2 - (\Gamma_b^2-1)\cos\theta_{\rm mis}$, yielding $\Gamma_{\rm rel} \simeq 1.3$ for $\Gamma_b = 15$ and $\theta_{\rm mis} = 3^\circ$. The target distribution is thus nearly isotropic in the jet 1 frame, and large values of $\sqrt{s}$ arise from wide-angle collisions between UHE protons rather than from the bulk relative motion of the jets.
\bibitem{footnoteRate} The prefactor $4\pi^2 = \frac{1}{2}(4\pi)(2\pi)$ arises as follows: integration over the solid angle of the isotropic projectile distribution yields $4\pi$, while integration over the azimuth of the angle-averaged target distribution around the projectile direction yields $2\pi$, leaving only the collision angle $\theta^{(1)}$; the factor $1/2$ avoids double counting, since each colliding pair appears twice in the double sum and integration over projectile and target. The spatial integration is performed over the host-galaxy element ${\rm d}r$, the interaction region being stationary in that frame; this choice leads to the ${\cal D}_D^3$ boost of Eq.~\eqref{obsSED}.
\bibitem{footnoteFlavor} Per-flavor sensitivities are multiplied by 3 for comparison with the all-flavor neutrino flux predicted here.
\bibitem{footnoteGRAND} Since GRAND sensitivity is reported as integrated fluence $F$~\cite{GRAND}, we convert it to point-source energy flux sensitivity assuming a nominal $10\,\mathrm{yr}$ observation time with a $10\%$ effective duty cycle, yielding an effective integration time $T_{\mathrm{eff}} = 1\,\mathrm{yr}$ ($F/T_{\mathrm{eff}}$).
\end{thebibliography}
\end{document}